\documentclass[prd,a4paper,twocolumn,superscriptaddress]{revtex4}
\usepackage{graphicx}
\begin{document}

\newcommand {\R}{{\mathcal R}}
\newcommand{\al}{\alpha}
\def\edth{\;\raise1.0pt\hbox{$'$}\hskip-6pt\partial\;}
\def\baredth{\;\overline{\raise1.0pt\hbox{$'$}\hskip-6pt
\partial}\;}
\def\bi#1{\hbox{\boldmath{$#1$}}}
\def\gsim{\raise2.90pt\hbox{$\scriptstyle
>$} \hspace{-6.4pt}
\lower.5pt\hbox{$\scriptscriptstyle
\sim$}\; }
\def\lsim{\raise2.90pt\hbox{$\scriptstyle
<$} \hspace{-6pt}\lower.5pt\hbox{$\scriptscriptstyle\sim$}\; }

\title{Measuring $\alpha$ in the Early Universe:\\
CMB Polarization, Reionization and the Fisher Matrix Analysis}

\author{G. Rocha}
\email[Electronic address: ]{graca@mrao.cam.ac.uk}
\affiliation{Astrophysics Group, Cavendish Laboratory,
Madingley Road, Cambridge CB3 0HE, United Kingdom}
\affiliation{Centro de Astrof\'{\i}sica da Universidade do Porto, R. das
Estrelas s/n, 4150-762 Porto, Portugal}
\author{R. Trotta}
\email[Electronic address: ]{roberto.trotta@physics.unige.ch}
\affiliation{D\'epartement de Physique Th\'eorique, Universit\'e de
Gen\`eve, 24 quai Ernest Ansermet, CH-1211 Gen\`eve 4, Switzerland}
\author{C.J.A.P. Martins}
\email[Electronic address: ]{C.J.A.P.Martins@damtp.cam.ac.uk}
\affiliation{Centro de Astrof\'{\i}sica da Universidade do Porto, R. das
Estrelas s/n, 4150-762 Porto, Portugal}
\affiliation{Department of Applied Mathematics and Theoretical Physics,
Centre for Mathematical Sciences,\\ University of Cambridge,
Wilberforce Road, Cambridge CB3 0WA, United Kingdom}
\affiliation{Institut d'Astrophysique de Paris, 98 bis Boulevard Arago,
75014 Paris, France}
\author{A. Melchiorri}
\email[Electronic address: ]{melch@astro.ox.ac.uk}
\affiliation{Department of Physics, Nuclear \& Astrophysics Laboratory,
University of Oxford, Keble Road, Oxford OX1 3RH, United Kingdom}
\affiliation{Universita' di Roma ``La Sapienza, Ple Aldo Moro 2, 00185, Rome, Italy}
\author{P. P. Avelino}
\email[Electronic address: ]{ppavelin@fc.up.pt}
\affiliation{Centro de F\'{\i}sica do Porto e
Departamento de F\'{\i}sica da Faculdade de Ci\^encias
da Universidade do Porto, Rua do Campo Alegre 687, 4169-007 Porto, 
Portugal}
\author{R. Bean}
\email[Electronic address: ]{rbean@astro.princeton.edu}
\affiliation{Department of Astrophysical Sciences, Princeton University,
Peyton Hall - Ivy Lane, Princeton NJ08544-1001, U.S.A.}
\author{P.T.P. Viana}
\email[Electronic address: ]{viana@astro.up.pt}
\affiliation{Centro de Astrof\'{\i}sica da Universidade do Porto, R. das
Estrelas s/n, 4150-762 Porto, Portugal}
\affiliation{Departamento de Matem\'atica Aplicada da Faculdade de Ci\^encias
da Universidade do Porto, Rua do Campo Alegre 687, 4169-007 Porto, Portugal}

\begin{abstract}
We present a detailed analysis of present and future Cosmic Microwave
Background constraints of the value of the fine-structure constant, $\alpha$.
We carry out a more detailed analysis of the WMAP first-year data, deriving
state-of-the-art constraints on $\alpha$ and discussing various other issues,
such as the possible hints for the running of the spectral index.
We find, at $95 \%$ C.L. that $0.95 <
\alpha_{\text{dec}} / \alpha_0 < 1.02$.
Setting $dn_S /dlnk=0$, yields 
$0.94< \alpha_{\text{dec}} / \alpha_0 < 1.01$ as previously reported.
We find that a lower value of $\alpha / \alpha_0$ makes a value
of $d n_S /dlnk = 0$ more compatible with the data.
We also perform a thorough Fisher Matrix Analysis (including both
temperature and polarization, as well as $\alpha$ and the optical depth
$\tau$), in order to estimate how future CMB experiments will be able to
constrain $\alpha$ and other cosmological parameters.
We find that Planck data alone can constrain $\tau$ with a accuracy of the order 4\% and that this constraint can be as small as 1.7\% for an ideal cosmic variance limited experiment. 
Constraints on $\alpha$ are of the order 0.3\% for Planck and can in principle be as small as 0.1\% using CMB data alone - tighter constraints will require further (non-CMB) priors.

\end{abstract}
\preprint{DAMTP-2002-53}
\maketitle

\section{Introduction}

The recent release of the Wilkinson Microwave Anisotropy Probe (WMAP)
first-year data \cite{Bennett,Hinshaw,Kogut,Verde} has pushed
cosmology into a new stage.
On one hand, it has quantitatively validated the broad features of
the `standard' cosmological model---the optimistically called
`concordance' model. But at the same time, it has also pushed
the borderline of research to new territory.
We now know that `dark components' make up the overwhelming majority
of the energy budget of the universe. Most of this is almost certainly
in some non-baryonic form,
for which there is at present no direct evidence or solid theoretical
explanation. One must therefore try to understand the nature of
this dark energy, or at least (as a first step) look for clues of
its origin.

It is clear that such an effort must be firmly grounded within
fundamental physics, and indeed that recent progress in fundamental physics
may shed new light on this issue. On the other hand, this is not a
one-way street. Cosmology and astrophysics are playing an increasingly
more important role as fundamental physics testbeds, since they
provide us with extreme conditions (that one has no hope of reproducing
in terrestrial laboratories) in which to carry out a plethora of tests and
search for new paradigms.
Perhaps the more illuminating example is that of multidimensional
cosmology. Currently preferred unification theories \cite{Polchinski,Damour1}
predict the existence of additional
space-time dimensions, which will have a number of possibly observable
consequences, including modifications in the gravitational laws on
very large (or very small) scales \cite{Will} and space-time variations of the
fundamental constants of nature \cite{Essay,Uzan}. 

There have been a number of recent reports of evidence
for a time variation of fundamental constants
\cite{Webb,Jenam,Murphy,Ivanchik}, and apart from their obvious 
direct impact if confirmed they are also crucial in a different, 
indirect way. They provide us with an important (and possibly even 
unique) opportunity to test
a number of fundamental physics models that might otherwise
be untestable. A case in point is that of string
theory \cite{Polchinski}. Indeed here the issue
is not \textit{if} such a theory predicts such variations, but
\textit{at what level} it does so, and hence if there is
any hope of detecting them in the near future (or if we have done
it already). Indeed, it has been argued \cite{Damour1,Damour2}. 
that even the results of Webb and collaborators \cite{Webb,Jenam,Murphy}
may be hard to explain in the simplest, best motivated models where the 
variation of alpha is driven by the spacetime variation of a very light 
scalar field. Playing devil's advocate, one could certainly conceive
that cosmological observations of this kind could one day prove
string theory wrong.

The most promising case, and the one that has been the subject of
most recent work (and speculation), is that of the fine-structure constant
$\alpha$, for which some fairly strong statistical
evidence of time variation at redshifts
$z\sim2-3$ already exists \cite{Webb,Jenam,Murphy}, together with weaker
(and somewhat more controversial) evidence from geophysical tests
using the Oklo natural nuclear reactor \cite{Fujii}.
Interesting and quite tight constraints can also be derived from
local laboratory tests \cite{Marion}, and indeed this is a context
where improvements of several orders of magnitude
can be expected in the coming years.

On the other hand, the theoretical expectation in the simplest, best motivated model is that
$\alpha$ should be a non-decreasing function of time \cite{Damour,Santiago,Barrow}. This is based on rather general and
simple assumptions, in particular that the cosmological dynamics of
the fine-structure constant is governed by a scalar filed whose
behavior is akin to that of a dilaton. If this is so, then it
is particularly important to try to constrain it at earlier
epochs, where any variations relative to the present-day value
should therefore be larger. In this regard, note that one of the 
interpretations of the Oklo results \cite{Fujii} is that $\alpha$ was
\textit{larger} at the Oklo epoch (effectively $z\sim0.1$) than today, whereas
the quasar results \cite{Webb,Jenam,Murphy} indicate that $\alpha$ was 
smaller at $z\sim2-3$ than today. Both results are not necessarily incompatible,
since they refer to two different cosmological epochs, and hence comparing
them necessarily requires specifying not only a \textit{background}
cosmological model but also a model for the variation of
the fine-structure constant with redshift, $\alpha=\alpha(z)$. However, if
both results are validated by future experiments, then the above 
theoretical expectation must clearly be wrong (with clear implications for
both the dilaton hypothesis and on a wider scale), which would be a perfect 
example of using astrophysics to learn about fundamental physics.

Cosmic microwave background (CMB)
anisotropies provide an ideal way of measuring the
fine-structure constant at high redshift, being mostly sensitive to the
epoch of decoupling, $z \sim 1100$
(one could also envisage searching for spatial variations at the
last scattering surface \cite{Sigu}). Here we continue our ongoing
work in this area \cite{Old,Avelino,Martins}, and particularly 
extend our most recent analysis \cite{Martinsw} of the
WMAP first-year data, providing updated constraints on
the value of $\alpha$ at decoupling, studying some crucial degeneracies
with other cosmological parameters and discussing what improvements can
be expected with forthcoming datasets.

We emphasize that in previous (pre-WMAP) work, CMB-based
constraints on $\alpha$ were obtained with the help of additional cosmological
datasets and priors. This has raised some eyebrows among skeptics, as different
datasets could possibly have different systematic errors that are
impossible to control and could conceivably conspire to produce the
results we quoted (statistically consistency with the value of $\alpha$
at decoupling being the same as today's, though with a slight preference
towards smaller values). Here, by contrast, we will present results of an
analysis of the WMAP dataset alone (we will only briefly discuss what
happens when other datasets are added). We also discuss how these
constraints can be improved in the future, especially when more precise CMB
polarization data is available. In particular, we show that
the existence of an early reionization epoch is a significant
help in further constraining $\alpha$, and indeed the prospects for measuring
$\alpha$ from the CMB are much better than if the optical depth $\tau$
was much smaller. 

Moreover, now that CMB polarization data is available, there
are two approaches one can take. One is to treat CMB temperature and
polarization as different datasets, and carry out independent analyses
(and, more to the point, cosmological parameter estimations), to check
if the results of the two are consistent. The other one is to combine the
two datasets, thus getting smaller errors on the parameters.
We will show that there are advantages to both approaches, and also that
the combination of the two can often by itself break many of the cosmological
degeneracies that plague this kind of analysis pipeline. On the other hand,
we will also show that in ideal circumstances (\textit{id est}, a cosmic
variance limited experiment) CMB polarization is much better than CMB
temperature in determining cosmological parameters. This result is not
new, and it is of course somewhat obvious, but it has never been quantified
in detail as will be done below.

On the other hand, because cosmic variance limited experiments are expensive
and experimentalists work with limited budgets, it is important to provide
detailed forecasts for future experiments. We provide detailed forecalsts
for the full (4-year) WMAP dataset, as well as for ESA's Planck Surveyor
(to be launched in 2007). It will be shown that Planck is almost
cosmic variance limited (taken into account the range of multipoles covered by this instrument) when it comes to CMB temperature, but far from
it for CMB polarization. Again this was previously known, but had not been
quantified. This, and the intrinsic superiority of CMB polarization
in measuring cosmological parameters, are therefore arguments for a post-Planck,
polarization-dedicated experiment.

\section{CMB Temperature and Polarization}

Following \cite{zaldarriaga1,kosowsky,waynehu1,waynehu2},
one can describe the CMB anisotropy field as a 2x2, $I_{ij}$, intensity tensor which is a function of direction on the sky $\vec{n}$ and 2 other directions perpendicular to $\hat{\bi{n}}$ which define its components 
${\hat{\bi{e_1}},\hat{\bi{e_2}}}$. 
The CMB radiation is expected to be polarised due to Thomson scattering of temperature anisotropies at the time when CMB photons last scattered. 
Polarised light is traditionally described via the Stokes parameters, $Q,U,V$, where $Q=(I_{11}-I_{22})/4$ and $U=I_{12}/2$, while the temperature 
anisotropy is given by $T=(I_{11}+I_{22})/4$ and $V$ can be ignored since it describes circular polarization which cannot be generated through Thomson scattering. 
Both Q and U depend of the choice of coordinate system in that they transform under a right handed rotation in the plane perpendicular to direction $\hat{\bi{n}}$ by an angle $\psi$ as:
\begin{eqnarray}
Q^{\prime}&=&Q\cos 2\psi  + U\sin 2\psi  \nonumber \\  
U^{\prime}&=&-Q\sin 2\psi  + U\cos 2\psi \,,
\label{QUtrans} 
\end{eqnarray}
where ${\bf \hat{e_1}}^{\prime}=\cos \psi{\bf \hat{e_1}}+\sin\psi{\bf 
\hat{e_2}}$ 
and ${\hat{\bf e_2}}^{\prime}=-\sin \psi{\bf \hat{e_1}}+\cos\psi{\bf 
\hat{e_2}}$.

In order to compute the rotationally invariant power spectrum a general method to analyse polarization over the whole sky is required. This is so because the calculation of the power spectrum involves the superposition of the different modes contributing to the perturbations. While it is simple to compute $Q$ and $U$ in the coordinate system where the wavevector defining the perturbation is aligned with the z axis, it is more complicated to do so when superimposing the different modes since one needs to rotate $Q$ and $U$ to a common coordinate frame before this superposition is done, and only in the small scale limit does this rotation have a simple expression \cite{uros}.

Most of the literature on the polarization of the CMB uses three alternative representations based on either 
the Newman-Penrose spin-weight 2 harmonics \cite{zaldarriaga1}, or a
coordinate representation of the tensor spherical 
harmonics \cite{kamionkowski1,kamionkowski2}, or the coordinate-independent, 
projected symmetric trace free (PSTF) tensor valued 
multipoles \cite{challinor}.
Here we follow the first by expanding the polarization in the sky in terms of spin-weighted harmonics which form a basis for tensor functions in the sky.
One starts by defining two other quantities $(Q\pm iU)'$: 
\begin{equation}
(Q\pm iU)'(\hat{\bi{n}})=e^{\mp 2i\psi}(Q\pm iU)(\hat{\bi{n}}).
\end{equation}
These quantities are then expanded in the appropriate
spin-weighted basis:
\begin{eqnarray}
T(\hat{\bi{n}})&=&\sum_{lm} a_{T,lm} Y_{lm}(\hat{\bi{n}}) \nonumber \\
(Q+iU)(\hat{\bi{n}})&=&\sum_{lm} 
a_{2,lm}\;_2Y_{lm}(\hat{\bi{n}}) \nonumber \\
(Q-iU)(\hat{\bi{n}})&=&\sum_{lm}
a_{-2,lm}\;_{-2}Y_{lm}(\hat{\bi{n}})\,,
\label{Pexpansion}
\end{eqnarray}
where $Y_{lm}$ are the spherical harmonics and ${}_2Y_{lm}$ are the so-called spin-2 spherical harmonics, which form a complete and orthonormal basis for spin-2 functions. A function  $\;_sf(\theta,\phi)$
defined on the sphere has spin-s if under
a right-handed rotation of ($\hat{{\bi e}}_1$,$\hat{{\bi e}}_2$)
by an  angle $\psi$  it transforms as 
$\;_s f^{\prime}(\theta,\phi)=e^{-is\psi}\;_sf(\theta,\phi)$.
Here we are interested in the polarizatin of the CMB which is a quantity of spin $\pm 2$.

$Q$ and $U$ are defined at a given direction $\hat{\bi{n}}$
with respect to the spherical coordinate system $(\hat{{\bf e}}_\theta,
\hat{{\bf e}}_\phi)$.
The expansion coefficients for the polarization variables
satisfy $a_{-2,lm}^*=a_{2,l-m}$. For temperature the relation is
$a_{T,lm}^*=a_{T,l-m}$, where

\begin{eqnarray}
a_{T,lm}&=&\int d\Omega\; Y_{lm}^{*}(\hat{\bi{n}}) T(\hat{\bi{n}})
\nonumber  \\  
a_{2,lm}&=&\int d\Omega \;_2Y_{lm}^{*}(\hat{\bi{n}}) (Q+iU)(\hat{\bi{n}})
\nonumber  \\
\nonumber \\  
a_{-2,lm}&=&\int d\Omega \;_{-2}Y_{lm}^{*}(\hat{\bi{n}}) (Q-iU)(\hat{\bi{n}}) 
\nonumber  \\  
\label{alm}
\end{eqnarray}
Usually one considers the following linear combinations:
\begin{eqnarray}
a_{E,lm}=-(a_{2,lm}+a_{-2,lm})/2 \nonumber \\ 
a_{B,lm}=i(a_{2,lm}-a_{-2,lm})/2\,.
\label{aeb}
\end{eqnarray}

The following rotationally invariant quantities then define the power spectra
\begin{eqnarray}
C_{Tl}&=&{1\over 2l+1}\sum_m \langle a_{T,lm}^{*} a_{T,lm}\rangle 
\nonumber \\
C_{El}&=&{1\over 2l+1}\sum_m \langle a_{E,lm}^{*} a_{E,lm}\rangle 
\nonumber \\
C_{Bl}&=&{1\over 2l+1}\sum_m \langle a_{B,lm}^{*} a_{B,lm}\rangle 
\nonumber \\
C_{Cl}&=&{1\over 2l+1}\sum_m \langle a_{T,lm}^{*}a_{E,lm}\rangle \,,
\label{Cls}
\end{eqnarray}
in terms of which,
\begin{eqnarray}
\langle a_{T,l^\prime m^\prime}^{*} a_{T,lm}\rangle&=&
C_{Tl} \delta_{l^\prime l} \delta_{m^\prime m} \nonumber \\
\langle a_{E,l^\prime m^\prime}^{*} a_{E,lm}\rangle&=&
C_{El} \delta_{l^\prime l} \delta_{m^\prime m} \nonumber \\
\langle a_{B,l^\prime m^\prime}^{*} a_{B,lm}\rangle&=&
C_{Bl} \delta_{l^\prime l} \delta_{m^\prime m} \nonumber \\
\langle a_{T,l^\prime m^\prime}^{*} a_{E,lm}\rangle&=&
C_{Cl} \delta_{l^\prime l} \delta_{m^\prime m} \nonumber \\
\langle a_{B,l^\prime m^\prime}^{*} a_{E,lm}\rangle&=&
\langle a_{B,l^\prime m^\prime}^{*} a_{T,lm}\rangle=0\,. 
\label{stat}
\end{eqnarray}

In real space one describes the polarization field in terms of two quantities that are scalars under rotation, E and B modes, defined as:
\begin{eqnarray}
\tilde{E}(\hat{{\bi n}})=\sum_{lm}\left[{(l+2)! \over (l-2)!}\right]^{1/2}
a_{E,lm}Y_{lm}(\hat{{\bi n}}) \nonumber \\  
\tilde{B}(\hat{\bi n})=\sum_{lm}\left[{(l+2)! \over (l-2)!}\right]^{1/2}
a_{B,lm}Y_{lm}(\hat{\bi n}) \,.
\label{EBexpansions} 
\end{eqnarray} 
These quantities are closely related to the rotationally invariant Laplacian 
of $Q$ and $U$. In multipole space the relation is as follows
\begin{equation}
a_{(\tilde{E},\tilde{B}),lm}=\left[{(l+2)! \over (l-2)!}\right]^{1/2}
a_{(E,B),lm}.
\label{eblm}
\end{equation}

While E remains unchanged under parity transformation, B changes its sign
(similar to the behaviour of  electric and magnetic fields).
This decomposition is also useful because the B mode is a direct signature 
of the presence of a background of gravitational waves, since it cannot 
be produced by density fluctuations \cite{zaldarriaga1,kamionkowski1}, 
Many models of inflation predict a significant gravity wave background.
These tensor fluctuations generated during inflation have their largest effects
on large angular scales and add in quadrature to the fluctuations generated by 
scalar modes. Whilst recent WMAP results placed limits on the amplitude of these tensor
modes one still lacks an experimental evidence for the presence of a stochastic background 
of gravitational waves. As mentioned above the detection of the pseudo-scalar field B 
would provide invaluable information about Inflation in that they reflect the presence 
of such a background. Therefore to fully characterize the CMB anisotropies only four 
power spectra are needed--those for T,E,B and the cross-correlation between T and E.
(Given that B has the opposite parity of E and T their cross-correlations with B vanishes.)

The first detection of polarization of the CMB was due to the DASI experiment \cite{dasi}, and more 
recently the WMAP experiment \cite{Kogut} has measured the TE cross-correlation power spectrum.
An important result from these is the existence of reionization at larger redshifts then expected 
from the Gunn-Petterson through, an issue that we will discuss at length below.

\section{The CMB, $\alpha$ and $\tau$}

The reason why the CMB is a good probe of
variations of the fine-structure constant is that these alter the
ionisation history of the universe \cite{steen,Kap,Old,vsl}.
The dominant effect is a change
in the redshift of recombination, due to a shift in the energy levels
(and, in particular, the binding energy) of Hydrogen. The Thomson
scattering cross-section is also changed for all particles, being
proportional to $\alpha^2$. A smaller effect
(which has so far been neglected) is expected to come from a change 
in the Helium abundance \cite{Trotta:2003xg}.

Increasing $\alpha$ increases the redshift
of last-scattering, which corresponds to a smaller sound horizon. Since the
position of the first Doppler peak ($\ell_{peak}$)
is inversely proportional to the sound horizon at last scattering, 
increasing $\alpha$ will produce a larger $\ell_{peak}$ \cite{Old}.
This larger redshift of last scattering also has the additional effect
of producing a larger early ISW effect, and hence a larger amplitude
of the first Doppler peak \cite{steen,Kap}. Finally, an increase
in $\alpha$ decreases the high-$\ell$ diffusion damping (which is
essentially due to the finite thickness of the last-scattering surface),
and thus increases the power on very small scales. 
These effects have been implemented in a modified CMBFAST algorithm which 
allows a varying $\alpha$ parameter \cite{Old,Avelino}. These follow
the extensive description given in \cite{steen,Kap}, with one important
exception that will be discussed below.

\begin{figure}
\includegraphics[width=3in,angle=0]{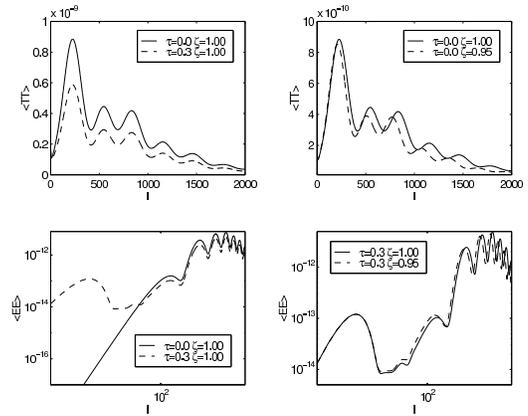}
\caption{\label{figcells} Contrasting the effects of varying
$\alpha$ (right)and reionization (left) on the CMB temperature (top) and polarization (bottom).
Here $\zeta=\al_{dec}/\al_0$. See the text for further details.}
\end{figure}

Fig.~\ref{figcells} illustrates the effect of $\alpha$ and $\tau$
on the CMB temperature and polarization power spectra. The CMB power
spectrum is, to a good approximation, insensitive to {\it how}
$\alpha$ varies from last scattering to today. Given the existing
observational constraints, one can therefore calculate the effect
of a varying $\alpha$ in both the temperature and polarization
power spectra by simply assuming two values for $\alpha$, one at
low redshift (effectively today's value, since any variation of
the magnitude of \cite{Webb} would have no noticeable effect) and
one around the epoch of decoupling, which may be different from
today's value. (In earlier works \cite{steen,Kap,Old,Battye} one
assumed a constant value of $\alpha$ throughout, \textit{id est}
the values at reionization and the present day were always the same.)

For the CMB temperature, reionization simply
changes the amplitude of the acoustic peaks, without affecting
their position and spacing (top left panel); a different value of
$\alpha$ at the last scattering, on the other hand, changes both
the amplitude and the position of the peaks (top right panel). 

\begin{figure}
\includegraphics[width=3.5in]{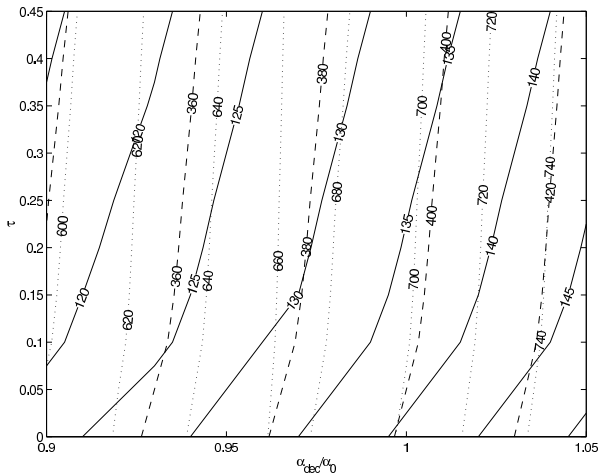}
\caption{\label{figpeaks} The separation in $\ell$ between the
reionization bump and the first (solid lines), second (dashed) and
third (dotted) peaks in the polarization spectrum, as a function
of $\alpha$ at decoupling and $\tau$. A (somewhat idealized) description of
how $\alpha$ and $\tau$ can be measured using CMB polarization. }
\end{figure}

The outstanding effect of reionization is to introduce a bump in the
polarization spectrum at large angular scales (lower left panel).
This bump is produced well after decoupling (at much lower
redshifts), when $\alpha$, if varying, is much closer to the
present day's value. If the value of
$\alpha$ at low redshift is different from that at decoupling, the
peaks in the polarization power spectrum at small angular scales
will be shifted sideways, while the reionization bump on large
angular scales won't (lower right panel). It follows that by
measuring the separation between the normal peaks and the bump,
one can measure both $\alpha$ and $\tau$, as illustrated in Fig.
\ref{figpeaks}. Thus we expect that the existence of an early
reionization epoch will, when more accurate cosmic microwave
background polarization data is available, lead to considerably
tighter constraints on $\alpha$.

A possible concern with the interpretation of our results is related 
to the implicit assumption of a sharp transition on the value of 
$\alpha$ happening sometime between recombination and the epoch 
of reionization. Hence, it is crucial to understand if this is a 
valid approximation. Appart from the value of $\alpha$ at the time 
of recombination the knowledge of its value at two other 
epochs is relevant as far as the CMB anisotropies are concerned. One 
such epoch is the period \textit{just before} recombination which is 
very important for the damping of CMB anisotropies on small angular scales. 
The other period is the epoch of reionization. In this work we effectively
assume that $\alpha$ is equal to $\alpha_{rec}$ before recombination and 
to $\alpha_0$ at the reionization epoch. 

A value of $\alpha$ different  
from $\alpha_0$ at the epoch of reionization will affect the CMB 
anisotropies through a change in the optical depth $\tau$, \textit{once a 
single cosmological model is assumed}. However, it is also well known that
$\tau$ is itself dependent on the cosmological model through its cosmological
parameters ($\Omega_m$ and $\Omega_\Lambda$ for example) as well as on the
cosmological density perturbations (in our case through the initial
power spectrum) \cite{pedrolidle}
. The exact dependence is difficult to determine
since there are several astrophysical uncertainties related to a number
of relevant non-linear physical processes which affect the accuracy
of reionization models. In general, this problem is solved by treating
$\tau$ as a free parameter (independent of the other cosmological
parameters and initial power spectrum), which accounts for the
relatively poor knowledge of the dependence of $\tau$ on the
cosmological model and in our case on the uncertainty about the exact
value of $\alpha$ during the reionization epoch. Hence, we find that
provided we treat $\tau$ as a free parameter the lack of a precise
knowledge of value of $\alpha$ during the epoch of reionization will 
not affect our results. 
In the present work, we assume that the universe was completely reionized in a
relatively small redshift interval (sudden reionization). A more refined modelling 
of the reionization history is not yet required by WMAP data, but will be necessary 
at noise levels appropriate for Planck and beyond \cite{Bruscoli02,Hu03,K03,Holder03}.
On the more practical side, there are of course
observational constraints on the value of $\alpha$ at redshifts of a
few \cite{Webb,Jenam,Murphy}, indicating that at that epoch the possible
changes relative to the present day are already very small (and would not
be detectable, on their own, through the CMB due to cosmic variance).

The knowledge of the value of $\alpha$ before 
recombination is also crucial for the details of the damping of 
small scale CMB anisotropies. 
Let's assume that the variation of $\alpha$ around the time of recombination is given by some functional, $f$:
$$ 
\frac{\alpha}{\alpha_{rec}}=f\left(\frac{1+z}{1+z_{rec}}\right)
$$
One can determine the dependence of the Silk damping scale \cite{kolbturner}
$$ 
R_S=\left(\int_{0}^{t_{dec}(\alpha)} dt \frac{\lambda_{\gamma}(\alpha)}{R^{2}(t)} \right)^\frac{1}{2}
$$
(where, $\lambda_{\gamma}$, is the photon mean free path)
on this functional $f$ and determine $\alpha_{eff}$ (relevant for the damping of
CMB anisotropies) as the constant value of $\alpha$ that gives the same 
Silk damping scale as the variable one. Even though we did not treat 
$\alpha_{eff}$ as another parameter in the present investigation (this
will  be done in future work) we expect that our constraints on 
$\alpha_{rec}$ should also be valid (to a good approximation) for $\alpha_{eff}$. 
This means that we are already able to constrain a 
combination of both $\alpha$ and $f$ at the time of recombination. 
Also, we see that we may be able to rule out particular models for the 
time variation of $\alpha$ on the basis of the details of such variation, 
even if the value of $\alpha$ at the time of recombination is not ruled 
out by our analysis.

Finally, we must emphasize that the effects discussed above are
direct effects of an $\alpha$ variation, and that indirect effects are
usually present as well since any variation of $\alpha$ is necessarily
coupled with the dynamics of the Universe \cite{Mota}. 
In this paper we take a pragmatic
approach and say that, since the CMB is quite insensitive to the details of
$\alpha$ variations from decoupling to the present day,
\textit{we do not in fact need to specify a redshift dependence for this
variation}---although we could have specified one if we so chose.

The price to pay would be that, since this coupling is very
dependent on the particular model we consider we would end up with 
very model-dependent constraints. Therefore, at this stage, and given the
lack of detailed and well-motivated cosmological models for $\alpha$
variations we prefer to focus on model-independent constraints, and hence
do not attempt to include this extra degree of freedom in our analysis.
Nevertheless, given some model-independent constraints one can always
translate them into constraints on the parameters of one's favourite
model. In fact we expect
that some models will be ruled out on the basis of the indirect effect
of a variation of $\alpha$ on the dynamics of the Universe rather than
the direct effects we described above. This is actually a simpler case
in which only the modifications to the background evolution
($a(t)$) would need to be taken into account in order to test the
model, with the direct effects of a varying $\alpha$ being negligible.

We conclude this section by emphasising that although a more detailed
analysis taking into account the expected variation of $\alpha$ with
time (and its direct and indirect implications for CMB anisotropies) for
specific models is certainly possible, our more general work can
easily be used to impose very strong constraints to more complex
varying $\alpha$ theories once the relevant variables are computed. 

\section{Up-to-date CMB constraints on $\alpha$ with WMAP}

We compare the recent WMAP temperature and cross-polarization dataset
with a set of flat cosmological models adopting the likelihood
estimator method described in \cite{Verde}. 
We restrict the analysis to flat universes. The models are
computed through a modified version of the CMBFAST code with
parameters sampled as follows: physical density in cold dark
matter $0.05 < \Omega_ch^2 < 0.20$ (step $0.01$), physical density
in baryons $0.010 < \Omega_bh^2 < 0.028$ (step $0.001$), 
$0.500 < \Omega_{\Lambda} < 0.950$ (step $0.025$), 0$0.900 <
\alpha_{\text{dec}} / \alpha_0 <1.050$ (step $0.005$). Here $h$ is
the Hubble parameter today, $H_0 \equiv 100h$ km s$^{-1}$
Mpc$^{-1}$ (determined by the flatness condition once the above 
parameters are fixed), while $\alpha_{\text{dec}}$ ($\alpha_0$) is 
the value of the fine structure constant at decoupling (today). 
We also vary the optical depth $\tau$ in the range $0.06-0.30$ 
(step $0.02$), the scalar spectral index of primordial fluctuations 
$0.880 < n_s < 1.08$ (step $0.005$) and its running 
$-0.15 < dn_s/dlnk < 0.05$ (step $0.01$) both evaluated at 
$k_0=0.002 Mpc^{-1}$ . 
We don't consider gravity waves or iso-curvature modes 
since these further modifications are not required by the WMAP data
(see e.g. \cite{Spergel}).
A different model for the dark energy from a cosmological constant
could also change our results, but again, is not suggested
by the WMAP data (see e.g. \cite{mmot}). An extra background
of relativistic particles is also well constrained by Big Bang 
Nucleosynthesis (see e.g. \cite{bhm}) and it will not be considered
here.

\begin{figure}
\includegraphics[width=3in]{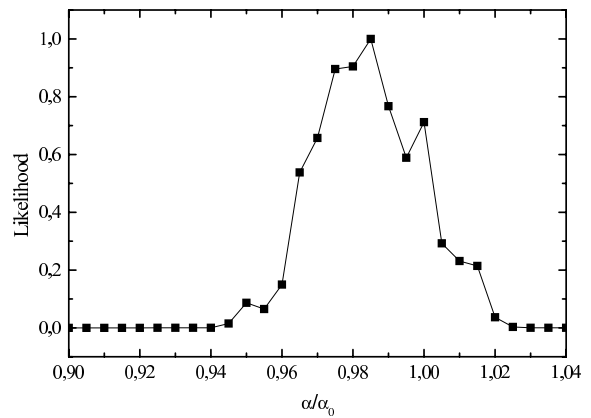}
\caption{\label{figalpha}
Likelihood distribution function for variations in the fine structure
constant obtained by an analysis of the WMAP data (TT+TE, one-year).}
\end{figure}

\begin{figure}
\includegraphics[width=3in]{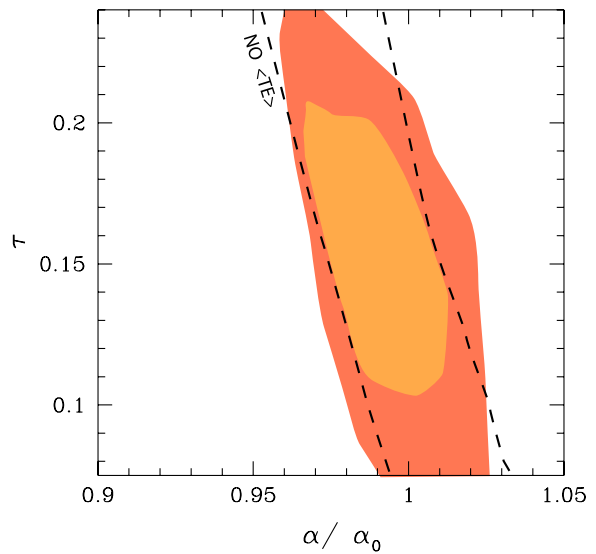}
\caption{\label{figalphavstau}
$2-$D Likelihood contour plot in the $\alpha / \alpha_0$ vs
$\tau$ plane for $2$ analysis: $<TT>$ only and $<TT>$+$<TE>$.
As we can see, the inclusion of polarization data, breaks the
degeneracy between these $2$ parameters.}
\end{figure}

\begin{figure}
\includegraphics[width=3in]{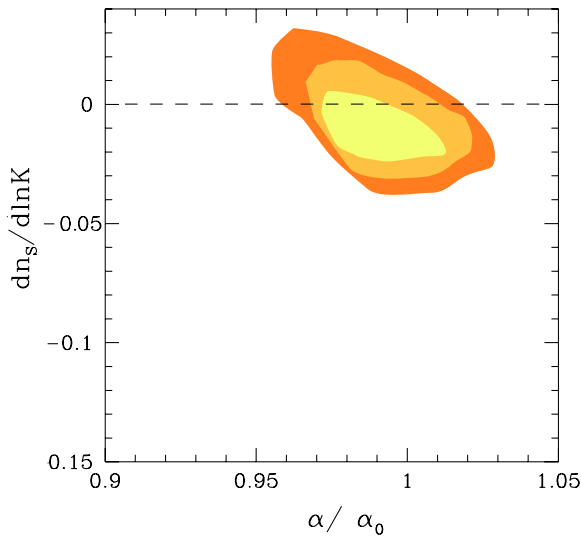}
\caption{\label{figalphavsdn}
$2-$D Likelihood contour plot in the $\alpha / \alpha_0$ vs
$dn_S / dlnk$ plane ($<TT>$+$<TE>$ one year). A zero scale dependence,
as expected in most of the inflationary models, is more consistent
with a value of $\alpha / \alpha_0 < 1$}
\end{figure}

The likelihood distribution function for $\alpha_{\text{dec}} / \alpha_0$,
obtained after marginalization over the remaining parameters, is
plotted in Figure \ref{figalpha}. We found, at $95 \%$ C.L. that $0.95 <
\alpha_{\text{dec}} / \alpha_0 < 1.02$, improving previous bounds,
(see \cite{Martins}) based on CMB and complementary datasets.
Setting $dn_S /dlnk=0$, yields 
$0.94< \alpha_{\text{dec}} / \alpha_0 < 1.01$ as already reported in 
(see \cite{Martinsw}).

It is interesting to consider the correlations between a 
$\alpha / \alpha_0$ and the other parameters in order to see 
how this modification to the standard model can change
our conclusions about cosmology.  

In Figure \ref{figalphavstau} we plot the $2-D$ likelihood contours in the
$\alpha / \alpha_0$ vs the optical depth  $\tau$ for $2$ different
analysis: using the temperature only WMAP data and including
the $<TE>$ cross spectrum temperature-polarization data.
As we can see, there is a clear degeneracy between these $2$
parameters if one consider just the $<TT>$ spectrum: 
increasing the optical depth, allows for an higher value
of the spectral index $n_S$ and a lower value of 
$\alpha / \alpha_0$ (again, see \cite{Martins}).
As we can see from  Figure  \ref{figalphavstau}, the inclusion of the
$<TE>$ data, is already able to partially break the degeneracy
between $\tau$ and $\alpha / \alpha_0$. However, as we explain below,
more detailed measurements of the polarization spectra are needed
to fully break this degeneracy.

One of the most unexpected results from the WMAP data is the hint 
for a scale-dependence of the spectral index $n_S$
(see e.g.  \cite{peiris}, \cite{kkmr}).
Such dependence is not predicted to be detectable in most of the viable 
single field inflationary model and, if confirmed, will therefore have strong
 consequences on the possibilities of reconstructing the inflationary 
potential. In Figure \ref{figalphavsdn} we plot a $2-D$ likelihood 
contour in the $\alpha / \alpha_0$ vs $d n_S /dlnk$ plane. As we can see,
a lower value of $\alpha / \alpha_0$ makes a value
of $d n_S /dlnk \sim 0$ more compatible with the data.
As already noticed in \cite{bms}, a modification of the recombination
scheme can therefore provide a possible explanation for the
high value of $dn_S /dlnk$ compatible with the WMAP data.


\section{Fisher Matrix Analysis Setup}

In our previous work \cite{Martins}, a Fisher Matrix Analysis was
carried out, using only the CMB temperature, in order to estimate
the precision with which cosmological parameters can be reconstructed 
in future experiments. Here we extend this analysis by including also
E-polarization measurements as well as the TE cross-correlation.
We consider the planned Planck satellite (HFI only) and an ideal experiment
which would measure both temperature and polarization to the cosmic variance
limit (in the following, 'CVL experiment`) for a range of multipoles, $l$, up to 2000. For illustration purposes, and
particularly as a way of checking that our method is producing credible
results, we will also present the FMA analysis for WMAP, and compare
the corresponding `predictions' with existing results.

The Fisher Matrix is a measure of the width and shape of the likelihood 
around its maximum and as such can also provide useful insight into the
degeneracies among different parameters, with minimal computational effort.
For a review of this technique, see \cite{fisher,tegmark,jungman1,jungman2,knox,zaldarriaga2,bond,efstathiou1,efstathiou2}.
In what follows we will present a brief description of our analysis procedure,
emphasizing the aspects that are new. We refer the reader to our previous 
work \cite{Martins} for further details.

We will assume that cosmological models are characterized by the 8 
dimensional parameter set
\begin{equation}
{\bf \Theta} = (\Omega_b h^2, \Omega_m h^2, \Omega_\Lambda h^2,
\R, n_s, Q, \tau, \al)\,,
\end{equation}
where  $\Omega_m = \Omega_c +\Omega_b$ is the energy density in
matter, $\Omega_\Lambda$ the energy density due to a
cosmological constant, and $h$ is a dependent
variable which denotes the Hubble parameter today, $H_0 \equiv 100h$ 
km s$^{-1}$ Mpc$^{-1}$. The quantity $\R \equiv \ell_{\rm ref}  / \ell$ is the
`shift' parameter (see \cite{melch:01,Bowen:01}
and references therein), which gives the position of the 
acoustic peaks with respect to a flat, $\Omega_\Lambda = 0$ reference model,
The shift parameter $\R$ depends on $\Omega_m$, on the curvature
$\Omega_{\kappa} \equiv 1 - \Omega_{\Lambda} - \Omega_m - \Omega_{\rm rad}$
through
\begin{eqnarray}
\label{eq:def_r}
\R &=& 2 \left( 1 - \frac{1}{\sqrt{1 + z_{\rm dec} }} \right) \nonumber \\
&& \times \frac{\sqrt{| \Omega_{\kappa}| }}{ \Omega_m}
\frac{1}{\chi(y)}
\left[ \sqrt{\Omega_{\rm rad} +
\frac{\Omega_m}{1 + z_{\rm dec} } } - \sqrt{\Omega_{\rm rad}} \right] ,
\end{eqnarray}
where $z_{\rm dec} $ is the redshift of decoupling, $\Omega_{\rm rad}$ is
the energy parameter due to radiation
($\Omega_{\rm rad}=4.13 \cdot 10^{-5}/h^2$ for photons and 3 neutrinos) and
\begin{eqnarray}
\label{eq:ydef}
y &=& \sqrt{|\Omega_{\kappa}|}\int_0^{z_{\rm dec} } \, dz\\
&& {[\Omega_{\rm rad} (1+z)^4 +
\Omega_m(1+z)^3+\Omega_{\kappa}(1+z)^2+\Omega_{\Lambda}]^{-1/2}}. \nonumber
\end{eqnarray}
The function $\chi(y)$ depends on the curvature of the universe and is
$y$, $\sin(y)$ or $\sinh(y)$ for flat, closed or open models,
respectively.
Inclusion of the shift parameter $\R$ into our set of parameters takes into
account the geometrical degeneracy between $\omega_\Lambda$ and
$\omega_m$ \cite{efstathiou1}. With our choice of the parameter set, $\R$ is an
independent variable, while the Hubble parameter $h$ becomes a dependent
one.

$n_s$ is the scalar spectral index and \mbox{$Q
= < \ell (\ell + 1) C_\ell > ^{1/2}$} denotes the overall
normalization, where the mean is taken over the multipole range
$2 \leq \ell \leq 2000$.

We assume  purely adiabatic initial conditions and
we do not allow for a tensor contribution.
In the FM approach, the likelihood distribution ${\cal L}$ for the
parameters $\bf \Theta$ is expanded to quadratic order around its maximum ${\cal L}_{m}$.
We denote this maximum likelihood (ML) point by $\bf \Theta_0$ and call
the corresponding model our ``ML model'', with parameters 
$\omega_b = 0.0200$, $\omega_m = 0.1310$,
$\omega_\Lambda = 0.2957$ (and $h = 0.65$), $\R = 0.9815$, $n_s =
1.00$, $Q = 1.00$, $\tau=0.20$ and $\al/\al_0=1.00$.
For the value of $z_{\rm dec}$ (which is
weakly dependent on $\omega_b$ and $\omega_{\rm tot}$) we
have used the fitting formula from \cite{HuandSugiyama}.
For the ML model we have $z_{\rm dec} = 1115.52$.

As mentioned above we also present the FMA for the WMAP best fit model as the fiducial model. (ie, 
$\omega_b = 0.0200$, $\omega_m = 0.1267$,
$\omega_\Lambda = 0.2957$, $\R = 0.9636$, $n_s =
0.99$, $Q = 1.00$, $\tau=0.17$ and $\al/\al_0=1.00$.)
Note that we will discuss cases with and without reionization (in
the latter case $\tau=0.0$) as well as with and without varying $\alpha$.

To compute the derivatives of the power spectrum with respect to a particular cosmological parameter one varies the considered parameter and keeps fixed the value of the others to their ML value. In particular given that we are not constraining our analysis to the case of a flat universe a variation in $\mathcal R$ is considered with all the other parameters fixed and equal to their ML value. Therefore such variation implies a variation of the dependent parameter $h$.

\begin{table}
\caption{\label{exppar} Experimental parameters for WMAP and Planck 
(nominal mission). Note that we express the 
sensitivities in $\mu$K.}
\begin{ruledtabular}
\begin{tabular}{|l|ccc|ccc|}
& \multicolumn{3}{c|}{WMAP}& \multicolumn{3}{c|}{Planck} \\\hline
$\nu$ (GHz) &  $40$  &  $60$  & $90$ & 
               $100$ &  $143$ & $217$  \\
$\theta_c$ (arcmin)&  
	$31.8$ & $21.0$  & $13.8$ &
	$10.7$ & $8.0$ & $5.5$ \\
$\sigma_cT$ ($\mu$K)  & 
	$19.8$  & $30.0$ & $45.6$ &   
	$5.4$  & $6.0$  & $13.1$  \\
$\sigma_{cE}$ ($\mu$K)  &
    $28.02$ & $42.43$ & $64.56$ &
    $n/a$  & $11.4$  & $26.7$  \\
$w^{-1}_c \cdot 10^{15}$ (K$^2$ ster) & 
	$33.6$  & $33.6$  & $33.6$  & 
	$0.215$ & $0.158$ & $0.350$  \\
$\ell_c$            & $254$ & $385$  & $586$  &
	$757$ & $1012$ & $1472$ \\\hline
$\ell_{\rm max}$ & \multicolumn{3}{c|}{$1000$} & \multicolumn{3}{c|}{$2000$} \\
$f_{\rm sky}$    & \multicolumn{3}{c|}{$0.80$} & \multicolumn{3}{c|}{$0.80$} \\
\end{tabular}
\end{ruledtabular}
\end{table}
%
In our previous work \cite{Martins} we assumed a flat fiducial model,
and differentiating around it requires computing open and closed models, 
which are calculated using different numerical techniques.
We have found that this can
limit the accuracy of the FMA. Here we instead
differentiate around a slightly closed model (as preferred by
WMAP) with $\Omega_{\rm{tot}} = 1.01$ to avoid extra sources of
numerical inaccuracies. We refer to \cite{Martins} for a
detailed description of the numerical technique used. The
experimental parameters used for the Planck analysis are in Table
\ref{exppar}. Note that we use the first 3 channels of the Planck High
Frequency Instrument (HFI) only. Adding the 3 channels of Planck's
Low Frequency Instrument leaves the expected errors unchanged:
therefore they can be used for other important tasks such as foreground 
removal and various consistency checks, leaving the HFI channels for direct
cosmological use. For the CVL experiment, we set the experimental noise to 
zero, and we use a total sky coverage $f_{\rm{sky}} = 1.00$. Although this is
never to be achieved in practice, the CVL experiment illustrates
the precision which can be obtained \textit{in principle} from CMB
temperature and E-polarization measurements.

If the errors $\Theta - \Theta_{0}$ about the ML model are small, a quadratic expansion around this ML leads to the expression,
\begin{equation}
 {\cal L} \approx {\cal L}_m \exp\left[-{1 \over 2} \label{eq:3}
\sum_{ij}{F}_{ij}
\delta \Theta_i \delta \Theta_j\right]
\end{equation}
where $F_{ij}$ is the Fisher matrix, given by derivatives of 
the CMB power spectrum with respect to the parameters ${\bf \Theta}$ 

In \cite{Martins} we computed the {\em Fisher information matrix} using temperature information alone. In this case for each $l$ a derivative of the temperature power spectrum with respect to the parameter under consideration is computed and then summed over all $l$, weighted by $Cov^{-1}(\hat{C}^{2}_{Tl})=\Delta C_\ell^2$, that is
\begin{equation}
F_{ij} = \sum_{\ell=2}^{\ell_{\rm max}} \frac{1}{\Delta C_\ell^2}
	       \frac{\partial C_\ell}{\partial \Theta_i}\frac{\partial C_\ell}{\partial \Theta_j} \vert_{\bf \Theta_0}\,.
\label{eq:fisher}
\end{equation}
The quantity $\Delta C_\ell$ is the standard deviation 
on the estimate of $C_{\ell}$: 
\begin{equation}
\Delta C_\ell^2 = \frac{2}{(2 \ell + 1) f_{\rm sky} }
( C_\ell + {B}_{\ell}^{-2})^2\,;
\end{equation}
the first term is the cosmic variance, arising from the fact
that we exchange an ensemble average with a spatial average. The second
term takes into account the expected error of the experimental
apparatus \cite{knox,efstathiou1}, 
\begin{equation}
{B}_{\ell}^2 = \sum_c w_c e^{- \ell (\ell +1)/\ell_c^2}\,.
\end{equation}
The sum runs over all channels of the experiment, with
the inverse weight per solid angle
$w_c^{-1} \equiv (\sigma_c \theta_c)^{-2}$ and 
$\ell_c \equiv \sqrt{8 \ln2}/\theta_c$, where
$\sigma_c$ is the sensitivity (expressed in $\mu$K) and 
$\theta_c$ is the FWHM of the beam (assuming a 
Gaussian profile) for each channel. Furthermore,
we can neglect the issues arising from point sources,
foreground removal and galactic plane contamination
assuming that once they have been taken into account
we are left with a ``clean'' fraction of the sky given by $f_{\rm sky}$.

In the more general case with polarization information included, instead of a single derivative we have a vector of four derivatives with the weighting given by the the inverse of the covariance matrix \cite{zaldarriaga1},
\begin{equation}
F_{ij}=\sum_l \sum_{X,Y}{\partial \hat{C}_{Xl} \over \partial \Theta_i}
{\rm Cov}^{-1}(\hat{C}_{Xl}\hat{C}_{Yl}){\partial \hat{C}_{Yl} \over \partial \Theta_j}\,,
\end{equation}
where $F_{ij}$ is the Fisher information or curvature 
matrix as above, $Cov^{-1}$ is the inverse of the covariance matrix,
$\Theta_i$ are the cosmological parameters we want to 
estimate and $X,Y$ stands for $T$ (temperature), $E,B$ (polarization modes), or
$C$ (cross-correlation of the power spectra for $T$ and $E$). 
For each $l$ one has to invert the covariance matrix and sum over $X$ and $Y$.
The diagonal terms of the covariance matrix between the different estimators 
are given by
\begin{eqnarray}
{\rm Cov }(\hat{C}_{Tl}^2)&=&\frac{2}{(2 \ell + 1) f_{\rm sky} }(\hat{C}_{Tl}+
{B}_{T\ell}^{-2})^2
\nonumber \\
{\rm Cov }(\hat{C}_{El}^2)&=&\frac{2}{(2 \ell + 1) f_{\rm sky}}(\hat{C}_{El}+
{B}_{P\ell}^{-2})^2
\nonumber \\
{\rm Cov }(\hat{C}_{Cl}^2)&=&\frac{1}{(2 \ell + 1) f_{\rm sky}}\left[\hat{C}_{Cl}^2+
(\hat{C}_{Tl}+{B}_{T\ell}^{-2})
(\hat{C}_{El}+{B}_{P\ell}^{-2})\right] \nonumber \\
{\rm Cov }(\hat{C}_{Bl}^2)&=&\frac{2}{(2 \ell + 1) f_{\rm sky} }(\hat{C}_{Bl}+
{B}_{P\ell}^{-2})^2.
\end{eqnarray}
The non-zero off diagonal terms are
\begin{eqnarray}
{\rm Cov }(\hat{C}_{Tl}\hat{C}_{El})&=&\frac{2}{(2 \ell + 1) f_{\rm sky} }\hat{C}_{Cl}^2
\nonumber \\
{\rm Cov }(\hat{C}_{Tl}\hat{C}_{Cl})&=&\frac{2}{(2 \ell + 1) f_{\rm sky}}\hat{C}_{Cl}
(\hat{C}_{Tl}+{B}_{T\ell}^{-2})
\nonumber \\
{\rm Cov }(\hat{C}_{El}\hat{C}_{Cl})&=&\frac{2}{(2 \ell + 1) f_{\rm sky}}\hat{C}_{Cl}
(\hat{C}_{El}+{B}_{P\ell}^{-2})\,,
\end{eqnarray}
where ${B}_{T\ell}^{-2}={B}_{\ell}^{-2}$ as above and ${B}_{P\ell}^2$ is
obtained using a similar expression but with the experimental specifications 
for the polarized channels.

For Gaussian fluctuations, the covariance matrix is then given by the
inverse of the Fisher matrix, $C = F^{-1}$ \cite{bond}. 
The $1\sigma$ error on the parameter $\Theta_i$ with all other parameters
marginalised is then given by $\sqrt{C_{ii}}$. If all other parameters
are held fixed to their ML values, the standard deviation on parameter $\Theta_i$ 
reduces to $\sqrt{1/F_{ii}}$ (conditional value). 
Other cases, in which some of the parameters are
held fixed and others are being marginalized over can easily be worked out.

In the case in which all parameters are being
estimated jointly, the joint error on parameter $i$ is
given by the projection on the $i$-th coordinate axis of the 
multi-dimensional hyper-ellipse which contains a fraction $\gamma$ of the 
joint likelihood. The equation of the hyper-ellipse is
\begin{equation}
({\bf \Theta - \Theta_0}) {\bf F } ({\bf \Theta - \Theta_0})^t =
q_{1-\gamma},
\end{equation}
where $q_{1-\gamma}$ is the quantile for the probability $1-\gamma$ for
a $\chi^2$ distribution with 6,7 and 8 degrees of freedom. For
$\gamma = 0.683$ ($1\sigma$ c.l.) we have for 6,7 and 8 degrees of freedom, $q_{1-\gamma} = 7.03$, $q_{1-\gamma} = 8.18$ and $q_{1-\gamma} = 9.30$, respectively.

As observed in \cite{Martins} the accuracy with which parameters can be determined depends on their true value as well as on the number of parameters considered. Note that the FMA \textit{assumes}
that the values of the parameters of the true model are 
in the vicinity of ${\bf \Theta_0}$. The validity
of the results therefore depends on this assumption, as well as 
on the assumption that the $a_{\ell m}$'s are independent
Gaussian random variables. If the FMA predicted errors are small
enough, the method is self-consistent and we can expect the FMA prediction 
to reproduce in a correct way the exact behaviour. This is indeed the case 
for the present analysis, with the notable exception of $\omega_\Lambda$, 
which as expected suffers from the geometrical degeneracy.

Also, special care must be taken when computing the derivatives of the power
spectrum with respect to the cosmological parameters. 
This differentiation strongly amplifies any numerical errors in the spectra,
leading to larger derivatives, which would artificially break degeneracies 
among parameters. In the present work we implement double--sided derivatives,
which reduce the truncation error from second order to third order terms.
The choice of the step size is a trade-off between truncation error and
numerical inaccuracy dominated cases. For an estimated numerical precision 
of the computed models of order $10^{-4}$, the step size should be 
approximately 5\% of the parameter value \cite{Numerical:92}, though
it turns out that for derivatives in direction of $\alpha$ and $n_s$ 
the step size can be chosen to be as small as 0.1\%. 
After several tests, we have chosen step sizes varying from 1\% to 5\%
for $\omega_b, \omega_m, \omega_\Lambda$  and $\R$. This choice gives
derivatives with an accuracy of about 0.5\%. The derivatives
with respect to $Q$ are exact, being the power spectrum itself.


\begin{table*}
\caption{\label{fmast}Fisher matrix analysis results for Standard model: expected $1\sigma$ errors for the WMAP and Planck satellites as well as for a CVL experiment. The column {\it marg.} gives the error with all
other parameters being marginalized over; in the column {\it fixed} the other
parameters are held fixed at their ML value; in the column {\it joint} all
parameters are being estimated jointly.}
\begin{ruledtabular}
\begin{tabular}{|c|c c c| c c c|c c c|}
Quantity &  \multicolumn{9}{c}{$1\sigma$ errors (\%)} \\\hline  
                & \multicolumn{3}{c|}{WMAP}           & \multicolumn{3}{c}{Planck HFI} & \multicolumn{3}{c}{CVL} \\ 
                        & marg. & fixed  & joint   & marg.  & fixed & joint  & marg.  & fixed & joint           \\\hline
	 & \multicolumn{9}{c}{Polarization} \\\hline 
$\omega_b$   &1437.41   &52.93     &4111.09      &6.40        &0.99       &18.31   &0.48        &0.25        &1.38 \\
$\omega_m$   &619.43    &31.47     &1771.62      &3.57        &0.33       &10.22   &0.70        &0.03        &2.01 \\
$\omega_\Lambda$ &1397.45  &980.08  &3996.79    &38.76        &34.40      &110.84  &11.28        &9.94       &32.27 \\
$n_s$     &260.43        &33.68     &744.83      &1.47        &0.91       &4.20    &0.30        &0.08        &0.86 \\
$Q$       &474.57        &25.13     &1357.31     &2.21        &0.45       &6.32    &0.24        &0.07        &0.68 \\
$\R$      &666.04        &22.10     &1904.92     &3.53        &0.30       &10.09   &0.66        &0.03        &1.88 \\ \hline
	 & \multicolumn{9}{c}{Temperature} \\\hline
$\omega_b$  &2.79        &1.26      &7.97      &0.82        &0.59        &2.36    &0.55        &0.38        &1.59 \\
$\omega_m$  &4.58        &0.83      &13.11     &1.44        &0.12        &4.12    &1.09        &0.08        &3.11 \\
$\omega_\Lambda$ &115.59  &86.53    &330.59    &91.65       &86.37       &262.11  &80.68       &77.25      &230.74 \\
$n_s$       &1.50        &0.52      &4.30      &0.48        &0.13        &1.36    &0.33        &0.07        &0.96 \\
$Q$         &0.80        &0.34      &2.29      &0.19        &0.10        &0.55    &0.17        &0.07        &0.48 \\
$\R$        &4.17        &0.73      &11.92     &1.41        &0.11        &4.03    &1.05        &0.07        &2.99 \\ \hline
	 & \multicolumn{9}{c}{Temperature and Polarization} \\\hline
$\omega_b$   &2.78        &1.26       &7.95      &0.77        &0.51        &2.20    &0.32        &0.21        &0.91 \\
$\omega_m$   &4.56        &0.83       &13.05     &1.16        &0.12        &3.32    &0.55        &0.03        &1.58 \\
$\omega_\Lambda$ &114.34  &86.09      &327.03    &31.79       &31.72       &90.92   &9.87        &9.49        &28.24\\
$n_s$        &1.50        &0.52       &4.28      &0.39        &0.13        &1.12    &0.20        &0.06        &0.57 \\
$Q$          &0.80        &0.34       &2.28      &0.18        &0.10        &0.52    &0.14        &0.05        &0.40 \\
$\R$         &4.15        &0.73       &11.86     &1.14        &0.10        &3.25    &0.52        &0.03        &1.49 \\
\end{tabular}
\end{ruledtabular}
\end{table*}


\begin{table*}
\caption{\label{fmaal}Fisher matrix analysis results for a model with a varying $\alpha$: expected $1\sigma$ errors for the WMAP and Planck satellites as well as for a CVL experiment. The column {\it marg.} gives the error with all
other parameters being marginalized over; in the column {\it fixed} the other
parameters are held fixed at their ML value; in the column {\it joint} all
parameters are being estimated jointly.}
\begin{ruledtabular}
\begin{tabular}{|c|c c c| c c c|c c c|}
Quantity &  \multicolumn{9}{c}{$1\sigma$ errors (\%)} \\\hline  
                & \multicolumn{3}{c|}{WMAP}           & \multicolumn{3}{c}{Planck HFI} & \multicolumn{3}{c}{CVL} \\ 
                        & marg. & fixed  & joint   & marg.  & fixed & joint  & marg.  & fixed & joint           \\\hline
	 & \multicolumn{9}{c}{Polarization} \\\hline 
$\omega_b$     &4109.93       &52.93    &11754.68    &6.42        &0.99       &18.36   &1.10        &0.25        &3.16 \\
$\omega_m$     &844.65        &31.47    &2415.75     &7.14        &0.33       &20.43   &1.64        &0.03        &4.69 \\
$\omega_\Lambda$ &1483.80     &980.08   &4243.77     &41.78       &34.40      &119.50  &12.03       &9.94        &34.41 \\
$n_s$          &365.06        &33.68    &1044.09     &3.90        &0.91       &11.16   &0.79        &0.08        &2.25 \\
$Q$            &2415.47       &25.13    &6908.40     &3.24        &0.45       &9.28    &0.24        &0.07        &0.69 \\
$\R$           &4847.40       &22.10    &13863.91    &10.13       &0.30       &28.98   &1.19        &0.03        &3.39 \\
$\alpha$       &887.24        &3.51     &2537.58     &2.62        &0.05       &7.50    &0.40        &$<0.01$        &1.15 \\ \hline
	 & \multicolumn{9}{c}{Temperature} \\\hline
$\omega_b$      &10.41        &1.26       &29.78    &0.97        &0.59        &2.78    &0.77        &0.38        &2.21 \\
$\omega_m$      &8.51         &0.83       &24.34    &2.54        &0.12        &7.27    &2.04        &0.08        &5.85 \\
$\omega_\Lambda$  &125.00     &86.53      &357.51   &107.64      &86.37       &307.85  &93.06       &77.25       &266.16 \\
$n_s$           &3.05         &0.52       &8.73     &1.32        &0.13        &3.76    &1.04        &0.07        &2.97 \\
$Q$             &2.11         &0.34       &6.05     &0.20        &0.10        &0.57    &0.17        &0.07        &0.50 \\
$\R$            &21.12        &0.73       &60.40    &1.50        &0.11        &4.29    &1.06        &0.07        &3.02 \\
$\alpha$        &4.64         &0.12       &13.27    &0.43        &0.02        &1.22    &0.31        &0.01        &0.88 \\ \hline
	 & \multicolumn{9}{c}{Temperature and Polarization} \\\hline
$\omega_b$      &10.00        &1.26       &28.60    &0.87        &0.51        &2.49    &0.38        &0.21        &1.09 \\
$\omega_m$      &8.23         &0.83       &23.54    &1.61        &0.12        &4.60    &0.67        &0.03        &1.90 \\
$\omega_\Lambda$ &123.13      &86.09      &352.17   &31.79       &31.72       &90.92   &9.96        &9.49        &28.49 \\
$n_s$           &2.97         &0.52       &8.48     &0.85        &0.13        &2.44    &0.32        &0.06        &0.91 \\
$Q$             &2.04         &0.34       &5.82     &0.18        &0.10        &0.53    &0.14        &0.05        &0.41 \\
$\R$            &20.34        &0.73       &58.18    &1.36        &0.10        &3.88    &0.60        &0.03        &1.72 \\
$\alpha$        &4.46         &0.12       &12.75    &0.31        &0.02        &0.88    &0.11        &$<0.01$        &0.32 \\
\end{tabular}
\end{ruledtabular}
\end{table*}


\section{FMA without reionization}

We will now start to describe the results of our analysis in detail. In order
to avoid confusion, we will begin in this chapter by describing the
results for the case $\tau=0$ (since most of the crucial degeneracies
can be understood in this case), and leave the more relevant case of
non-zero $\tau$ for the following chapter.
While it may seem pointless after WMAP to discuss the
cases without (or with very little) reionization, we shall see that a lot
can be learned by comparing the results for the various cases.

\subsection{Analysis results: The FMA forecast}

Tables \ref{fmast}--\ref{fmaevnotauTT} summarize the results of our FMA for 
WMAP, Planck and a CVL experiment. We consider the cases of models with and
without a varying $\alpha$ being included in the analysis, for $\tau=0$.
We also consider the use of temperature information alone
(TT), E-polarization alone (EE) and both channels (EE+TT) jointly.

Table \ref{fmast} shows the $1\sigma$ errors on each of the parameters of 
our FMA for a `standard model', that is with no reionization or variation of $\alpha$. The inclusion of polarization data does indeed increase the accuracy on each parameter for Planck and for a CVL experiment.
For the Planck mission the polarization data helps to better constrain each of the parameters though the increase in accuracy is only of the order 10\% in 
most cases. The error in $\omega_{\Lambda}$ is still large, and larger than those of the other parameters. Indeed, this error is almost insensitive to the
experimental details when only temperature is considered in the analysis, 
which of course is a manifestation of the so-called geometrical 
degeneracy \cite{efstathiou1,efstathiou2}. 

The existence of this nearly exact degeneracy limits in a fundamental way the accuracy on measurements of the Hubble constant as well as of the curvature of the universe obtained with the CMB observations, and hence limits the accuracy on $\omega_{m}$ and $\omega_{\Lambda}$. This degeneracy can only be removed when constraints on the geometry of the universe from other complementary observations, such as Type Ia supernova or gravitational 
lensing, are jointly considered \cite{efstathiou1,efstathiou2}. Our plots show that actually using polarization data the confidence contours can narrow significantly on the $ \omega_{\Lambda}$ axis. 
This case is very different from other degeneracies between parameters which actually can be broken with good enough CMB data and by probing a larger set of angular scales ie an enlarged range of multipoles $l$, as well as using the CMB polarised data. 

The geometrical degeneracy gives rise to almost identical CMB anisotropies in universes with different background geometries but identical matter content, lines of constant $\mathcal R$ are directions of degeneracy. This degeneracy along $\delta(\omega_{m}^{-1/2} \mathcal R)=0$ results in a linear relation between $\delta \omega_{k}$ and $\delta \omega_{\Lambda}$, with coefficients that depend on the fiducial model. 

This is why we used the $\mathcal R$ parameter to replace $\omega_{k}$ in our fisher analysis instead of the $\omega_{D}$ parameter of \cite{efstathiou1,efstathiou2}. 

The accuracy on the parameter $\mathcal R$ is related to the ability of fixing the positions of the Doppler peaks. Hence Planck is expected to determine $\mathcal R$ with high accuracy given that it samples the Doppler peak region almost entirely. 
Indeed this is the case with the error reducing from $4\%$ for WMAP to $1\%$ for Planck and to $0.5\%$ for a CVL experiment (see Table~II).

Table \ref{fmaal} shows the $1\sigma$ errors on each of the parameters of our
FMA for a model with a time-varying $\alpha$. While the inclusion of a varying
$\alpha$ as a parameter (with the nominal value equal to that of the standard
model) has no noticeable effect on the accuracy of the other parameters for a
CVL experiment, for Planck and most notoriously for WMAP this is not the case (compare Table \ref{fmast} with Table \ref{fmaal}). For these two satellite
missions the accuracy of most of the other parameters is reduced by inclusion 
of this extra parameter as should be expected (for allowing an extra degree of
freedom). The same trend as before is observed with the inclusion of
polarization data.

From our WMAP predictions one would expect to be able to constrain $\alpha$ to 
about 5\% accuracy at $1 \sigma$ while the actual analysis presented in previous section 
gives an accuracy of the order of $7\%$ at $2 \sigma$. This is in reasonable agreement with our prediction with the discrepancy being due to the effect of a $\tau \neq 0$ (see next section).
On the other hand, the results of our forecast are that Planck and a CVL
experiment will be able to constrain variations in $\alpha$ with an
accuracy of  0.3\% and 0.1\% respectively ($1\sigma$ c.l., all other 
parameters marginalized). If all parameters are being estimated 
simultaneously, then these limits increase to about 0.9\% and 0.3\% respectively. This is therefore the best that one can hope to do with the
CMB alone---it is somewhat below the $10^{-5}$ level of the claimed
detection of a variation using quasar absorption
systems \cite{Webb,Jenam,Murphy}, but it is also at a much higher redshift,
where any variations relative to the present day are expected to be
larger than at $z\sim3$. Therefore, for specific models such limits can be
at least as constraining as those at low redshift. On the other hand,
there \textit{is} a way of doing better than this, which is to combine
CMB data with other observables---this is the approach we already
took in \cite{Avelino,Martins}, for example.

From these tables we conclude that for WMAP the inclusion of polarization
information does not improve significantly the accuracy on each of the
parameters, since its accuracy from polarization data alone is expected
to be worse than that from temperature alone by a factor of
$\simeq 10^{2}- 10^{3}$. With Planck though there is room for improvement, 
with the accuracy from polarization alone at most only a factor 10 poorer 
than from temperature. 
Also for this case a better accuracy on $\omega_{\Lambda}$ is obtained using polarization data alone vs using temperature data alone, for both cases with and without inclusion of a varying $\alpha$. 
For the CVL experiment the polarization makes a real difference, with the accuracy of polarization alone being \textit{slightly
better} than that of the temperature alone. Combining the two typically
increases the accuracy on most parameters by a factor of order 2. As
expected this is most noticeably so for $\omega_{\Lambda}$. 
Assuming that the improvement was only owing to the use of independent sets 
of data we should expect an improvement by at least a factor of $\sqrt{2}$. 


\begin{figure*}
\includegraphics[width=3.5in]{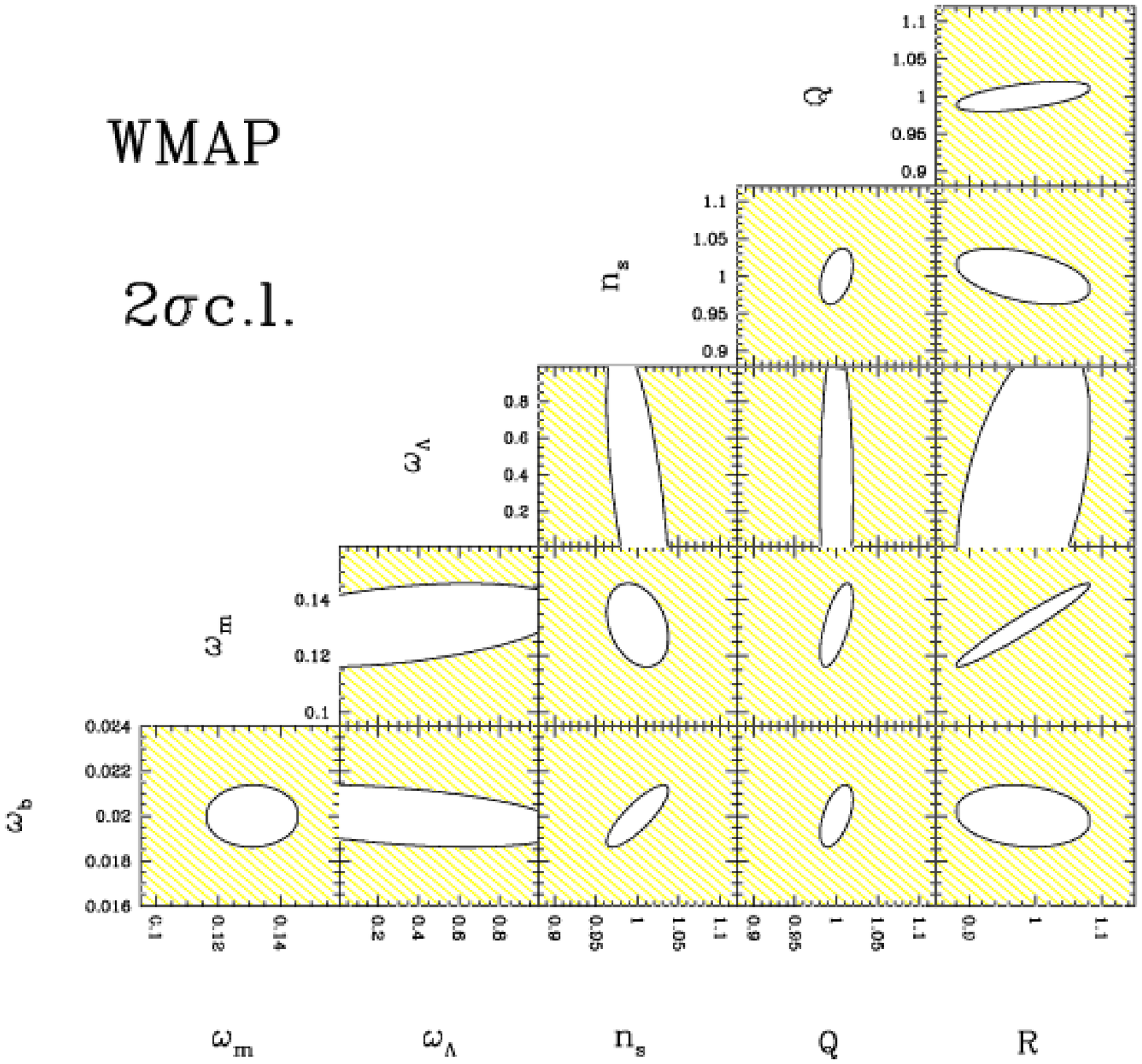}
\includegraphics[width=3.5in]{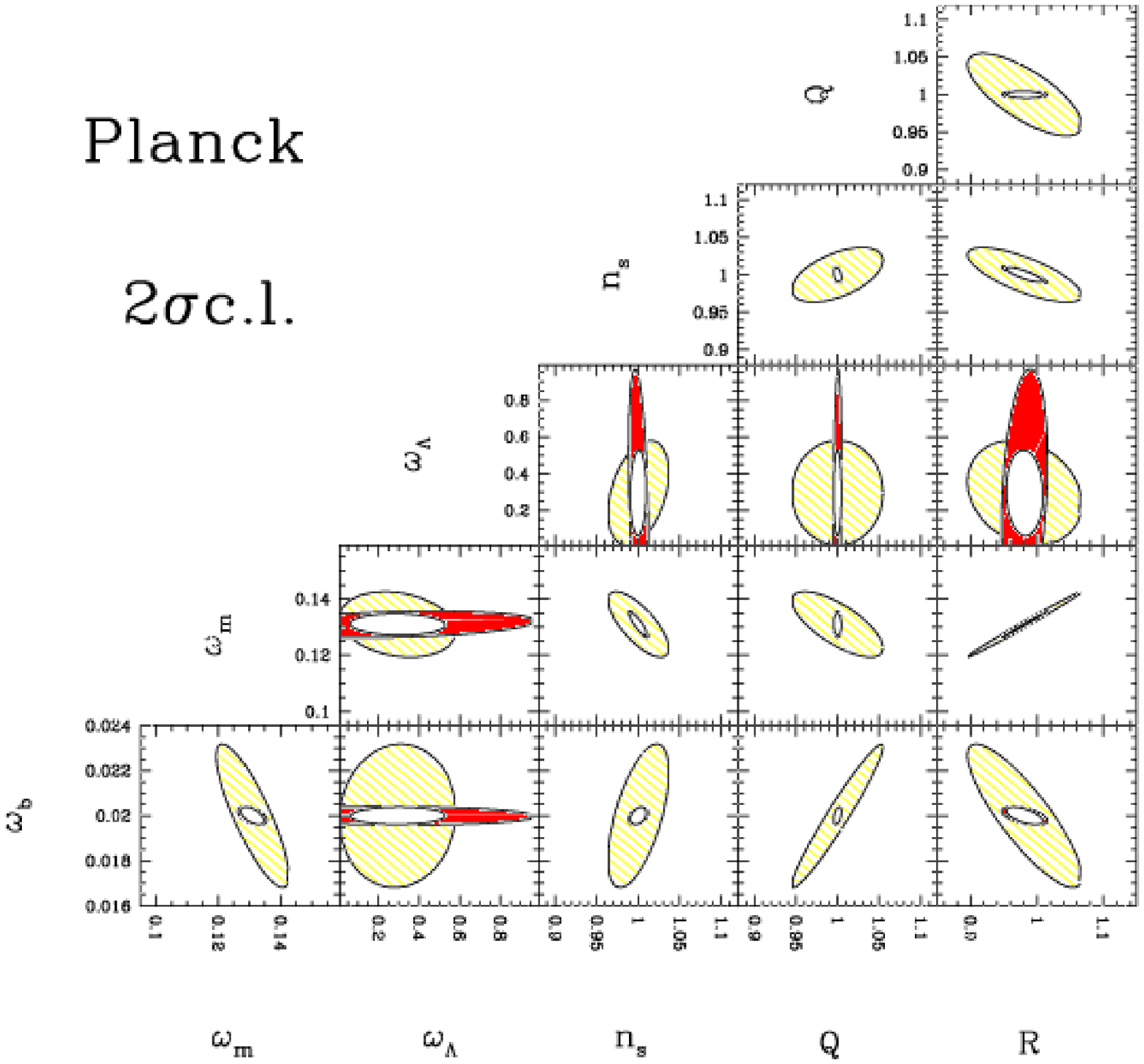}
\includegraphics[width=3.5in]{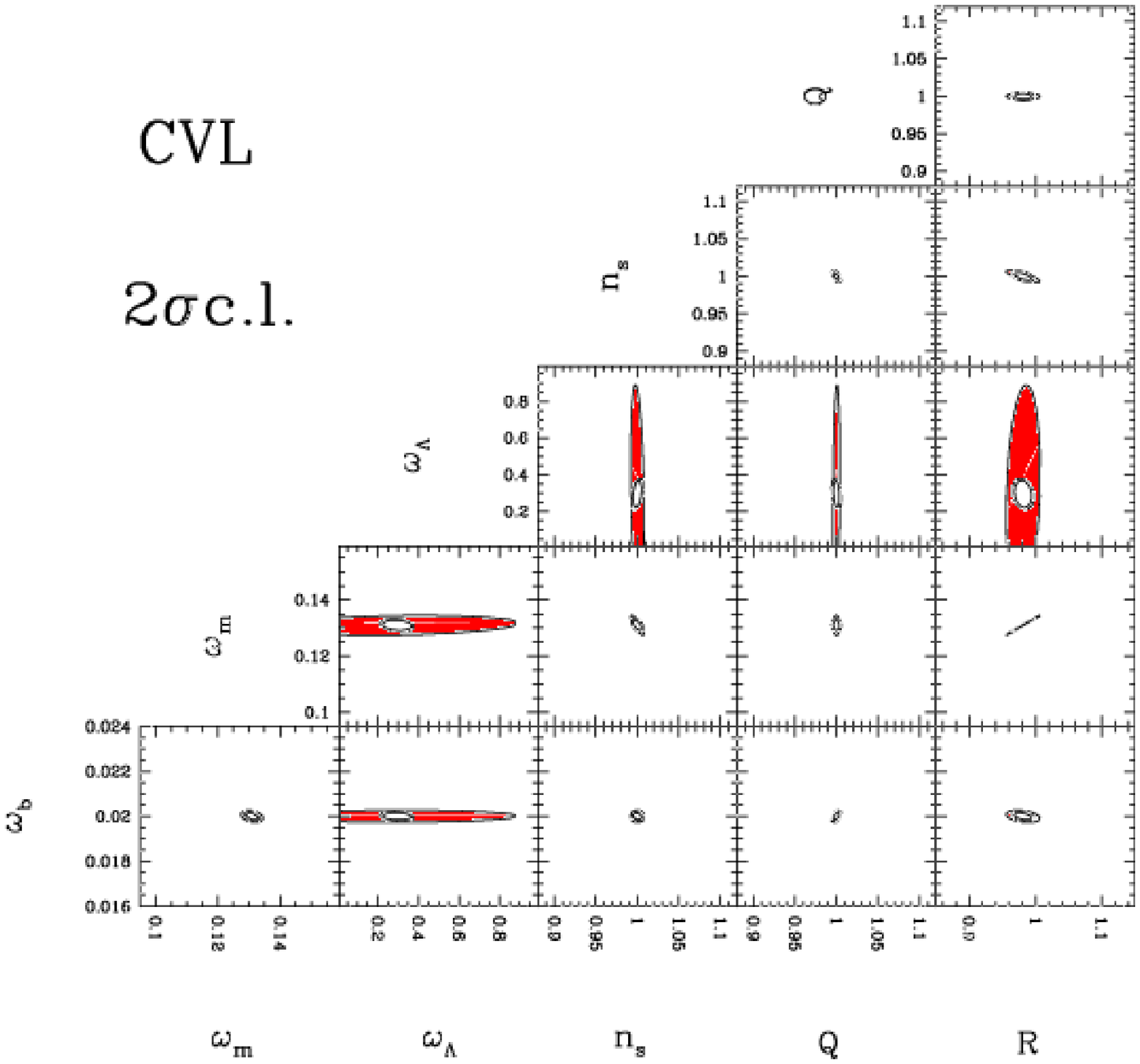}
\caption{\label{figellipse1}Ellipses containing $95.4\%$ ($2\sigma$) of
joint confidence (all other parameters marginalized) using temperature alone (red), E-polarization alone (yellow), and both jointly (white), for a standard model.
In the WMAP case the errors from E only are very large, hence the contours for T coincide almost exactly with the temperature-polarization combined case. In the CVL case it is the E contours that almost coincide with the combined ones. 
  }
\end{figure*}

\begin{figure*}
\includegraphics[width=3.5in]{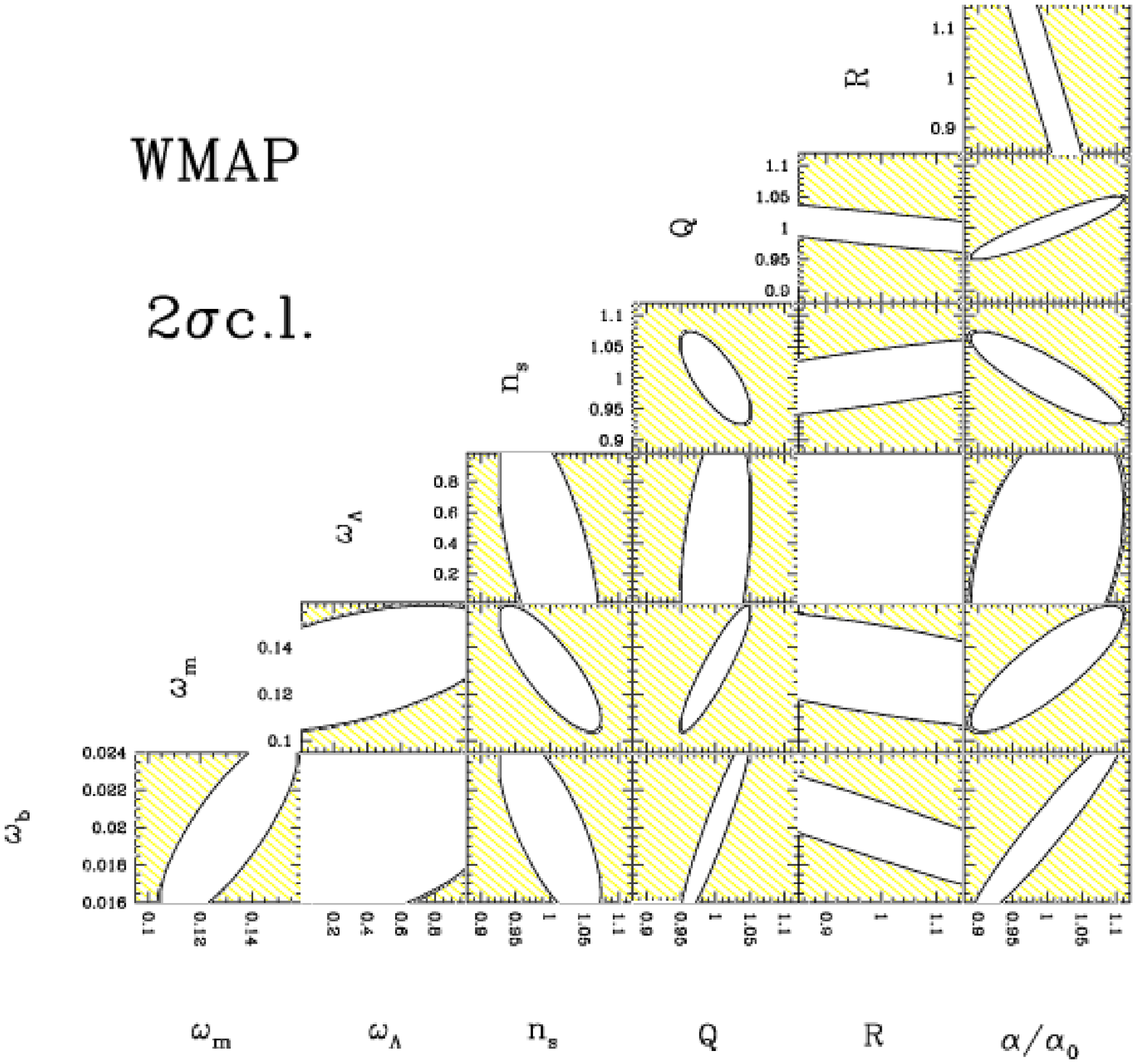}
\includegraphics[width=3.5in]{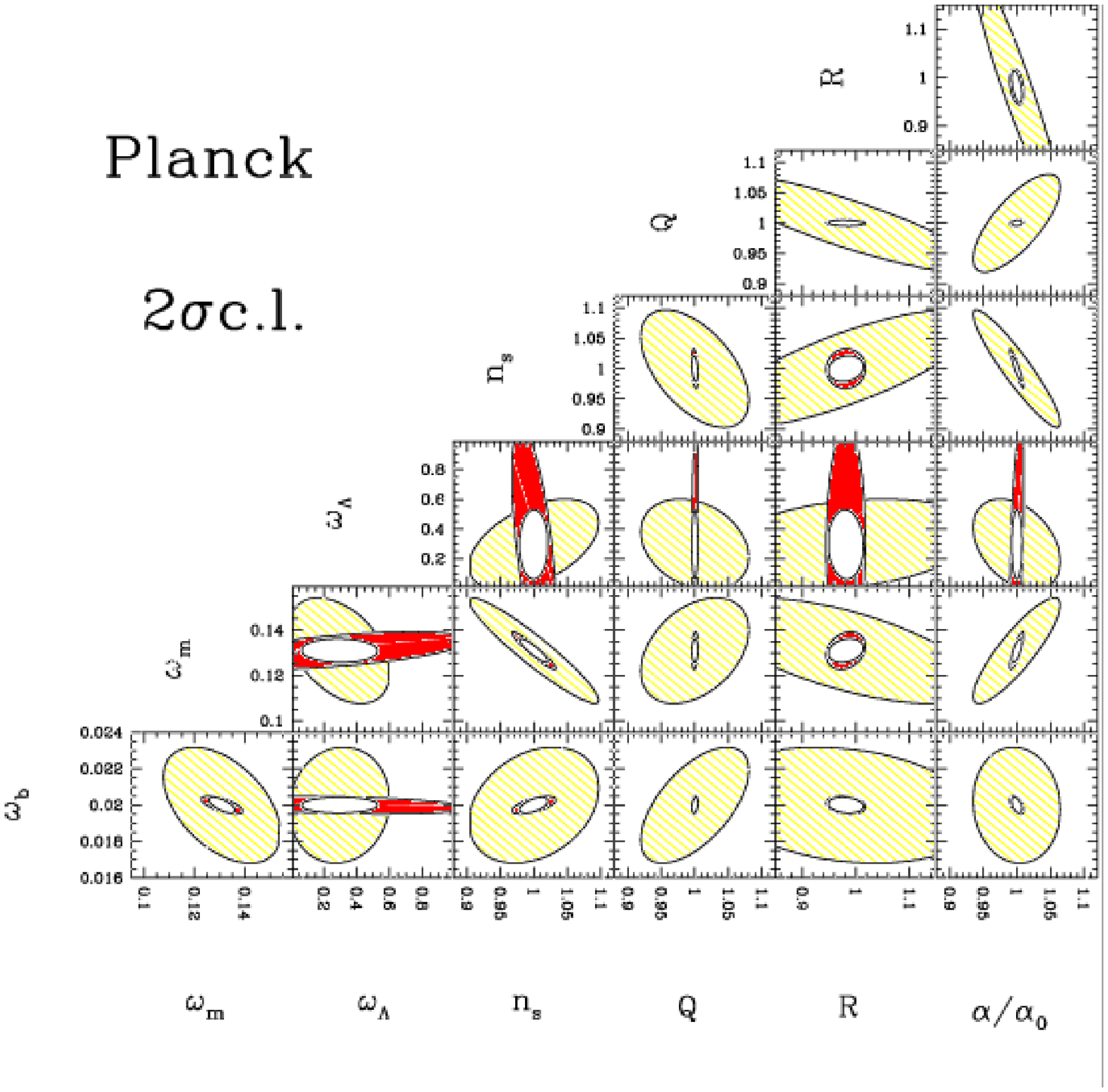}
\includegraphics[width=3.5in]{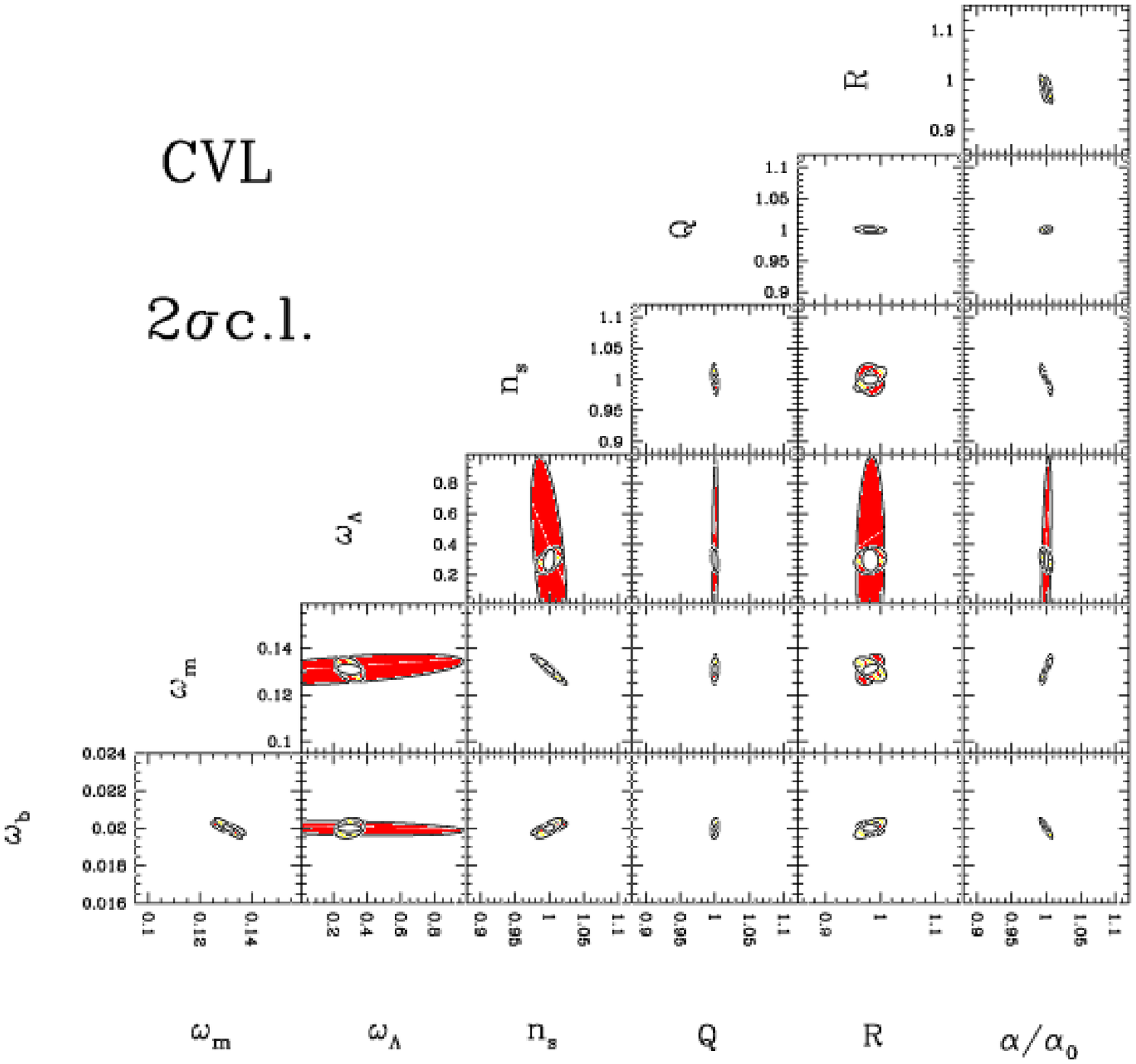}
\caption{\label{figellipse2}Ellipses containing $95.4\%$ ($2\sigma$) of
joint confidence (all other parameters marginalized) using temperature alone (red), E-polarization alone (yellow), and both jointly (white), for a model with varying $\alpha$.
In the WMAP case the errors from E only are very large, hence the contours for T coincide almost exactly with the temperature-polarization combined case. In the CVL case it is the E contours that almost coincide with the combined ones. 
}
\end{figure*}



\begin{table*}
\caption{\label{fmaevnotau}For a model with a varying $\alpha$ and the case Temperature and Polarization considered jointly. In the lines we display the components
of the eigenvectors of the FM for WMAP,
Planck and a CVL experiment. The quantity $1/\sqrt{\lambda_i}$ is proportional to the error along
the principal direction ${\bf u^{(i)}}$. For each principal direction, an
asterisk marks the largest cosmological parameters contribution, a dagger
the second largest. Not only Planck has errors smaller by a factor of about
5 on average, but also the alignment of the principal directions with the
axis defined by the physical parameter is better than WMAP in 6 cases out of 7.}
\begin{ruledtabular}
\begin{tabular}{|c c c c c c c c c|}
\multicolumn{9}{|c|}{WMAP} \\\hline
Direction $i$ & $1/\sqrt{\lambda_i}$ & $\omega_b$ & $\omega_m$ &
$\omega_\Lambda$ & $n_s$ & $Q$ & $\R$  & $\alpha$ \\\hline

1& 2.50E-04&     9.9446E-01* &    -9.9203E-02$\dag$  &    -2.5224E-05  &    -2.7487E-02  &    -3.9411E-03  &     1.2295E-02  &     1.6954E-02  \\
2& 8.84E-04&    8.1778E-02  &     7.0553E-01* &    -5.6359E-04  &    -6.8131E-02  &     2.4777E-02  &    -1.1338E-01  &    -6.9096E-01$\dag$ \\
3& 2.24E-03&     4.8801E-02  &     5.2913E-01 $\dag$&     9.3752E-04  &     2.6766E-01  &    -6.3566E-01* &     4.0924E-02  &     4.9016E-01  \\ 
4& 1.24E-02&     4.2341E-02  &     2.5947E-01  &     1.2292E-02  &     6.5656E-01$\dag$ &     6.6964E-01* &     4.5581E-02  &     2.2174E-01  \\
5& 1.48E-02&    1.0147E-02  &     3.7938E-01  &    -3.5290E-02  &    -6.9349E-01* &     3.7432E-01  &     2.0829E-01  &     4.3623E-01$\dag$  \\
6& 1.94E-01&    -9.0774E-03  &    -2.9295E-02  &     2.2193E-01$\dag$  &     8.9661E-02  &    -7.8874E-02  &     9.4671E-01* &    -1.9819E-01  \\
7&  3.71E-01&    1.9270E-03  &     1.7036E-02  &     9.7435E-01* &    -5.4121E-02  &     2.3700E-02  &    -2.0877E-01$\dag$  &     5.7273E-02  \\ \hline
\multicolumn{9}{|c|}{Planck} \\\hline
Direction $i$ & $1/\sqrt{\lambda_i}$ & $\omega_b$ & $\omega_m$ &
$\omega_\Lambda$ & $n_s$ & $Q$ & $\R$ & $\alpha$ \\\hline
1&  9.02E-05&    7.9666E-01* &    -4.4311E-01$\dag$  &    -1.2864E-05  &     4.6149E-03  &    -1.4650E-02  &     6.7622E-02  &     4.0518E-01  \\
2& 1.38E-04&     6.0235E-01* &     5.7873E-01$\dag$ &     1.1892E-05  &    -3.3913E-02  &    -5.8211E-02  &    -8.4462E-02  &    -5.3905E-01 \\
3& 4.80E-04&     2.5914E-02  &     6.1004E-01$\dag$ &    -1.5285E-05  &     3.4825E-01  &     3.4725E-01  &     2.8006E-02  &     6.2011E-01* \\ 
4&  1.88E-03&    4.0978E-02  &    -2.0888E-01  &     2.2619E-04  &    -7.9733E-02  &     9.3426E-01* &     2.5167E-03  &    -2.7474E-01$\dag$  \\
5& 8.88E-03&     1.2289E-02  &    -2.2979E-01  &    -3.4989E-03  &     9.1281E-01* &    -4.7146E-02  &    -2.2075E-01  &    -2.5072E-01$\dag$  \\
6& 1.36E-02&    -1.1477E-03  &     1.1923E-02  &     7.0929E-03  &     1.9486E-01$\dag$  &    -2.7260E-02  &     9.6887E-01* &    -1.4961E-01  \\
7& 9.40E-02&     4.5352E-05  &    -8.4463E-04  &     9.9997E-01* &     1.8356E-03  &    -1.7712E-04  &    -7.6431E-03$\dag$  &     2.6714E-04  \\ \hline
\multicolumn{9}{|c|}{CVL} \\\hline
Direction $i$ & $1/\sqrt{\lambda_i}$ & $\omega_b$ & $\omega_m$ &
$\omega_\Lambda$ & $n_s$ & $Q$ & $\R$ & $\alpha$ \\\hline
1& 2.67E-05&    -1.2198E-01  &     7.5184E-01* &     1.6953E-05  &    -4.5292E-03  &    -1.5331E-03  &    -1.0829E-01  &    -6.3883E-01$\dag$ \\
2& 4.30E-05&     9.8787E-01* &     1.5297E-01$\dag$  &    -4.0577E-06  &     2.3058E-02  &    -2.0123E-03  &    -1.1111E-02  &    -6.8706E-03  \\
3& 2.26E-04&    -8.5658E-02  &     5.3126E-01$\dag$ &    -1.6190E-04  &     3.8153E-01  &     4.0338E-01  &     2.5197E-02  &     6.3365E-01* \\ 
4& 1.30E-03&     4.2889E-02  &    -2.9019E-01  &     4.2863E-03  &     6.5704E-02  &     8.8528E-01* &     1.3375E-02  &    -3.5457E-01$\dag$  \\
5& 3.31E-03&     7.1120E-03  &    -2.0415E-01  &    -3.3636E-02  &     9.1855E-01* &    -2.2722E-01  &    -8.6918E-02  &    -2.3286E-01$\dag$  \\
6&  5.95E-03&   -2.8965E-05  &     5.6352E-02  &     1.1741E-02  &     7.0270E-02  &    -4.2538E-02  &     9.8973E-01* &    -1.0183E-01$\dag$  \\
7& 2.95E-02&     4.7963E-05  &    -6.2146E-03  &     9.9936E-01* &     2.9871E-02$\dag$  &    -1.0880E-02  &    -1.4605E-02  &    -5.0067E-03  \\
\end{tabular}
\end{ruledtabular}
\end{table*}


\begin{table*}
\caption{\label{fmaevnotauTT}For a model with a varying $\alpha$ and the case (TT) ie Temperature only. In the lines we display the components
of the eigenvectors of the FM for WMAP,
Planck and a CVL experiment. The quantity $1/\sqrt{\lambda_i}$ is proportional to the error along
the principal direction ${\bf u^{(i)}}$. For each principal direction, an
asterisk marks the largest cosmological parameters contribution, a dagger
the second largest. Not only Planck has errors smaller by a factor of about
4 on average, but also the alignment of the principal directions with the
axis defined by the physical parameter is better than WMAP in 6 cases out of 7.}
\begin{ruledtabular}
\begin{tabular}{|c c c c c c c c c|}
\multicolumn{9}{|c|}{WMAP} \\\hline
Direction $i$ & $1/\sqrt{\lambda_i}$ & $\omega_b$ & $\omega_m$ &
$\omega_\Lambda$ & $n_s$ & $Q$ & $\R$  & $\alpha$ \\\hline
1& 2.50E-04&     9.9447E-01* &    -9.9159E-02$\dag$  &    -2.5230E-05  &    -2.7515E-02  &    -3.9234E-03  &     1.2276E-02  &     1.6802E-02  \\
2&  8.84E-04&     8.1646E-02  &     7.0565E-01* &    -5.6626E-04  &    -6.8099E-02  &     2.4756E-02  &    -1.1338E-01  &    -6.9086E-01$\dag$\\
3&  2.24E-03&    4.8844E-02  &     5.2886E-01$\dag$ &     9.4022E-04  &     2.6766E-01  &    -6.3596E-01* &     4.0937E-02  &     4.9006E-01  \\
4& 1.24E-02&     4.2256E-02  &     2.5530E-01  &     1.2657E-02  &     6.6444E-01$\dag$ &     6.6515E-01* &     4.4102E-02  &     2.1685E-01  \\
5& 1.49E-02&     1.0648E-02  &     3.8232E-01  &    -3.5272E-02  &    -6.8593E-01* &     3.8154E-01  &     2.0973E-01  &     4.3865E-01$\dag$  \\
6& 2.00E-01&    -9.0958E-03  &    -2.9575E-02  &     2.3865E-01$\dag$  &     8.8766E-02  &    -7.9309E-02  &     9.4276E-01* &    -1.9779E-01  \\
7& 3.78E-01&     2.0990E-03  &     1.7737E-02  &     9.7038E-01* &    -5.5730E-02  &     2.5328E-02  &    -2.2492E-01$\dag$  &     6.0883E-02  \\ \hline
\multicolumn{9}{|c|}{Planck} \\\hline
Direction $i$ & $1/\sqrt{\lambda_i}$ & $\omega_b$ & $\omega_m$ &
$\omega_\Lambda$ & $n_s$ & $Q$ & $\R$ & $\alpha$ \\\hline
1& 1.01E-04&     7.2972E-01* &    -5.0661E-01$\dag$ &     1.3138E-06  &     6.7011E-03  &    -6.9279E-03  &     7.5433E-02  &     4.5284E-01  \\
2&  1.54E-04&    6.8066E-01* &     5.0680E-01 &    -1.8992E-05  &    -4.2986E-02  &    -6.6152E-02  &    -7.8097E-02  &    -5.1724E-01$\dag$\\
3& 4.94E-04&     5.4883E-02  &     6.2059E-01* &     2.1300E-05  &     3.4975E-01  &     3.5572E-01  &     2.4796E-02  &     6.0198E-01$\dag$\\
4& 1.95E-03&     3.3117E-02  &    -2.1287E-01  &    -8.9072E-04  &    -9.9243E-02  &     9.3131E-01* &     3.4607E-03  &    -2.7638E-01$\dag$  \\
5&  1.14E-02&    9.3798E-03  &    -2.3387E-01  &     2.6011E-02  &     9.2519E-01* &    -3.5345E-02  &    -1.1636E-01  &    -2.7161E-01$\dag$  \\
6&  1.49E-02&   -2.3014E-03  &     3.6400E-02  &     2.0129E-03  &     9.6754E-02  &    -2.1077E-02  &     9.8694E-01* &    -1.2173E-01$\dag$ \\
7&   3.18E-01&  -1.9911E-04  &     5.8193E-03  &     9.9966E-01* &    -2.4365E-02$\dag$  &     1.7831E-03  &     1.0413E-03  &     7.0427E-03  \\ \hline
\multicolumn{9}{|c|}{CVL} \\\hline
Direction $i$ & $1/\sqrt{\lambda_i}$ & $\omega_b$ & $\omega_m$ &
$\omega_\Lambda$ & $n_s$ & $Q$ & $\R$ & $\alpha$ \\\hline
1&  5.86E-05&    6.7177E-01* &    -4.8930E-01  &     8.7005E-07  &     3.9437E-02  &     2.7795E-02  &     8.1011E-02  &     5.4811E-01$\dag$\\
2&  1.16E-04&    7.3379E-01* &     5.3949E-01$\dag$&    -1.2978E-05  &    -2.8027E-05  &    -2.4037E-02  &    -7.2166E-02  &    -4.0585E-01  \\
3&  2.91E-04&   -9.4755E-02  &     6.0914E-01$\dag$ &     1.7574E-05  &     3.4556E-01  &     3.4843E-01  &     1.1703E-02  &     6.1564E-01* \\
4&  1.71E-03&    3.4674E-02  &    -2.0806E-01  &    -8.4424E-04  &    -8.2888E-02  &     9.3517E-01* &     1.4441E-02  &    -2.7182E-01$\dag$  \\
5& 9.14E-03&     9.1126E-03  &    -1.9125E-01  &     2.0128E-02  &     8.9154E-01* &    -5.1488E-02  &     2.8905E-01$\dag$  &    -2.8615E-01  \\
6& 1.05E-02&    -3.6709E-03  &     1.3640E-01  &    -9.7678E-03  &    -2.7727E-01$\dag$ &    -7.0379E-03  &     9.5096E-01* &     6.0169E-03  \\
7&  2.75E-01&   -1.7944E-04  &     5.0041E-03  &     9.9975E-01* &    -2.0734E-02$\dag$  &     1.7511E-03  &     3.4827E-03  &     5.5738E-03  \\
\end{tabular}
\end{ruledtabular}
\end{table*}


\subsection{Analysis results: Confidence contours}

In order to provide better intuition for the various effects involved,
we show in Figs. \ref{figellipse1}-\ref{figellipse2} joint 2D confidence
contours for all pairs of parameters (all remaining parameters marginalized)
for the cases shown in Tables \ref{fmast}-\ref{fmaal} respectively (that is, 
the cases $\tau=0$  without and with a varying $\alpha$).
For each case we show plots corresponding to our three experiments
(WMAP, Planck and CVL), and contours for TT only, EE only and all combined.
Note that all contours are $2 \sigma$.
To notice that in the WMAP case the errors from E only are very large, hence the contours for T coincide almost exactly with the temperature-polarization combined case. In the CVL case it is the E contours that almost coincide with the combined ones. 

Again, starting with the standard model in Fig. \ref{figellipse1} we can
observe the expected degeneracies between parameters, as previously
discussed in \cite {efstathiou1,efstathiou2}. These degeneracies among parameters limit 
our ability to disentangle one parameter from another, using CMB observations
alone. The search for means to break such degeneracies is therefore of 
extreme importance.

The contour plots for WMAP exhibit the degeneracy directions in the planes ($\omega_{\Lambda}$,$\mathcal R$), ($n_{s}$,$\mathcal R$), for example $\mathcal R$ suffers strong degeneracy
with $\omega_{m}$, $\omega_{\Lambda}$. A correlation between $\omega_{\Lambda}$
and both $n_{s}$ and $Q$  is also noticeable. The contour plot in the plane
($\omega_{\Lambda}$,$\mathcal R$) prevents a good constraint of both parameters in
agreement with results tabulated in Table \ref{fmast}. 
For both Planck and a CVL experiment the direction of
degeneracy for polarization alone is almost orthogonal to this direction 
while the direction for temperature alone corresponds to $\mathcal R=constant$. The
degeneracy direction on the ($\omega_{m}$,$\mathcal R$) plane is defined by  
$\delta (\omega_{m}^{1/2} \mathcal R$)=0.

The contour plots for Planck are perhaps the perfect example of a case where 
the degeneracy directions between $\mathcal R$ and $\omega_{\Lambda}$ are different 
and almost orthogonal for Temperature and Polarization alone. This therefore
explains why the joint use of T and E data helps to break degeneracies. For
example the degeneracy between $\mathcal R$ and $\omega_{b}$ present when polarization 
is considered alone, disappears when temperature information is included.
It is interesting to notice, when comparing WMAP and Planck plots, that the 
joint use of T and E does not necessarily break degeneracies between the
parameters, whilst narrowing down the width of the contour plots without affecting the degeneracy directions.

For the CVL experiment the effect of polarization is to better constrain
all parameters in particular $\omega_{\Lambda}$, helping to narrow down the
range of allowed values in the $\omega_{\Lambda}$ direction as compared with
Temperature alone. 
For instance in the plane 
($n_{s}$,$\omega_{\Lambda}$) the direction $n_{s}$ is well constrained but 
there is no discriminatory power on the $\omega_{\Lambda}$ direction until
polarization data is included. For all but the 2D planes containing $\omega_{\Lambda}$, 
the contours are narrowed to give better constraints 
to each of the parameters. This is due to the exact degeneracy mentioned 
above: more accurate CMB measurements simply narrow the likelihood contours
around the degeneracy lines on the ($\omega_{\Lambda}$, $\omega_{k}$) plane
\cite{efstathiou1,efstathiou2}.   

Fig. \ref{figellipse1} also shows that $\omega_{b}$ and $\omega_{m}$ are
slightly anticorrelated for the Planck experiment.
For the WMAP experiment the plot shows a degeneracy between $\omega_{m}$ and
$\omega_{\Lambda}$. If we restrict ourselves to spatially flat models there is 
a relationship between these two parameters that will result in similar 
position of the Doppler peaks. The degeneracy direction can be obtained by
differentiating $l_{D}$, the location of the maximum of the first Doppler 
peak \cite{efstathiou1,efstathiou2} 
These degeneracy lines in the $\omega_{c}$ - $\omega_{\Lambda}$ plane are given by (assuming that 
$\omega_{b}$ is held fixed in the expression of $l_{D}$):

\begin{eqnarray}
\omega_{c}=(\omega_{c})_{t}+b \omega_{\Lambda};    
b=-\frac{(\partial l_{D}/ \partial \omega_{\Lambda})_{t}}{(\partial l_{D}/ \partial \omega_{c})_{t}}
\end{eqnarray}
 
Unlike the geometrical degeneracy, this is not exact. Both the
height and the amplitude of the peaks depend upon the parameter $\omega_{m}$,
hence an experiment such as Planck which probes high multipoles will be able 
to break this degeneracy. This is clearly visible in Fig. \ref{figellipse1} 
for both Planck and a CVL experiment (compare with the case for WMAP).

Similarly the condition of constant height of the first Doppler peak 
determines the degeneracies among $\omega_{b}$, $\omega_{c}$ $n_{s}$ and $Q$. 
Both WMAP and Planck are sensitive to higher multipoles than the first Doppler 
peak. The other peaks help to pin down the value of $\omega_{b}$ and 
therefore these degeneracies can actually be broken. The plots for 
WMAP show a mild degeneracy in the ($n_{s}$,$\omega_{b}$) plane for the 
EE+TT+ET joint analysis, which seems to be lifted for the Planck experiment.

In our previous works \cite{Old,Avelino,Martins} we observed a 
degeneracy between $\alpha$ and some of the other parameters,
most notably $w_{b}$, $n_{s}$ and $\R$. Our previous FMA analysis with
temperature information alone \cite{Martins} showed that these degeneracies
could be removed by using higher multipole measurements, e.g., from Planck.
The question we want to address here is whether the use of polarization data
allows further improvements.

As previously pointed out, a variation in $\alpha$ affects both the location 
and height of the Doppler peaks, hence this parameter will be correlated with
parameters that determine the peak structure. Therefore, from the previous
discussion on degeneracies among parameters for a standard model, one can 
anticipate the degeneracies exhibited in Fig. \ref{figellipse2} in the 
planes ($\alpha$,$n_{s}$), ($\alpha$,$\mathcal R$), ($\alpha$,$Q$),
($\alpha$,$\omega_{b}$) and ($\alpha$,$\omega_{m}$).

In our previous work \cite{Martins} we showed that using temperature alone the degeneracies of $\alpha$ with $\omega_{b}$ and $\alpha$ with $n_{s}$ are lifted as we move from WMAP to Planck when higher multipoles measurements can break it.  

All the degeneracy directions for these pairs of parameters for the WMAP 
joint analysis (which actually is dominated by the temperature data alone) 
are approximately preserved by using polarization data alone for the 
Planck experiment. A joint analysis of temperature and polarization helps 
to narrow down the confidence contours without necessarily breaking the degeneracy.

With the inclusion of the new parameter $\alpha$ the WMAP contour plots get wider 
as compared with Fig. \ref{figellipse1}, while leaving almost unchanged the degeneracy directions in most planes of pairs of parameters. 
For Planck the contour plots are still wider whilst the degeneracy directions
for polarization alone change for some of the parameters. For example, the
direction of degeneracy between the ($\mathcal R$,$n_{s}$) changes when compared with
Fig. \ref{figellipse1}, which is due to the presence of the degeneracy between
$\alpha$ and $n_{s}$ which is almost orthogonal to the direction of degeneracy
in the plane ($\alpha$,$\omega_{m}$). Another changed direction of degeneracy 
is that of ($\omega_{b}$,$\mathcal R$), with wider contour plots. The degeneracy 
present in the WMAP plot for the plane ($\alpha$,$\omega_{b}$) seems to be
broken with Planck data. Notice the strong degeneracy between $\alpha$ and $\mathcal R$
which still persists when using jointly temperature and polarization data. 

Using Temperature and Polarization data jointly seems either to help to break
some of the degeneracies or at least to narrow down the contours without 
lifting the degeneracy, in particular for those cases where the degeneracy 
directions for each of the temperature and polarization are different 
(in some cases almost orthogonal see for example the planes containing $\omega_{\Lambda}$ as one of the parameters).

For the CVL experiment most of the plots remain unchanged when compared with 
no inclusion of $\alpha$, with the temperature alone contour plot slightly 
wider in the ($n_{s}$,$\omega_{\Lambda}$) plane. A large range of 
possibilities along the $\omega_{\Lambda}$ direction still remains as 
expected from the exact geometrical degeneracy mentioned above.

\subsection{Analysis results: Principal directions}

The power of an experiment can be roughly quantified by looking at the 
eigenvalues $\lambda_i$ and eigenvectors ${\bf u^{(i)}}$ of its FM: 
The error along the direction in parameter space defined by ${\bf u^{(i)}}$ 
(principal direction) is proportional to $\lambda_i^{-1/2}$.
It can be measured by assessing how the principal components mix 
inflationary variables (such as $n_{s}$) with physical cosmic densities. 
The accuracy on the former is typically limited by
cosmic variance (the derivatives of $C_{l}$ with respect to these variables 
has large amplitude for low multipoles; the accuracy on the latter is 
set by the accuracy with which the $C_{l}$ is measured at high multipoles 
(the derivatives of the angular power spectrum with respect to these 
variables is larger for $l\sim2000 -3000$).

But we are interested in determining the errors on the physical 
parameters rather then on their linear combinations along the principal
directions. Therefore in the ideal case we want the principal directions 
to be as much aligned as possible to the coordinate system 
defined by the physical parameters. We display in Table \ref{fmaevnotau}
eigenvalues and eigenvectors of the FM for WMAP and Planck and a CVL 
experiment. Planck's errors, as measured by the inverse square root of the
eigenvalues, are smaller by a factor of about $6$ on average that those for WMAP (to be compared with a factor of $4$ 
using temperature alone obtained in our previous analysis \cite{Martins}) 
While a CVL experiment's errors are smaller by a factor of about $3$ on average than those for Planck.


For 5 of the 7
eigenvectors Planck also obtains a better alignment of the principal
directions with the axis of the physical parameters.
This is established by comparing the ratios between the largest (marked
with an asterisk in Tables \ref{fmaevnotau}) and the second largest (marked with
a dagger) cosmological parameters' contribution to the principal
directions.
This is of course in a slightly different
form the statement that Planck will measure the cosmological parameters
with less correlations among them.
It is to be noticed that for Planck direction 7 is mostly aligned with $\omega_{\Lambda}$.
While $\alpha$ is the second largest parameter contribution to two of the principal directions for both WMAP and Planck, this is the case for four principal directions for a CVL experiment, and is also the largest parameter contribution to two and one of the principal directions for Planck and a CVL experiment respectively.

For comparison we also display in Table \ref{fmaevnotauTT}
eigenvalues and eigenvectors of the FM for WMAP and Planck and a CVL 
experiment using Temperature information alone.
Comparing Tables \ref{fmaevnotau} and \ref{fmaevnotauTT} we conclude that for WMAP the largest and second largest parameter contribution to the principal direction are exactly the same. 
On the other hand for Planck 2 of the principal directions change namely direction 7, whose main contribution is from $\omega_{\Lambda}$ and $n_{s}$ when Temperature information alone is used while when Polarization is included the second largest contribution comes now from $\mathcal R$. 
For direction 3 the largest and second largest contribution are interchanged (arising from $\omega_{m}$ and $\alpha$) when polarization is included.
Finally for a CVL experiment for direction 1 the second largest contribution from $\alpha$ is replaced by $\omega_{m}$ when polarization is included. For direction 2 both largest contribution change from $\omega_{b}$ to $\omega_{m}$ and that from $\omega_{m}$ (second largest) to $\alpha$. The major contributions for the remaining directions remain the same  while the second largest contribution changes for all of them.
Only for 2 and 3 of the 7 eigenvectors Planck and a CVL experiment respectively obtain a better alignment of the principal
directions with the axis of the physical parameters (with the other directions equally aligned), when polarization is included. 

Therefore we conclude that indeed polarization does not necessarily help to further break degeneracies between parameters when no information on reionization or tensor component of the CMB is included.


\begin{table*}
\caption{\label{fmasttau}Fisher matrix analysis results for a standard model with inclusion of reionization ($\tau=0.20$): expected $1\sigma$ errors for the WMAP and Planck satellites as well as for a CVL experiment. The column {\it marg.} gives the error with all
other parameters being marginalized over; in the column {\it fixed} the other
parameters are held fixed at their ML value; in the column {\it joint} all
parameters are being estimated jointly.}
\begin{ruledtabular}
\begin{tabular}{|c|c c c| c c c|c c c|}
Quantity &  \multicolumn{9}{c}{$1\sigma$ errors (\%)} \\\hline  
                & \multicolumn{3}{c|}{WMAP}           & \multicolumn{3}{c}{Planck HFI} & \multicolumn{3}{c}{CVL} \\ 
                        & marg. & fixed  & joint   & marg.  & fixed & joint  & marg.  & fixed & joint           \\\hline
	 & \multicolumn{9}{c}{Polarization} \\\hline 
$\omega_b$     &223.67       &22.18      &639.70    &6.21       &1.11      &17.75    &0.48     &0.25    &1.38 \\  
$\omega_m$     &104.48       &22.12      &298.81    &3.37       &0.39      &9.64     &0.70     &0.03    &1.99 \\
$\omega_\Lambda$ &1231.56    &113.78     &3522.35   &37.37      &22.87     &106.89   &11.40    &9.99    &32.61 \\
$n_s$          &107.77       &5.31       &308.22    &1.53       &0.96      &4.38     &0.30     &0.08    &0.86 \\
$Q$            &139.04       &18.38      &397.68    &2.23       &0.51      &6.38     &0.24     &0.07    &0.67 \\
$\R$           &91.43        &20.44      &261.50    &3.33       &0.35      &9.52     &0.65     &0.03    &1.86 \\
$\tau$         &156.71       &9.64       &448.22    &5.74       &2.78      &16.42    &1.81     &1.52    &5.18  \\ \hline
	 & \multicolumn{9}{c}{Temperature} \\\hline
$\omega_b$     &10.59        &1.35       &30.28     &0.86       &0.60      &2.46     &0.57     &0.38    &1.64 \\
$\omega_m$     &13.54        &0.88       &38.72     &1.51       &0.13      &4.31     &1.10     &0.08    &3.14 \\
$\omega_\Lambda$ &114.06     &96.36      &326.22    &110.15     &96.15     &315.03   &98.15    &86.00   &280.72 \\
$n_s$          &8.64         &0.53       &24.72     &0.54       &0.13      &1.56     &0.36     &0.07    &1.04 \\
$Q$            &1.46         &0.36       &4.19      &0.20       &0.11      &0.56     &0.17     &0.07    &0.50 \\
$\R$           &13.98        &0.78       &39.98     &1.47       &0.12      &4.21     &1.05     &0.07    &3.01 \\
$\tau$         &107.58       &13.26      &307.68    &16.50      &8.28      &47.20    &14.02    &5.89    &40.09 \\ \hline
	 & \multicolumn{9}{c}{Temperature and Polarization} \\\hline
$\omega_b$     &3.10         &1.34       &8.86      &0.80       &0.53      &2.30    &0.32      &0.21     &0.92 \\
$\omega_m$     &5.09         &0.88       &14.56     &1.24       &0.12      &3.55    &0.55      &0.03     &1.58  \\
$\omega_\Lambda$ &89.62      &72.75      &256.33    &30.58      &22.04     &87.46   &10.72     &9.85     &30.65 \\
$n_s$          &1.66         &0.52       &4.76      &0.43       &0.13      &1.23    &0.20      &0.05     &0.58  \\
$Q$            &0.96         &0.36       &2.74      &0.19       &0.10      &0.53    &0.14      &0.05     &0.41  \\
$\R$           &4.49         &0.78       &12.85     &1.22       &0.11      &3.48    &0.52      &0.03     &1.49 \\
$\tau$         &12.38        &7.90       &35.41     &4.04       &2.65      &11.56   &1.73      &1.48     &4.96 \\ 
\end{tabular}
\end{ruledtabular}
\end{table*}

\begin{table*}
\caption{\label{fmaaltau}Fisher matrix analysis results for a model with varying $\alpha$ and inclusion of reionization ($\tau=0.20$): expected $1\sigma$ errors for the WMAP and Planck satellites as well as for a CVL experiment. The column {\it marg.} gives the error with all
other parameters being marginalized over; in the column {\it fixed} the other
parameters are held fixed at their ML value; in the column {\it joint} all
parameters are being estimated jointly.}
\begin{ruledtabular}
\begin{tabular}{|c|c c c| c c c|c c c|}
Quantity &  \multicolumn{9}{c}{$1\sigma$ errors (\%)} \\\hline  
             & \multicolumn{3}{c|}{WMAP}           & \multicolumn{3}{c}{Planck HFI} & \multicolumn{3}{c}{CVL} \\ 
                        & marg. & fixed  & joint   & marg.  & fixed & joint  & marg.  & fixed & joint           \\\hline
	 & \multicolumn{9}{c}{Polarization} \\\hline
$\omega_b$     &281.91       &22.18      &806.27    &6.46       &1.11       &18.47    &1.09        &0.25        &3.12 \\
$\omega_m$     &446.89       &22.12      &1278.15   &7.75       &0.39       &22.17    &1.61        &0.03        &4.60 \\
$\omega_\Lambda$  &1248.94   &113.78     &3572.04   &41.61      &22.87      &119.01   &11.60       &9.99        &33.17 \\
$n_s$          &126.90       &5.31       &362.93    &4.14       &0.96       &11.85    &0.77        &0.08        &2.22 \\ 
$Q$            &200.97       &18.38      &574.78    &2.99       &0.51       &8.55     &0.24        &0.07        &0.68 \\ 
$\R$           &254.76       &20.44      &728.63    &9.56       &0.35       &27.33    &1.19        &0.03        &3.40 \\ 
$\alpha$       &111.52       &3.74       &318.96    &2.66       &0.06       &7.62     &0.40        &$<0.01$        &1.14 \\ 
$\tau$         &275.13       &9.64       &786.88    &8.81       &2.78       &25.19    &2.26        &1.52        &6.45 \\ \hline 
	 & \multicolumn{9}{c}{Temperature} \\\hline
$\omega_b$     &13.56        &1.35       &38.78     &1.09       &0.60       &3.12     &0.83        &0.38        &2.37  \\       
$\omega_m$     &17.73        &0.88       &50.71     &3.76       &0.13       &10.74    &2.64        &0.08        &7.55  \\        
$\omega_\Lambda$ &137.68     &96.36      &393.77    &111.61     &96.15      &319.21   &98.97       &86.00       &283.05 \\       
$n_s$          &10.10        &0.53       &28.88     &2.18       &0.13       &6.24     &1.49        &0.07        &4.26  \\        
$Q$            &2.41         &0.36       &6.89      &0.20       &0.11       &0.57     &0.18        &0.07        &0.50  \\       
$\R$           &23.86        &0.78       &68.25     &1.58       &0.12       &4.53     &1.06        &0.07        &3.04  \\       
$\alpha$       &5.16         &0.13       &14.76     &0.66       &0.02       &1.88     &0.41        &0.01        &1.18  \\     
$\tau$         &111.97       &13.26      &320.24    &26.93      &8.28       &77.02    &20.32       &5.89        &58.11 \\ \hline 
	 & \multicolumn{9}{c}{Temperature and Polarization} \\\hline
$\omega_b$     &7.37         &1.34       &21.07     &0.91        &0.53      &2.61     &0.38        &0.21        &1.09  \\
$\omega_m$     &6.94         &0.88       &19.85     &1.81        &0.12      &5.17     &0.67        &0.03        &1.91 \\
$\omega_\Lambda$  &89.69     &72.75      &256.51    &30.89       &22.04     &88.36    &10.79       &9.85        &30.85\\
$n_s$          &2.32         &0.52       &6.65      &0.97        &0.13      &2.77     &0.33        &0.05        &0.93 \\
$Q$            &1.63         &0.36       &4.67      &0.19        &0.10      &0.54     &0.14        &0.05        &0.41\\
$\R$           &14.22        &0.78       &40.68     &1.43        &0.11      &4.08     &0.60        &0.03        &1.72 \\
$\alpha$       &3.03         &0.13       &8.68      &0.34        &0.02      &0.97     &0.11        &$<0.01$        &0.32 \\
$\tau$         &12.67        &7.90       &36.23     &4.48        &2.65      &12.80    &1.80        &1.48        &5.15 \\
\end{tabular}
\end{ruledtabular}
\end{table*}


\begin{figure*}
\includegraphics[width=3.5in]{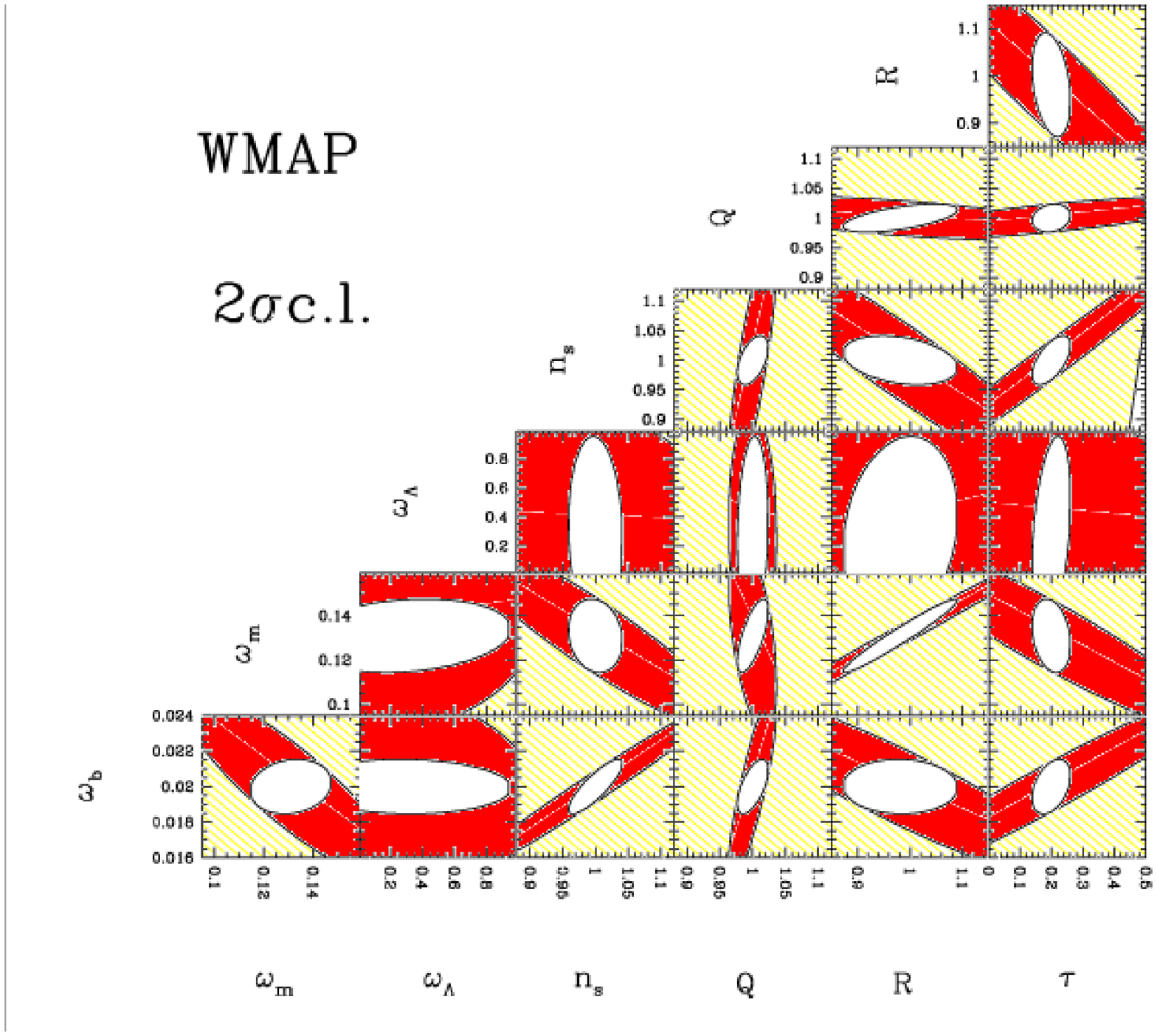}
\includegraphics[width=3.5in]{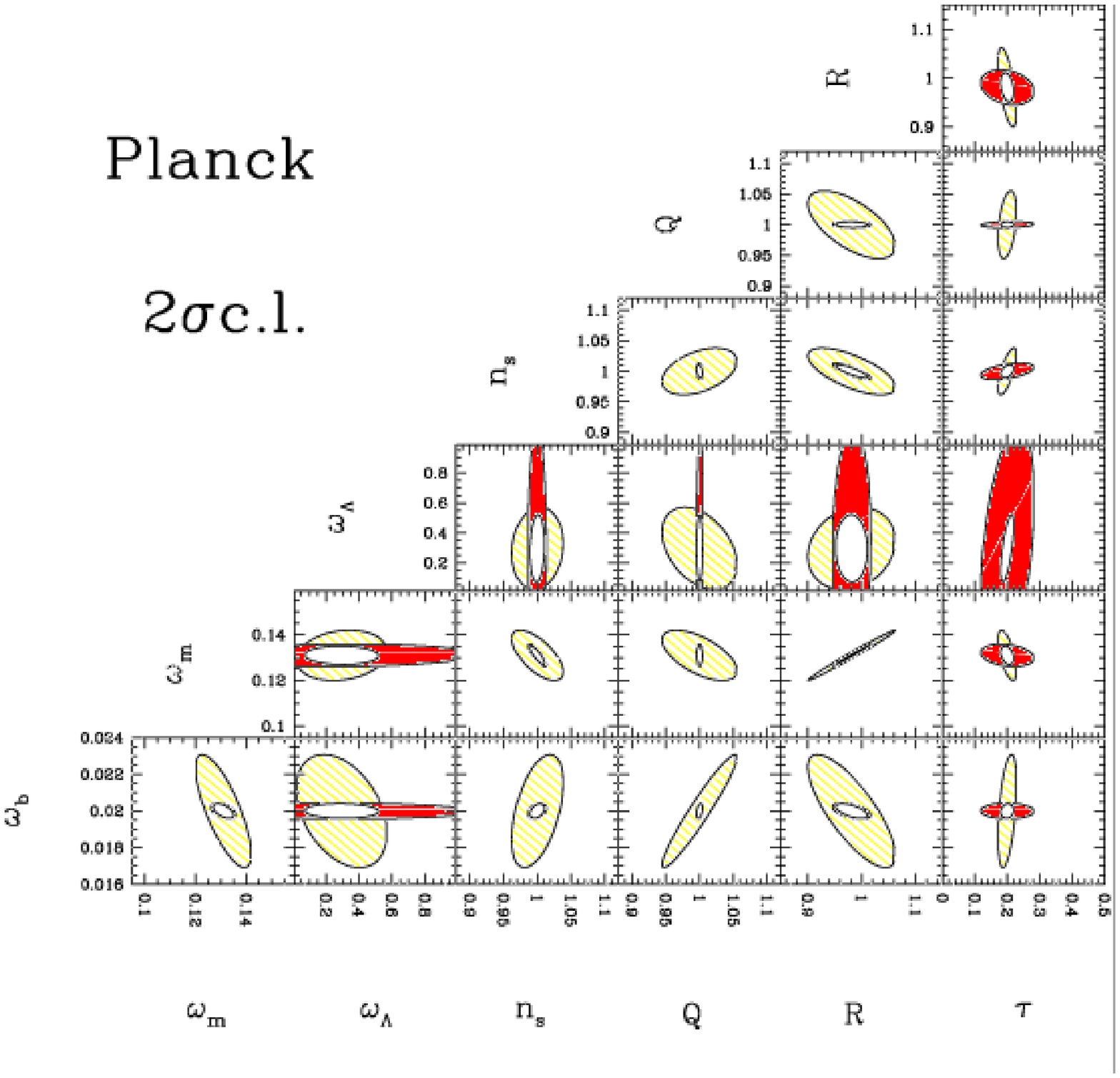}
\includegraphics[width=3.5in]{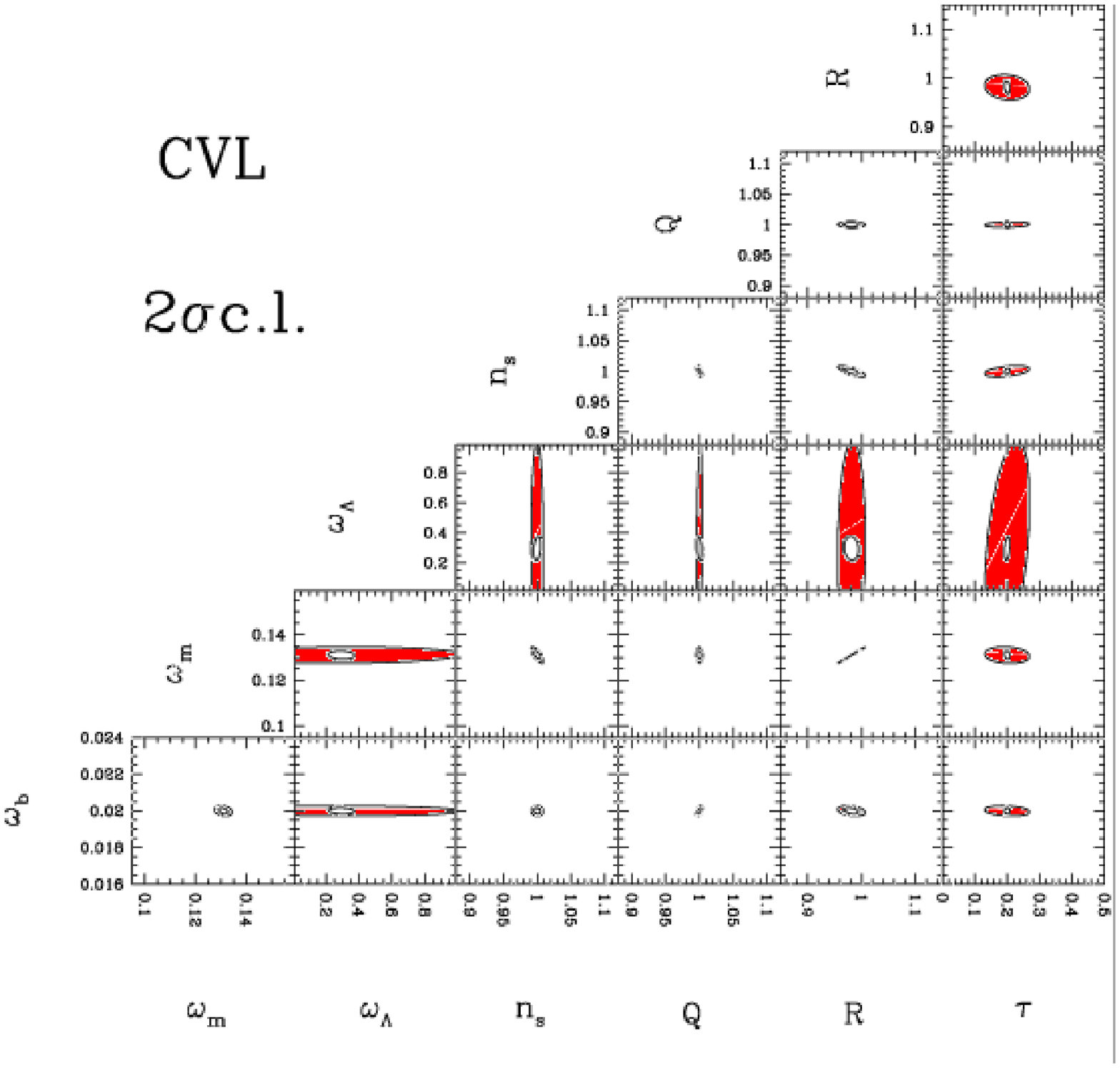}
\caption{\label{figellipse3}Ellipses containing $95.4\%$ ($2\sigma$) of
joint confidence (all other parameters marginalized) using temperature alone (red), E-polarization alone (yellow), and both jointly (white), for a  standard model with inclusion of reionization ($\tau=0.20$).
}
\end{figure*}

\begin{figure*}
\includegraphics[width=3.5in]{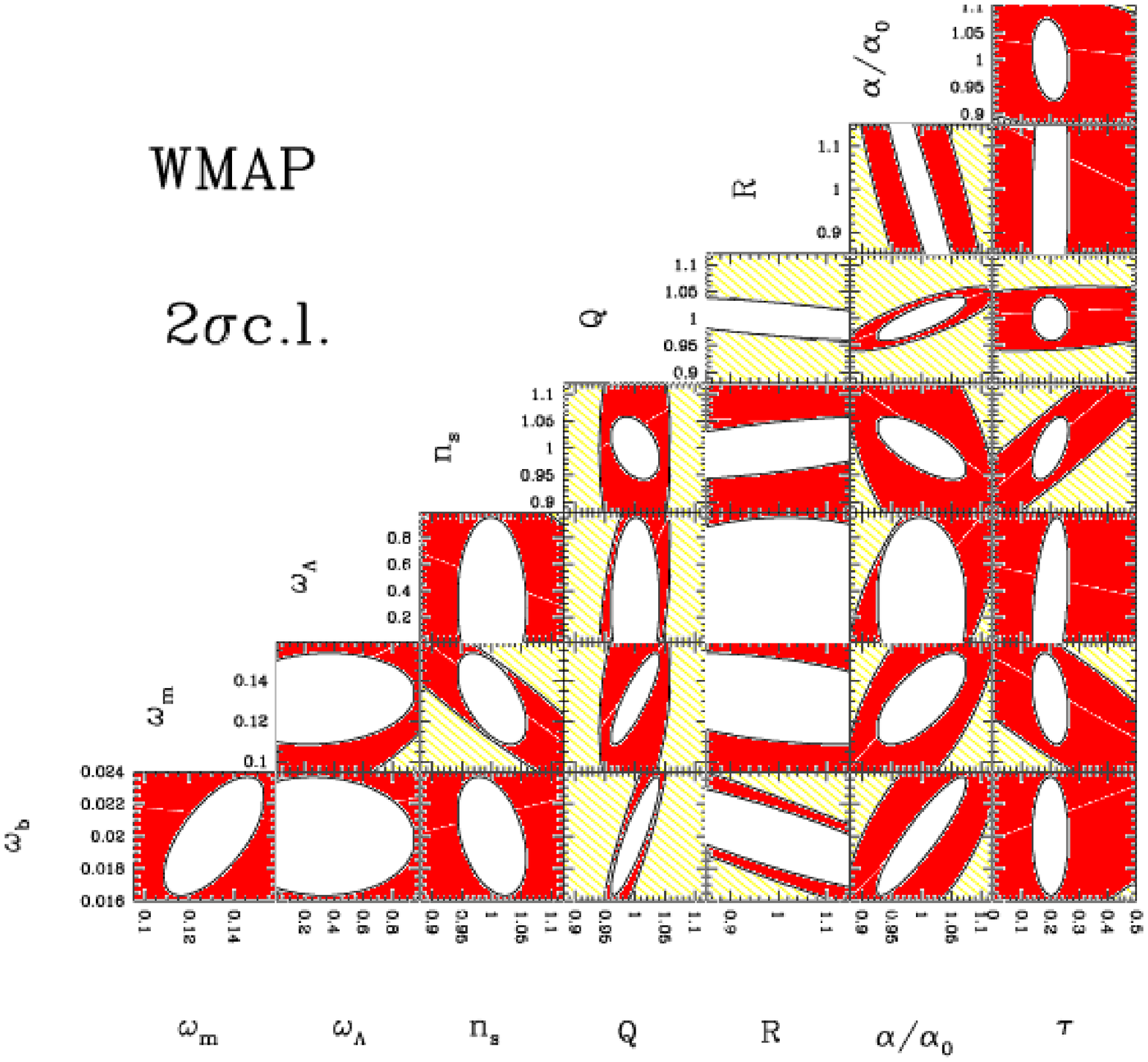}
\includegraphics[width=3.5in]{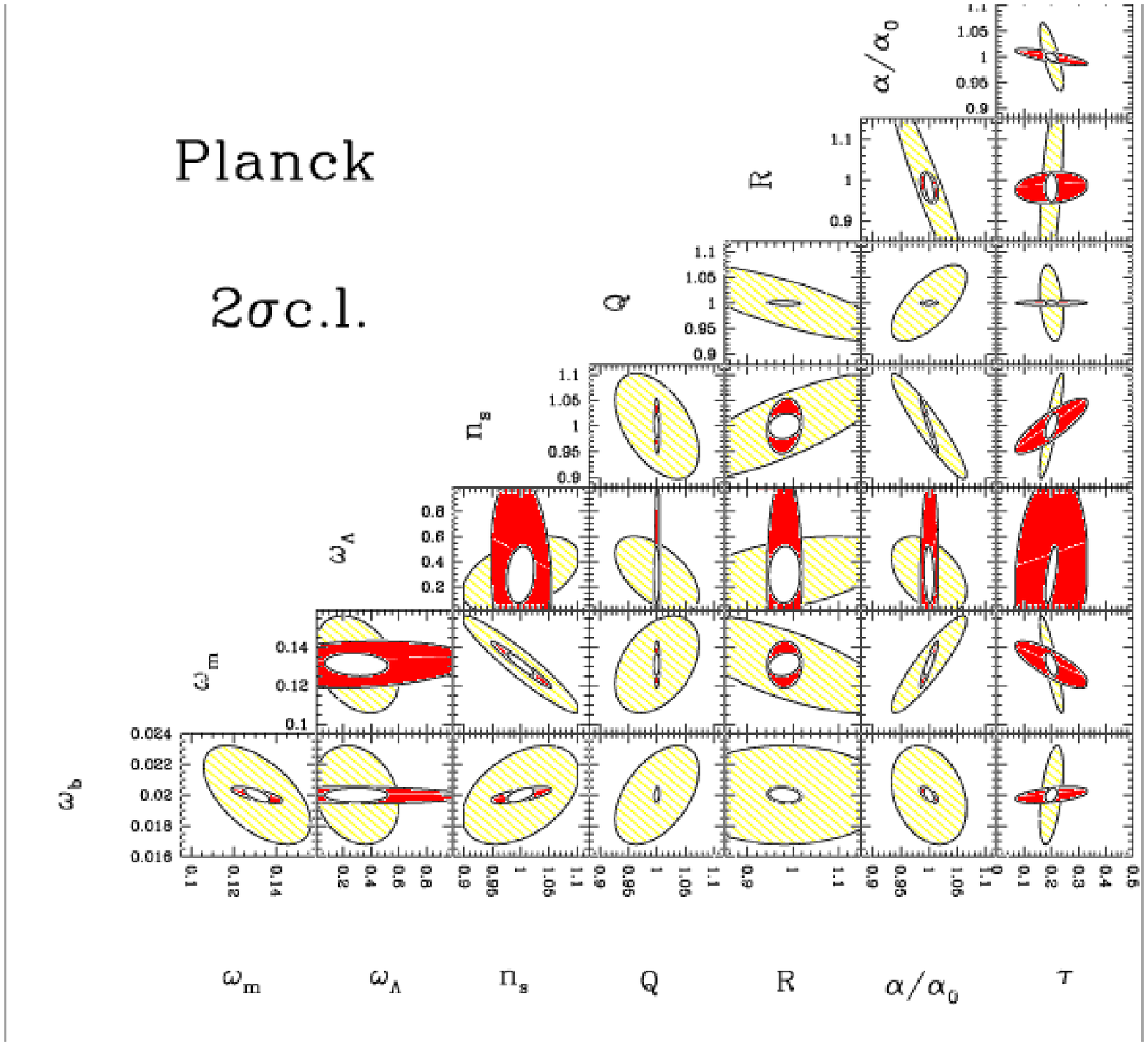}
\includegraphics[width=3.5in]{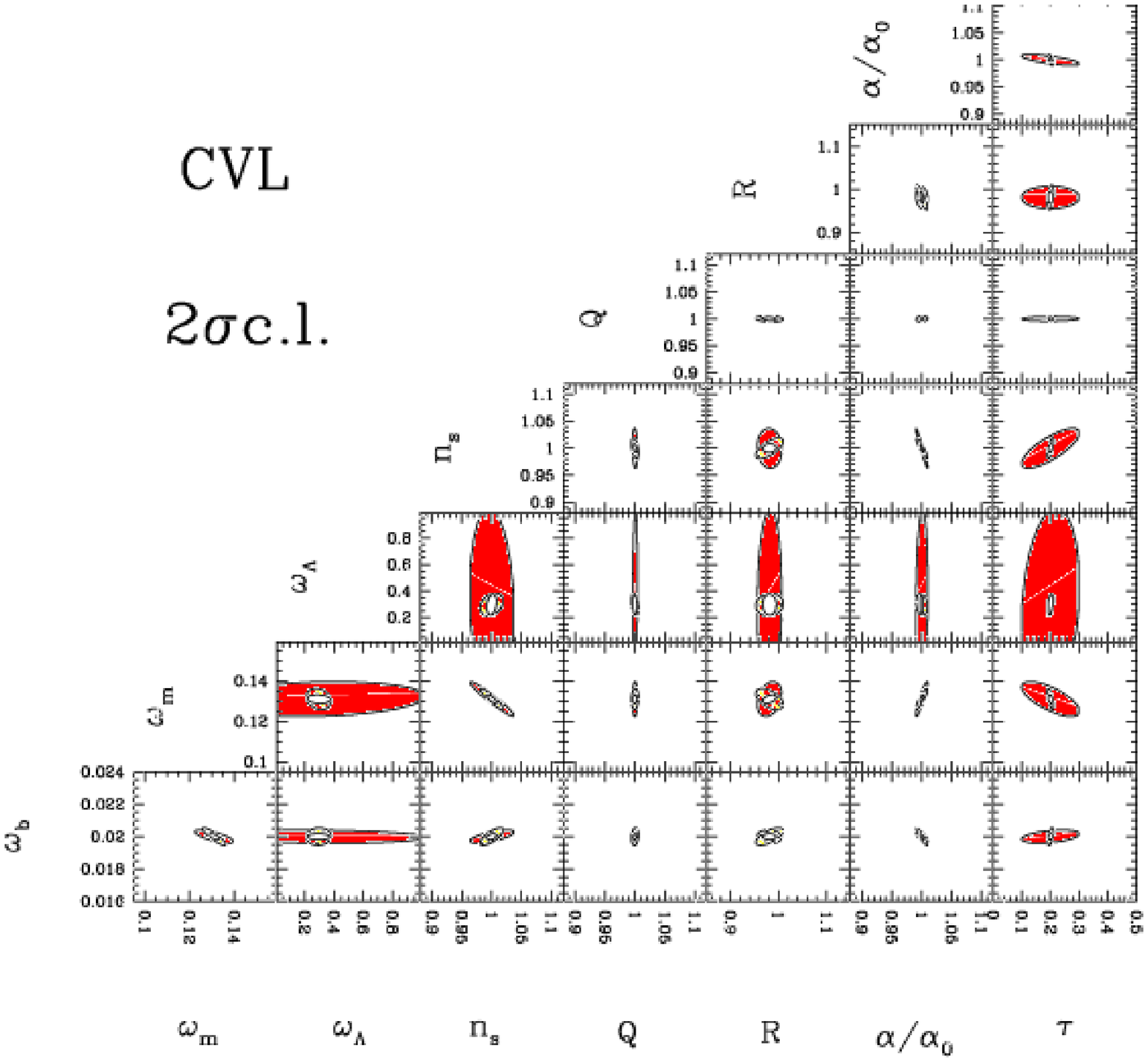}
\caption{\label{figellipse4}Ellipses containing $95.4\%$ ($2\sigma$) of
joint confidence (all other parameters marginalized) using temperature alone (red), E-polarization alone (yellow), and both jointly (white), for a model with varying $\alpha$ and inclusion of reionization ($\tau=0.20$).
}
\end{figure*}



\begin{table*}
\caption{\label{fmaev}For a model with a varying $\alpha$ and inclusion of reionization ($\tau=0.20$) and the case Temperature and Polarization considered jointly. In the lines we display the components
of the eigenvectors of the FM for WMAP,
Planck and a CVL experiment. The quantity $1/\sqrt{\lambda_i}$ is proportional to the error along
the principal direction ${\bf u^{(i)}}$. For each principal direction, an
asterisk marks the largest cosmological parameters contribution, a dagger
the second largest. Planck has errors smaller by a factor of about
5 on average than WMAP.}
\begin{ruledtabular}
\begin{tabular}{|c c c c c c c c c c|}
\multicolumn{10}{|c|}{WMAP} \\\hline
Direction $i$ & $1/\sqrt{\lambda_i}$ & $\omega_b$ & $\omega_m$ &
$\omega_\Lambda$ & $n_s$ & $Q$ & $\R$  & $\alpha$&  $\tau$ \\\hline
1&2.67E-04&     9.9485E-01* &    -9.5907E-02$\dag$  &     3.4445E-06  &    -2.9885E-02  &     1.8838E-03  &     1.0970E-02  &     7.0563E-03  &     1.4101E-03  \\
2&9.34E-04&     7.1608E-02  &     7.0264E-01* &    -5.5848E-04  &    -7.4249E-02  &     2.9712E-02  &    -1.1371E-01  &    -6.9403E-01$\dag$ &     1.2867E-02  \\
3& 2.37E-03&    5.6495E-02  &     5.3116E-01$\dag$ &     7.4496E-04  &     2.6323E-01  &    -6.4191E-01* &     3.8949E-02  &     4.8141E-01  &    -7.5954E-03  \\
4& 1.11E-02&    1.5030E-02  &    -1.1183E-01  &     2.6785E-02  &     7.5467E-01* &     8.1770E-02  &    -1.0330E-01  &    -1.8322E-01  &    -6.0506E-01$\dag$ \\
5&1.56E-02&     4.0281E-02  &     4.2554E-01  &    -1.1637E-02  &     1.8055E-01  &     7.5707E-01* &     1.4249E-01  &     4.2651E-01$\dag$  &     9.5864E-02  \\
6& 2.87E-02&    4.3952E-03  &    -1.4321E-01  &    -1.9274E-02  &     5.5866E-01$\dag$&    -3.4689E-02  &    -1.1323E-01  &    -1.7258E-01  &     7.8942E-01* \\
7&1.43E-01&    -8.9145E-03  &    -2.9306E-02  &    -6.9614E-02  &     1.0039E-01  &    -7.7565E-02  &     9.6787E-01* &    -2.0293E-01$\dag$  &     1.3044E-02  \\
8& 2.66E-01&   -4.7667E-04  &     3.1536E-03  &     9.9696E-01* &    -5.9578E-04  &     1.0494E-03  &     6.9739E-02$\dag$ &    -8.3540E-03  &     3.3561E-02  \\ \hline
\multicolumn{10}{|c|}{Planck} \\\hline
Direction $i$ & $1/\sqrt{\lambda_i}$ & $\omega_b$ & $\omega_m$ &
$\omega_\Lambda$ & $n_s$ & $Q$ & $\R$ & $\alpha$&  $\tau$ \\\hline
1&9.52E-05&     8.0730E-01* &    -4.3681E-01$\dag$  &     7.2721E-05  &     2.1375E-03  &    -1.7179E-02  &     6.5966E-02  &     3.9091E-01  &    -2.5317E-04  \\
2& 1.44E-04&    5.8762E-01* &     5.8285E-01$\dag$&     1.1644E-04  &    -3.6480E-02  &    -5.9816E-02  &    -8.5667E-02  &    -5.5021E-01 &     2.1552E-03  \\
3& 5.11E-04&    3.5099E-02  &     6.0865E-01$\dag$ &    -3.5134E-05  &     3.5326E-01  &     3.4997E-01  &     2.8062E-02  &     6.1633E-01* &    -1.9763E-02  \\
4& 1.89E-03&    4.0108E-02  &    -2.1450E-01  &     2.1539E-03  &    -6.8889E-02  &     9.3126E-01* &    -1.9318E-03  &    -2.8091E-01$\dag$  &    -3.8245E-02  \\
5&4.95E-03&    -4.2534E-03  &     1.0972E-01  &    -4.8443E-02  &    -4.2498E-01$\dag$  &     6.7120E-02  &     9.4317E-02  &     1.2132E-01  &     8.8140E-01* \\
6&1.10E-02&     1.0598E-02  &    -2.0227E-01  &    -4.5823E-02  &     8.0262E-01* &    -3.4226E-02  &    -2.1882E-01  &    -2.1656E-01  &     4.6553E-01$\dag$  \\
7& 1.43E-02&   -1.1429E-03  &     6.9110E-03  &    -1.3268E-02  &     2.0966E-01$\dag$  &    -2.6773E-02  &     9.6472E-01* &    -1.5502E-01  &     1.9638E-02  \\
8& 9.16E-02&    5.2242E-05  &    -3.4224E-03  &     9.9768E-01* &     1.9182E-02  &    -6.5898E-04  &     7.3693E-03  &    -5.4534E-03  &     6.4522E-02$\dag$  \\ \hline
\multicolumn{10}{|c|}{CVL} \\\hline
Direction $i$ & $1/\sqrt{\lambda_i}$ & $\omega_b$ & $\omega_m$ &
$\omega_\Lambda$ & $n_s$ & $Q$ & $\R$ & $\alpha$&  $\tau$ \\\hline
1& 2.67E-05&   -1.2266E-01  &     7.5163E-01* &     8.1216E-06  &    -4.6840E-03  &    -1.6753E-03  &    -1.0827E-01  &    -6.3895E-01$\dag$ &     1.9959E-04  \\
2&4.30E-05&     9.8772E-01* &     1.5389E-01$\dag$  &     4.2633E-05  &     2.3505E-02  &    -1.5836E-03  &    -1.1188E-02  &    -6.8644E-03  &    -5.1213E-04  \\
3& 2.27E-04&   -8.6274E-02  &     5.2862E-01$\dag$ &    -2.5551E-04  &     3.8678E-01  &     4.0618E-01  &     2.3505E-02  &     6.3051E-01* &    -2.1039E-02  \\
4& 1.28E-03&    4.3348E-02  &    -2.9931E-01$\dag$  &     5.8579E-03  &     9.2070E-02  &     8.7287E-01* &     6.9630E-03  &    -3.6458E-01  &    -7.1777E-02  \\
5&2.70E-03&    -2.8246E-03  &     1.2912E-01  &    -1.2398E-02  &    -6.1385E-01$\dag$&     2.2772E-01  &     8.4123E-02  &     1.4231E-01  &     7.2606E-01* \\
6& 3.93E-03&    5.6917E-03  &    -1.4889E-01  &    -4.9020E-02  &     6.7866E-01$\dag$ &    -1.4028E-01  &    -2.2646E-02  &    -1.7680E-01  &     6.8071E-01* \\
7& 5.96E-03&   -1.3903E-04  &     5.9101E-02  &     5.2914E-03  &     5.7747E-02  &    -3.8594E-02  &     9.8992E-01* &    -9.8529E-02$\dag$  &    -4.4876E-02  \\
8& 3.19E-02&   -7.2460E-05  &    -4.1402E-03  &     9.9869E-01* &     2.4943E-02  &    -8.8701E-03  &    -5.3456E-03  &    -4.0841E-03  &     4.3079E-02$\dag$  \\
\end{tabular}
\end{ruledtabular}
\end{table*}


%
\begin{table*}
\caption{\label{fmaevTT}For a model with a varying $\alpha$ and inclusion of reionization ($\tau=0.20$) and the case (TT) ie Temperature only. In the lines we display the components
of the eigenvectors of the FM for WMAP,
Planck and a CVL experiment. The quantity $1/\sqrt{\lambda_i}$ is proportional to the error along
the principal direction ${\bf u^{(i)}}$. For each principal direction, an
asterisk marks the largest cosmological parameters contribution, a dagger
the second largest. Planck has errors smaller by a factor of about
5 on average than WMAP.}
\begin{ruledtabular}
\begin{tabular}{|c c c c c c c c c c|}
\multicolumn{10}{|c|}{WMAP} \\\hline
Direction $i$ & $1/\sqrt{\lambda_i}$ & $\omega_b$ & $\omega_m$ &
$\omega_\Lambda$ & $n_s$ & $Q$ & $\R$  & $\alpha$&  $\tau$ \\\hline
1& 2.68E-04&     9.9485E-01* &    -9.5954E-02$\dag$  &    -3.8023E-05  &    -2.9988E-02  &     1.9066E-03  &     1.1031E-02  &     6.9603E-03  &     1.6935E-03  \\
2& 9.35E-04&     7.1581E-02  &     7.0263E-01* &    -5.4546E-04  &    -7.3932E-02  &     2.9625E-02  &    -1.1370E-01  &    -6.9410E-01$\dag$ &     1.1876E-02  \\
3& 2.37E-03&     5.6582E-02  &     5.3096E-01$\dag$ &     7.0218E-04  &     2.6304E-01  &    -6.4218E-01* &     3.9069E-02  &     4.8138E-01  &    -6.9515E-03  \\
4& 1.23E-02&     2.0318E-02  &    -8.8391E-02  &     2.0436E-02  &     8.6583E-01* &     1.5967E-01  &    -1.0780E-01  &    -1.6237E-01  &    -4.2215E-01$\dag$  \\
5&  1.58E-02&    3.8004E-02  &     4.4637E-01  &    -1.3439E-02  &     5.1628E-02  &     7.4470E-01* &     1.6739E-01  &     4.5598E-01$\dag$  &     7.7154E-02  \\
6&  1.72E-01&   -4.5506E-03  &    -6.9729E-02  &     3.1859E-01  &     3.0219E-01  &    -5.4673E-02  &     6.8576E-01* &    -2.0902E-01  &     5.3419E-01$\dag$ \\
7& 2.71E-01&     8.7518E-03  &    -4.7970E-02  &    -1.4708E-02  &     2.5961E-01  &     5.4289E-02  &    -6.4792E-01$\dag$ &     4.5317E-02  &     7.1078E-01* \\ 
8& 4.26E-01&     1.8060E-03  &     3.0947E-02  &     9.4746E-01* &    -1.1576E-01  &     2.6839E-02  &    -2.3604E-01$\dag$  &     8.0201E-02  &    -1.5838E-01  \\ \hline
\multicolumn{10}{|c|}{Planck} \\\hline
Direction $i$ & $1/\sqrt{\lambda_i}$ & $\omega_b$ & $\omega_m$ &
$\omega_\Lambda$ & $n_s$ & $Q$ & $\R$ & $\alpha$&  $\tau$ \\\hline
1& 1.05E-04&     7.4727E-01* &    -4.9695E-01$\dag$  &    -5.0448E-07  &     3.3342E-03  &    -1.0958E-02  &     7.3355E-02  &     4.3488E-01  &     4.5857E-04  \\
2& 1.57E-04&     6.6077E-01* &     5.1936E-01&    -1.9827E-05  &    -4.4896E-02  &    -6.7337E-02  &    -7.9760E-02  &    -5.2983E-01$\dag$ &     3.8563E-03  \\
3& 5.25E-04&     6.1332E-02  &     6.1587E-01* &     1.2590E-05  &     3.5518E-01  &     3.5940E-01  &     2.5334E-02  &     6.0046E-01$\dag$ &    -2.0469E-02  \\
4& 1.96E-03&     3.3384E-02  &    -2.1678E-01  &    -7.5551E-04  &    -9.5447E-02  &     9.2928E-01* &     1.9606E-03  &    -2.8126E-01$\dag$  &    -1.0701E-02  \\
5& 1.04E-02&     9.3302E-03  &    -2.2248E-01  &     1.7817E-02  &     8.6225E-01* &    -4.7210E-02  &    -8.6676E-02  &    -2.6306E-01  &    -3.5733E-01$\dag$  \\
6&  1.55E-02&   -3.0476E-03  &     4.7288E-02  &     1.5170E-03  &     4.6850E-02  &    -1.9767E-02  &     9.8892E-01* &    -1.0832E-01$\dag$ &    -7.3922E-02  \\
7&  5.73E-02&    1.9693E-03  &    -7.2588E-02  &    -1.4866E-02  &     3.4196E-01$\dag$  &    -8.5608E-04  &     4.6134E-02  &    -9.7738E-02  &     9.3053E-01* \\ 
8&  3.30E-01&   -9.4432E-05  &     2.6522E-03  &     9.9973E-01* &    -1.0430E-02  &     1.5550E-03  &     7.2972E-04  &     3.1688E-03  &     2.0310E-02$\dag$  \\ \hline
\multicolumn{10}{|c|}{CVL} \\\hline
Direction $i$ & $1/\sqrt{\lambda_i}$ & $\omega_b$ & $\omega_m$ &
$\omega_\Lambda$ & $n_s$ & $Q$ & $\R$ & $\alpha$&  $\tau$ \\\hline
1& 5.85E-05&     6.7166E-01* &    -4.8908E-01  &     2.0754E-07  &     4.0071E-02  &     2.8188E-02  &     8.0901E-02  &     5.4839E-01$\dag$ &    -1.5833E-03  \\
2& 1.16E-04&     7.3358E-01* &     5.4117E-01$\dag$&    -1.3289E-05  &     5.7146E-04  &    -2.3329E-02  &    -7.2058E-02  &    -4.0405E-01  &     1.3803E-03  \\
3& 2.93E-04&    -9.6933E-02  &     6.0597E-01$\dag$ &     1.5493E-05  &     3.5035E-01  &     3.5105E-01  &     1.0230E-02  &     6.1395E-01* &    -1.9331E-02  \\
4&  1.72E-03&    3.5113E-02  &    -2.1156E-01  &    -7.1277E-04  &    -7.6886E-02  &     9.3352E-01* &     1.4161E-02  &    -2.7618E-01$\dag$  &    -1.2973E-02  \\
5&  8.45E-03&    9.6096E-03  &    -1.9881E-01  &     1.4790E-02  &     8.6840E-01* &    -6.2132E-02  &     1.6748E-01  &    -2.7495E-01  &    -3.1390E-01$\dag$  \\
6&  1.05E-02&   -2.6411E-03  &     1.1073E-01  &    -4.8547E-03  &    -1.5346E-01$\dag$  &    -1.0584E-02  &     9.7974E-01* &    -3.0626E-02  &     5.6666E-02  \\
7&  4.19E-02&    1.9014E-03  &    -6.4709E-02  &    -2.4494E-02  &     3.0336E-01$\dag$  &     2.7857E-05  &    -2.5345E-03  &    -7.9100E-02  &     9.4706E-01* \\
8&  2.93E-01&   -7.2266E-05  &     1.7408E-03  &     9.9958E-01* &    -6.2208E-03  &     1.5285E-03  &     2.2272E-03  &     1.7693E-03  &     2.8118E-02$\dag$  \\
\end{tabular}
\end{ruledtabular}
\end{table*}


\begin{table*}
\caption{\label{fmasttaulow}Fisher matrix analysis results for a standard model with inclusion of reionization (for $\tau=0.02$): expected $1\sigma$ errors for the WMAP and Planck satellites as well as for a CVL experiment. The column {\it marg.} gives the error with all
other parameters being marginalized over; in the column {\it fixed} the other
parameters are held fixed at their ML value; in the column {\it joint} all
parameters are being estimated jointly.}
\begin{ruledtabular}
\begin{tabular}{|c|c c c| c c c|c c c|}
Quantity &  \multicolumn{9}{c}{$1\sigma$ errors (\%)} \\\hline  
             & \multicolumn{3}{c|}{WMAP}           & \multicolumn{3}{c}{Planck HFI} & \multicolumn{3}{c}{CVL} \\ 
                        & marg. & fixed  & joint   & marg.  & fixed & joint  & marg.  & fixed & joint           \\\hline
	 & \multicolumn{9}{c}{Polarization} \\\hline

$\omega_b$       &241.44   &50.73    &690.54      &6.36   &1.01    &18.18    &0.48   &0.25    &1.38 \\  
$\omega_m$       &99.44    &31.71    &284.39      &3.55   &0.34    &10.14    &0.70   &0.03    &2.01 \\  
$\omega_\Lambda$ &1201.35  &719.21   &3435.95     &39.02  &33.98   &111.61   &11.55  &10.20   &33.05 \\   
$n_s$            &125.97   &19.26    &360.29      &1.48   &0.91    &4.22     &0.30   &0.08    &0.86 \\  
$Q$              &151.63   &25.09    &433.68      &2.20   &0.45    &6.30     &0.24   &0.07    &0.68 \\  
$\R$             &87.25    &22.00    &249.55      &3.50   &0.31    &10.01    &0.66   &0.03    &1.89 \\  
$\tau$           &228.76   &63.74    &654.28      &11.45  &10.29   &32.75    &4.23   &4.10    &12.10 \\ \hline
	 & \multicolumn{9}{c}{Temperature} \\\hline
$\omega_b$       &6.00     &1.27    &17.16     &0.83     &0.59     &2.37     &0.56    &0.38   &1.59 \\
$\omega_m$       &8.63     &0.83    &24.69     &1.47     &0.13     &4.20     &1.09    &0.08   &3.12 \\
$\omega_\Lambda$ &173.23   &89.11   &495.44    &94.22    &88.94    &269.48   &83.32   &79.55  &238.30 \\
$n_s$            &4.42     &0.52    &12.64     &0.50     &0.13     &1.43     &0.34    &0.07   &0.98 \\
$Q$              &0.90     &0.35    &2.58      &0.19     &0.10     &0.55     &0.17    &0.07   &0.49 \\
$\R$             &8.78     &0.74    &25.10     &1.43     &0.11     &4.10     &1.05    &0.07   &3.00 \\
$\tau$           &659.96   &195.96  &1887.52   &163.30   &126.81   &467.05   &132.38  &96.66  &378.61 \\ \hline
	 & \multicolumn{9}{c}{Temperature and Polarization} \\\hline
$\omega_b$      &2.77     &1.26    &7.93     &0.77       &0.51     &2.21     &0.32    &0.21   &0.91 \\
$\omega_m$      &4.54     &0.83    &12.99    &1.17       &0.12     &3.34     &0.55    &0.03   &1.58 \\
$\omega_\Lambda$ &109.71  &87.68   &313.79   &32.15      &31.29    &91.95    &10.36   &9.88   &29.63 \\
$n_s$           &1.47     &0.52    &4.21     &0.39       &0.13     &1.13     &0.20    &0.06   &0.57 \\
$Q$             &0.81     &0.35    &2.33     &0.18       &0.10     &0.52     &0.14    &0.05   &0.41 \\
$\R$            &4.10     &0.74    &11.72    &1.14       &0.11     &3.27     &0.52    &0.03   &1.49 \\
$\tau$          &63.32    &60.36   &181.09   &10.38      &10.06    &29.69    &3.87    &3.81   &11.07 \\ 
\end{tabular}
\end{ruledtabular}
\end{table*}


\begin{table*}
\caption{\label{fmaaltaulow}Fisher matrix analysis results for a model with varying $\alpha$ and inclusion of reionization (for $\tau=0.02$): expected $1\sigma$ errors for the WMAP and Planck satellites as well as for a CVL experiment. The column {\it marg.} gives the error with all
other parameters being marginalized over; in the column {\it fixed} the other
parameters are held fixed at their ML value; in the column {\it joint} all
parameters are being estimated jointly.}
\begin{ruledtabular}
\begin{tabular}{|c|c c c| c c c|c c c|}
Quantity &  \multicolumn{9}{c}{$1\sigma$ errors (\%)} \\\hline  
             & \multicolumn{3}{c|}{WMAP}           & \multicolumn{3}{c}{Planck HFI} & \multicolumn{3}{c}{CVL} \\ 
                        & marg. & fixed  & joint   & marg.  & fixed & joint  & marg.  & fixed & joint           \\\hline
	 & \multicolumn{9}{c}{Polarization} \\\hline
$\omega_b$     &569.33       &50.73   &1628.32    &6.41     &1.01    &18.32    &1.11    &0.25   &3.17 \\
$\omega_m$     &716.71       &31.71   &2049.84    &7.22     &0.34    &20.66    &1.65    &0.03   &4.71 \\
$\omega_\Lambda$  &1439.68   &719.21  &4117.59    &42.43    &33.98   &121.36   &12.22   &10.20  &34.96 \\
$n_s$          &299.32       &19.26   &856.07     &3.91     &0.91    &11.19    &0.79    &0.08   &2.25 \\
$Q$            &174.27       &25.09   &498.41     &3.15     &0.45    &9.00     &0.24    &0.07   &0.69 \\ 
$\R$           &419.62       &22.00   &1200.15    &9.87     &0.31    &28.23    &1.19    &0.03   &3.40 \\ 
$\alpha$       &192.47       &3.57    &550.48     &2.59     &0.05    &7.42     &0.40    &$<0.01$   &1.15 \\ 
$\tau$         &875.90       &63.74   &2505.14    &15.15    &10.29   &43.34    &4.73    &4.10   &13.52 \\ \hline
	 & \multicolumn{9}{c}{Temperature} \\\hline
$\omega_b$       &14.24     &1.27     &40.73      &1.02    &0.59    &2.92      &0.79    &0.38   &2.27 \\
$\omega_m$       &9.93      &0.83     &28.41      &2.94    &0.13    &8.42      &2.23    &0.08   &6.37 \\
$\omega_\Lambda$ &173.24    &89.11    &495.49     &108.85  &88.94   &311.31    &93.56   &79.55  &267.59 \\
$n_s$            &4.59      &0.52     &13.12      &1.58    &0.13    &4.51      &1.16    &0.07   &3.32 \\
$Q$              &2.44      &0.35     &6.99       &0.20    &0.10    &0.56      &0.17    &0.07   &0.50 \\
$\R$             &26.80     &0.74     &76.65      &1.51    &0.11    &4.31      &1.06    &0.07   &3.03 \\
$\alpha$         &5.00      &0.12     &14.31      &0.49    &0.02    &1.41      &0.34    &0.01   &0.96 \\
$\tau$           &710.55    &195.96   &2032.22    & 193.10 &126.81  &552.27    &148.41  &96.66  &424.46 \\ \hline
	 & \multicolumn{9}{c}{Temperature and Polarization} \\\hline
$\omega_b$      &9.59       &1.26     &27.43     &0.87    &0.51     &2.50      &0.38    &0.21   &1.10 \\
$\omega_m$      &8.25       &0.83     &23.59     &1.63    &0.12     &4.65      &0.67    &0.03   &1.91 \\ 
$\omega_\Lambda$ &120.16    &87.68    &343.67    &32.15   &31.29    &91.95     &10.45   &9.88   &29.89 \\
$n_s$           &2.97       &0.52     &8.51      &0.86    &0.13     &2.47      &0.32    &0.06   &0.92 \\
$Q$             &1.99       &0.35     &5.69      &0.19    &0.10     &0.53      &0.14    &0.05   &0.41 \\
$\R$            &19.47      &0.74     &55.69     &1.36    &0.11     &3.90      &0.60    &0.03   &1.72 \\
$\alpha$        &4.32       &0.12     &12.34     &0.31    &0.02     &0.89      &0.11    &$<0.01$   &0.32 \\
$\tau$          &64.65      &60.36    &184.91    &10.52   &10.06    &30.09     &3.91    &3.81   &11.18 \\
\end{tabular}
\end{ruledtabular}
\end{table*}


\section{FMA with reionization}

The existence of a period when the intergalactic medium was reionized as 
well as its driving mechanism are still to be understood.
One  possible way of studying this phase is via the CMB polarization 
anisotropy. The optical depth to electrons of the CMB photons enhances 
the polarization signal at large angular scales (see Fig. \ref{figcells}) introducing 
a bump in the polarization spectrum at small multipoles. On the other hand
reionization decreases the amplitude of the acoustic peaks on the temperature
power spectrum at intermediate and small angular scales 
This signal has now been detected by WMAP via the
temperature polarization cross power-spectrum \cite{Kogut}.

In the absence of polarization observations, the optical depth to Thomson
scattering is degenerate with the amplitude of the fluctuations, $Q$ (with
$Qe^{\tau} = constant$).
From previous Fisher Matrix Analysis for a standard model, e.g. \cite{zaldarriaga2}, one expects 
that the inclusion of polarization measurements will help to better constrain 
some of the cosmological parameters, by probing the ionization history of the universe, hence constraining $\tau$ and breaking degeneracies of this with other parameters. We will now repeat the analysis of the previous
chapter for the case $\tau\neq0$.

\subsection{Analysis results: The FMA forecast}

Tables \ref{fmasttau}--\ref{fmaaltaulow} summarize the results of our FMA for 
WMAP, Planck and a CVL experiment. We consider the cases of models with and
without a varying $\alpha$ being included in the analysis, and also two
values of the optical depth, $\tau=0.2$ (close to the one preferred by WMAP)
and $\tau=0.02$.
We also consider the use of temperature information alone
(TT), E-polarization alone (EE) and both channels (EE+TT) jointly.
To show that our FMA fiducial model is close enough to the WMAP best fit model to produce similar FMA results, we display in 
Table \ref{fmaalwmap} the results of our FMA using as fiducial model the WMAP best fit model.

For the sake of completeness we also consider the case (TE) alone as well as (EE+TE) and (EE+TT+TE) for WMAP 4-years.  
Table \ref{fmaalwmap4} display the results of our FMA for WMAP 4-years using the WMAP fiducial model.
The FMA predictions for WMAP - 4years are to be compared with the recent WMAP 1-year results.

The errors in most of the other cosmological parameters are unaffected by the
presence of reionization \textit{if} one has both temperature and
polarization data. If one has just one of them then the accuracy is
quite different, and also it will depend on whether has high or low $\tau$.
This is because different degeneracies may be dominant in each case, while
combining temperature and polarization information helps break such
degeneracies.

The inclusion of the new parameter $\tau$ for a standard model 
reduces the
accuracy in other parameters for all but the CVL experiment (and in this case for all but $\omega_{\Lambda}$) as can be seen from a comparison of 
Table \ref{fmasttau} with Table \ref{fmast}.

Comparing Tables \ref{fmasttau} (for $\tau=0.20$) and  \ref{fmasttaulow} (for $\tau=0.02$) an immediate effect of considering a large value of $\tau$ is to increase the accuracy on $\tau$ itself. For example the case with temperature and polarization information used jointly, the accuracy on the other parameters is not necessarily reduced by considering a larger value of $\tau$ while its accuracy remains almost the same for a CVL experiment.   
Whilst comparing Tables \ref{fmaaltau} (for $\tau=0.20$) and \ref{fmaaltaulow} (for $\tau=0.02$) the effect of a large value of $\tau$, considering the case temperature and polarization used jointly, for WMAP is to increase the accuracy on  most of the parameters particularly noticeable for the parameters $\alpha$ and $\tau$; for Planck only the accuracy on $\tau$ is improved while the other parameters have slightly worse accuracy; finally for a CVL experiment
 the accuracy is the same for all but $\omega_{\Lambda}$ which is slightly worse, and $\tau$ which is much better.
It is interesting to note that while for WMAP a large value of $\tau$ does indeed help to improve the accuracy on most parameters, for Planck and a CVL experiment the accuracy is improved using polarization data alone but the inverse is true using temperature data alone. Hence it is not surprising the results obtained when one considers temperature and polarization jointly. 

As we go from Table \ref{fmaalwmap} to Table \ref{fmaalwmap4} the accuracy on all parameters increases as should be expected. For the WMAP - 4 years one predicts an accuracy of 3\% and 11\% on $\alpha$ and $\tau$ respectively as opposed to 4\% and 14\% respectively, for the 2-year mission.  

The results of our forecast are 
that WMAP (2-years mission) is able to constrain $\tau$ with accuracy of the order 13\%, 
which is approximately two times better than the current precision  obtained from the WMAP 1-year
observations, of the order of 23\%. While our FMA predictions for WMAP - 4 years, gives an accuracy of the order 10\% using all (TT+EE+TE) temperature, polarization and temperature-polarization cross correlation information. 

Planck and a CVL experiment can constrain $\alpha$ with accuracies of 
the order 0.3\% and 0.1\% respectively and $\tau$ with accuracies of the order 4.5\% and 1.8\% respectively.

For WMAP the accuracy on $\tau$ from polarization data alone is worse by a factor of 2 than from temperature alone. On the other hand, for Planck and the CVL
experiment the accuracy from polarization is better by a factor of 3 and 8 respectively, than from temperature alone.
While the accuracy on $\alpha$ from polarization alone is worse by a factor of the order 22 and 4 than from temperature alone for WMAP and Planck respectively. For a CVL experiment the accuracies are similar for both polarization and temperature data alone.

The accuracy on $\tau$ obtained with Planck using Temperature data alone 
is roughly the same as a CVL experiment. This suggests that Planck is indeed 
a cosmic variance limited experiment with respect to Temperature. The 
inclusion of polarization the accuracy for the CVL experiment is improved by 
a factor of 4 when compared to Planck satellite.

\subsection{Analysis results: Confidence contours}

As before, we show in Figs. \ref{figellipse3}-\ref{figellipse4} all joint 2D
confidence contours (all remaining parameters marginalized).
As previously in the WMAP case the errors from E only are very large, hence the contours for T coincide almost exactly with the temperature-polarization combined case. In the CVL case it is the E contours that almost coincide with the combined ones.  

From Fig. \ref{figellipse3} without $\alpha$, we can infer a good agreement between our
predictions and WMAP observations. Particularly striking is the good 
agreement for the contour plots in the ($n_{s}$,$\tau$) plane which 
clearly exhibits the observed degeneracy \cite{Spergel}.
For Planck the inclusion of polarization data helps to break degeneracies 
in particular between $\tau$ and the other parameters for example with 
$n_{s}$. For a CVL experiment the contours are further narrowed with the 
joint temperature polarization analysis in agreement with the tabulated
accuracies on $\tau$.

Again, looking at Fig. \ref{figellipse4} with $\alpha$, our predictions for the contour 
plots in the plane ($\tau$,$n_{s}$) are in close agreement with the 
observed degeneracy \cite{Verde}. This same plot shows that the 
degeneracy direction between $\alpha$ and $n_{s}$ is almost orthogonal to 
that between $\tau$ and $n_{s}$, 
The net result of this is a better accuracy on $\alpha$ when the 
parameter $\tau$ is included (compare Tables \ref{fmaal} and \ref{fmaaltau})
while the accuracy on $\tau$ itself remains almost unchanged with inclusion 
of $\alpha$ (compare Tables \ref{fmasttau} and \ref{fmaaltau}). 
This is in agreement with our discussion in section~III, and quantitatively 
explains why our $\alpha$ mechanism (summarized in Fig. \ref{figpeaks}) works. 
The accuracy 
on $n_{s}$ is similar to that obtained without $\tau$ (compare Tables
\ref{fmaal} and \ref{fmaaltau}) but gets worse with inclusion of $\alpha$
(compare Tables \ref{fmasttau} and \ref{fmaaltau}).

In other words the inclusion of 
reionization helps to lift most of the degeneracies when using information 
from both the temperature and polarization jointly hence increasing the accuracies for the cases of interest , ie, $\alpha$ and $\tau$.


\begin{table*}
\caption{\label{fmaalwmap}Fisher matrix analysis results for a model with varying $\alpha$ and inclusion of reionization (for WMAP best fit model as the fisher analysis fiducial model, $\tau=0.17$): expected $1\sigma$ errors for the WMAP and Planck satellites as well as for a CVL experiment. The column {\it marg.} gives the error with all
other parameters being marginalized over; in the column {\it fixed} the other
parameters are held fixed at their ML value; in the column {\it joint} all
parameters are being estimated jointly.}
\begin{ruledtabular}
\begin{tabular}{|c|c c c| c c c|c c c|}
Quantity &  \multicolumn{9}{c}{$1\sigma$ errors (\%)} \\\hline  
             & \multicolumn{3}{c|}{WMAP}           & \multicolumn{3}{c}{Planck HFI} & \multicolumn{3}{c}{CVL} \\ 
                        & marg. & fixed  & joint   & marg.  & fixed & joint  & marg.  & fixed & joint           \\\hline
	 & \multicolumn{9}{c}{Polarization} \\\hline
$\omega_b$  & 285.33     &  26.18    &  816.08    &    5.84     &   0.87     &  16.70    &   0.96      &  0.12     &   2.73 \\   
$\omega_m$  & 445.06     &  28.16    & 1272.90    &    7.48     &   0.46     &  21.41    &   1.40      &  0.03     &   4.00 \\   
$\omega_\Lambda$  & 184.17     & 144.61    & 3386.80    &   44.12     &  24.08     & 126.18    &  12.83      &  9.33     &  36.70  \\  
$n_s$      & 161.11     &   6.14    &  460.78    &    4.22     &   1.00     &  12.08    &   0.71      &  0.08     &   2.04 \\   
$Q$        & 191.24     &  21.06    &  546.95    &    2.91     &   0.55     &   8.32    &   0.25      &  0.07     &   0.73 \\   
$\R$       & 221.83     &  21.69    &  634.44    &    8.81     &   0.35     &  25.19    &   0.79      &  0.02     &   2.26 \\   
$\alpha$   & 113.11     &   4.52    &  323.49    &    2.61     &   0.07     &   7.48    &   0.32      &  0.00     &   0.91 \\   
$\tau$     & 336.62     &  11.25    &  962.75    &    9.25     &   3.05     &  26.45    &   2.32      &  1.30     &   6.63 \\ \hline 
 	 & \multicolumn{9}{c}{Temperature} \\\hline 
$\omega_b$  &  18.50     &   0.98    &   52.91    &    0.98     &   0.35     &   2.80    &   0.73      &  0.24     &   2.08  \\  
$\omega_m$  &  17.89     &   0.94    &   51.17    &    3.30     &   0.14     &   9.45    &   2.31      &  0.08     &   6.60  \\  
$\omega_\Lambda$  & 149.92     &  83.49    &  428.77    &  107.48     &  83.30     & 307.39    &  94.61      & 74.50     & 270.59  \\  
$n_s$      &   9.50     &   0.54    &   27.17    &    2.07     &   0.14     &   5.91    &   1.42      &  0.07     &   4.06  \\  
$Q$        &   3.27     &   0.37    &    9.36    &    0.21     &   0.11     &   0.60    &   0.19      &  0.07     &   0.53  \\  
$\R$       &  34.95     &   0.72    &   99.97    &    1.34     &   0.10     &   3.84    &   0.86      &  0.06     &   2.45  \\  
$\alpha$   &   7.95     &   0.13    &   22.75    &    0.59     &   0.02     &   1.69    &   0.37      &  0.01     &   1.06  \\  
$\tau$     & 119.62     &  17.00    &  342.11    &   32.86     &   9.93     &  93.98    &  25.31      &  6.84     &  72.38  \\ \hline 
	 & \multicolumn{9}{c}{Temperature and Polarization} \\\hline 
$\omega_b$  &   9.15     &   0.98    &   26.18    &    0.84     &   0.32     &   2.39         &    0.37   &     0.11   &     1.07    \\  
$\omega_m$  &   7.55     &   0.94    &   21.58    &    1.62     &   0.13     &   4.65         &    0.61   &     0.03   &     1.75    \\  
$\omega_\Lambda$ &  95.34     &  71.51    &  272.68    &   32.24     &  22.94     &  92.22    &   11.80   &     9.21   &    33.76    \\  
$n_s$     &   2.58     &   0.54    &    7.39    &    0.93     &   0.14     &   2.67           &    0.33   &     0.05   &     0.94    \\  
$Q$       &   1.77     &   0.37    &    5.06    &    0.19     &   0.11     &   0.56           &    0.15   &     0.05   &     0.43    \\  
$\R$      &  17.55     &   0.71    &   50.19    &    1.19     &   0.10     &   3.42           &    0.49   &     0.02   &     1.40    \\  
$\alpha$  &   3.89     &   0.13    &   11.12    &    0.31     &   0.02     &   0.88           &    0.10   &    $<0.01$   &     0.30    \\  
$\tau$    &  13.57     &   9.49    &   38.81    &    4.71     &   2.92     &  13.48           &    1.81   &     1.28   &     5.18    \\  
\end{tabular}
\end{ruledtabular}
\end{table*}

\begin{table*}
\caption{\label{fmaalwmap4}Fisher matrix analysis results for a standard model with inclusion of reionization (for WMAP best fit model as the fisher analysis fiducial model, $\tau=0.17$): expected $1\sigma$ errors for the WMAP - 4 years experiment. The column {\it marg.} gives the error with all
other parameters being marginalized over; in the column {\it fixed} the other
parameters are held fixed at their ML value; in the column {\it joint} all
parameters are being estimated jointly.}
\begin{ruledtabular}
\begin{tabular}{|c|c c c| c c c|}
Quantity &  \multicolumn{6}{c}{$1\sigma$ errors (\%)} \\  
         & \multicolumn{6}{c}{WMAP - 4 years}  \\ \hline
         & marg.  & fixed   & joint   & marg.  & fixed   & joint  \\ \hline
	 & \multicolumn{3}{c}{Polarization (EE)}   & \multicolumn{3}{c}{Temperature (TT)} \\\hline
$\omega_b$ &    110.64   &    16.58  &    316.44   &        7.33 &    0.81  &     20.96  \\  
$\omega_m$ &     49.48   &    17.16  &    141.52   &        8.91 &    0.77  &     25.49  \\ 
$\omega_\Lambda$  &    622.34   &    97.58  &   1779.93   &      113.30 &   83.39  &    324.06  \\ 
$n_s$     &     69.43   &     4.89  &    198.58   &        6.68 &    0.53  &     19.11  \\ 
$Q$       &     79.22   &    13.51  &    226.58   &        0.90 &    0.32  &      2.58  \\ 
$\R$      &     46.52   &    13.04  &    133.06   &        9.25 &    0.59  &     26.47  \\ 
$\tau$    &    100.84   &     8.21  &    288.40   &      102.72 &   16.70  &    293.79  \\ \hline
& \multicolumn{3}{c}{Temp+Pol (TT+EE)}            &  \multicolumn{3}{c}{All (TT+EE+TE)} \\\hline
$\omega_b$    &    2.14 &    0.80  &      6.11    &       2.13  &   0.80   &     6.08   \\     
$\omega_m$    &    3.09 &    0.77  &      8.85    &       3.08  &   0.77   &     8.81   \\     
$\omega_\Lambda$  &  90.70 &   63.84  &    259.41 &      86.97  &  62.69   &   248.75   \\   
$n_s$     &        1.46 &    0.52  &      4.18    &       1.45  &   0.52   &     4.15   \\      
$Q$       &        0.52 &    0.32  &      1.48    &       0.52  &   0.32   &     1.48   \\
$\R$      &        2.86 &    0.59  &      8.17    &       2.84  &   0.59   &     8.12   \\  
$\tau$    &       10.52 &    7.45  &     30.08    &       10.41  &   7.44   &    29.78   \\ \hline 
\end{tabular}
\end{ruledtabular}
\end{table*}


\begin{table*}
\caption{\label{fmaalwmap4}Fisher matrix analysis results for a model with varying $\alpha$ and inclusion of reionization (for WMAP best fit model as the fisher analysis fiducial model, $\tau=0.17$): expected $1\sigma$ errors for the WMAP - 4 years experiment. The column {\it marg.} gives the error with all
other parameters being marginalized over; in the column {\it fixed} the other
parameters are held fixed at their ML value; in the column {\it joint} all
parameters are being estimated jointly.}
\begin{ruledtabular}
\begin{tabular}{|c|c c c| c c c|}
Quantity &  \multicolumn{6}{c}{$1\sigma$ errors (\%)} \\  
         & \multicolumn{6}{c}{WMAP - 4 years}  \\ \hline
         & marg.  & fixed   & joint   & marg.  & fixed   & joint  \\ \hline
	 & \multicolumn{3}{c}{Polarization (EE)}   & \multicolumn{3}{c}{Temperature (TT)} \\\hline
$\omega_b$     &   173.74   &    16.58  &    496.91    &      14.09   &     0.81   &    40.30 \\ 
$\omega_m$     &   260.62   &    17.16  &    745.40    &      13.76   &     0.77   &    39.36  \\
$\omega_\Lambda$  &   637.28   &    97.58  &   1822.66    &     133.73   &    83.39   &   382.47 \\ 
$n_s$      &   108.18   &     4.89  &    309.41    &       7.86   &     0.53   &    22.47  \\
$Q$        &    96.60   &    13.51  &    276.30    &       2.33   &     0.32   &     6.67  \\
$\R$       &   133.23   &    13.04  &    381.04    &      26.29   &     0.59   &    75.19  \\
$\alpha$   &    69.10   &     2.48  &    197.62    &       5.83   &     0.12   &    16.66  \\
$\tau$     &   228.69   &     8.21  &    654.07    &     103.86   &    16.70   &   297.05  \\ \hline
         &\multicolumn{3}{c}{Temp+Pol (TT+EE)}               & \multicolumn{3}{c}{All (TT+EE+TE)} \\\hline 
$\omega_b$    &   7.50  &      0.80  &     21.44   &        7.41 &       0.80 &      21.18 \\ 
$\omega_m$    &   5.48  &      0.77  &     15.66   &        5.46 &       0.77 &      15.62  \\ 
$\omega_\Lambda$   &  91.57  &     63.84  &    261.91 &    87.48 &      62.69 &     250.20 \\ 
$n_s$      &      2.03  &      0.52  &      5.82   &        2.03 &       0.52 &       5.81   \\
$Q$        &      1.31  &      0.32  &      3.73   &        1.30 &       0.32 &       3.71   \\
$\R$       &     14.34  &      0.59  &     41.01   &       14.17 &       0.59 &      40.53   \\
$\alpha$   &      3.08  &      0.11  &      8.80   &        3.05 &       0.11 &       8.71   \\
$\tau$     &     10.65  &      7.45  &     30.46   &       10.52 &       7.44 &      30.08   \\ \hline
\end{tabular}
\end{ruledtabular}
\end{table*}


\begin{figure*}
\includegraphics[width=3.5in]{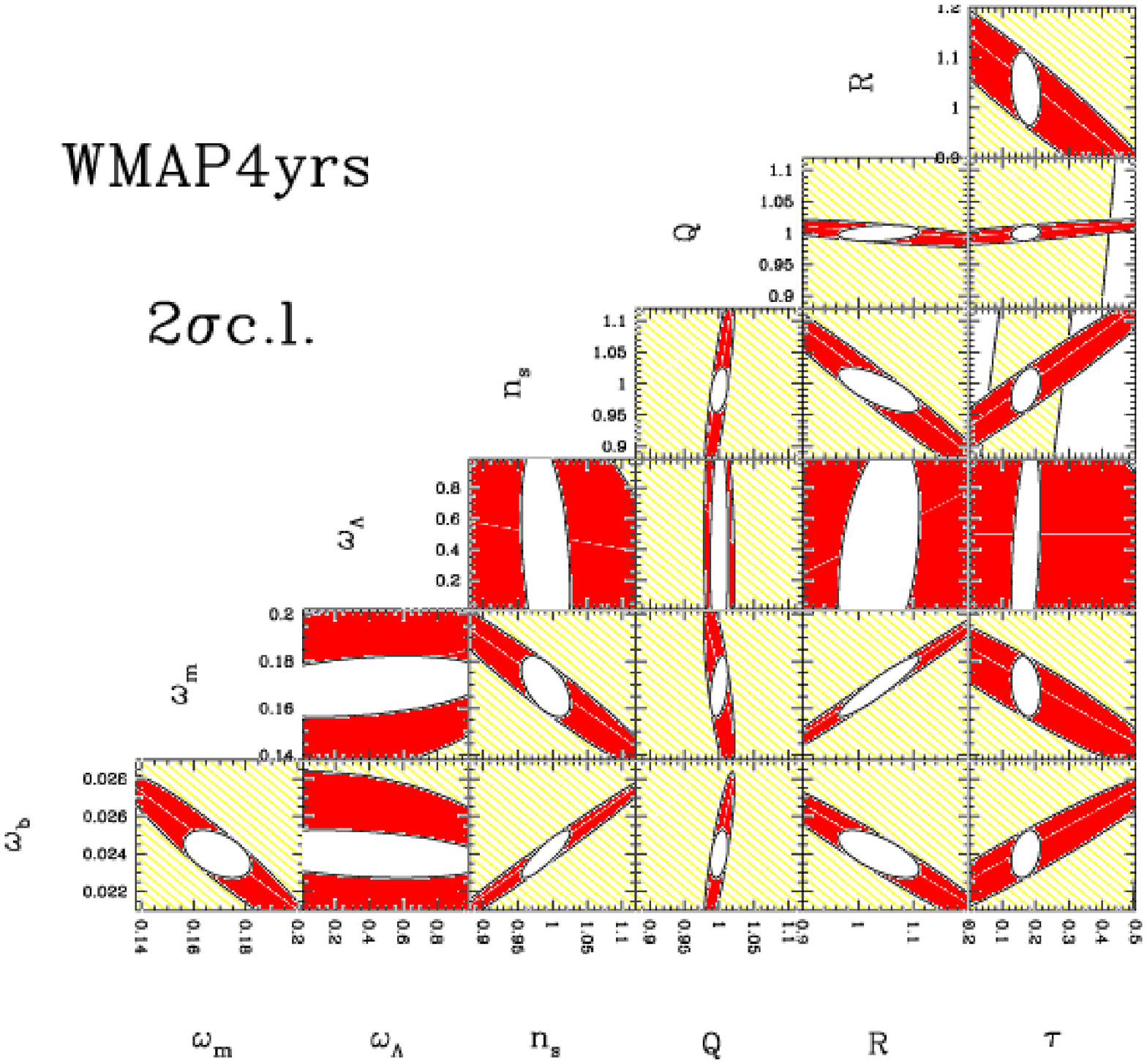}
\includegraphics[width=3.5in]{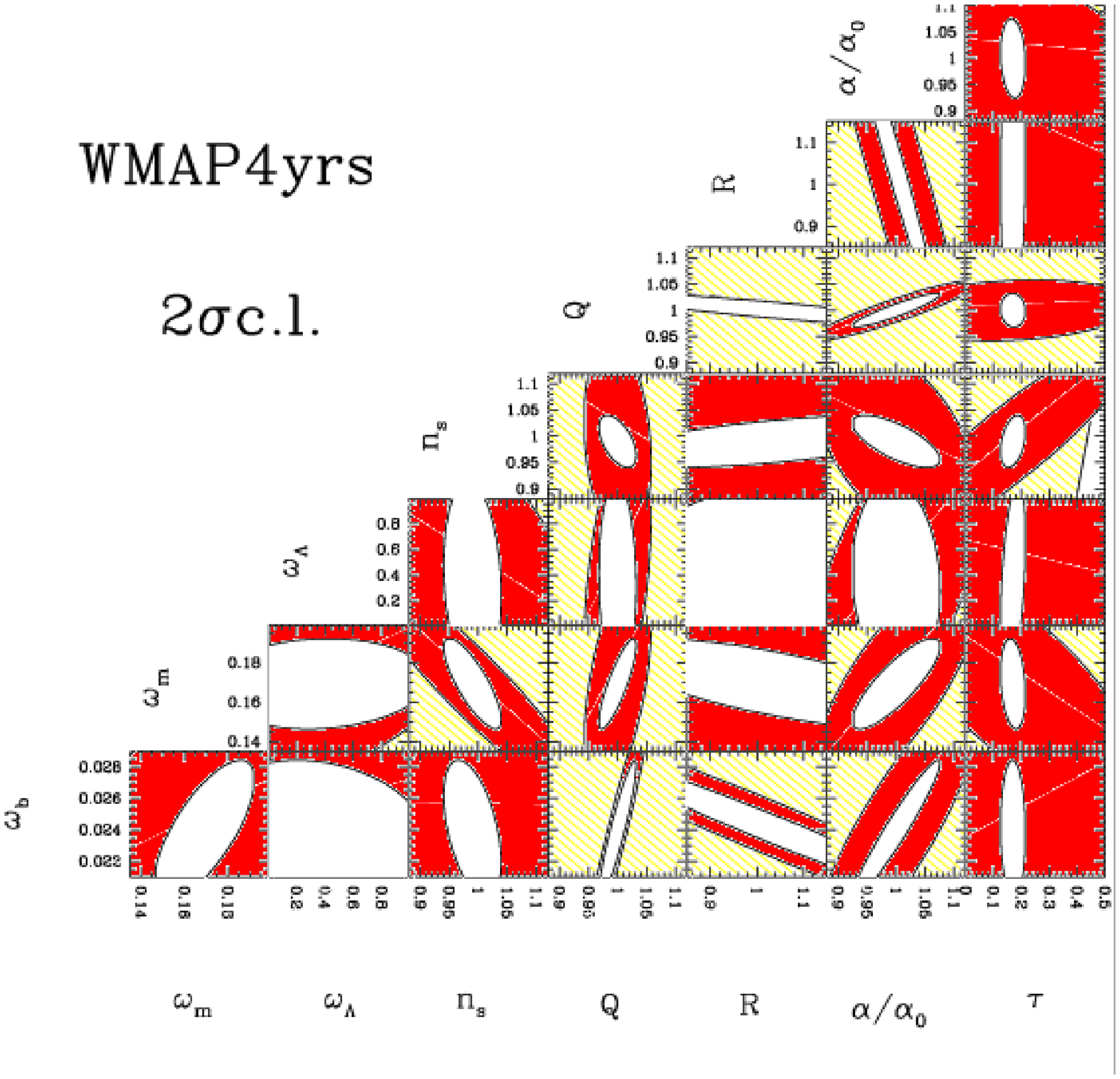}
\caption{\label{figellipse4}Ellipses containing $95.4\%$ ($2\sigma$) of
joint confidence (all other parameters marginalized) for the WMAP - 4years, using temperature alone (red), E-polarization alone (yellow), and both jointly (white).
}
\end{figure*}


\subsection{Analysis results: Principal directions}

Our previous discussion of principal directions changes completely when reionization is included, as 
polarization data helps to better constrain the fine structure constant 
and removes the existing degeneracies between $\alpha$ and $\tau$ see Table \ref{fmaev}

In Table \ref{fmaev} we display the eigenvectors and eigenvalues for WMAP, Planck and a CVL experiment when reionization is included (with $\tau=0.20$) for Temperature and Polarization considered jointly.

 Planck's errors, as measured by the inverse square root of the
eigenvalues, are smaller by a factor of about $5$ on average that those for WMAP.
In the case of a CVL experiment's errors are smaller by a factor of about $3$ on average than those for Planck.

The physical parameter $\tau$ is the largest parameter contribution to the principal direction 6 for both WMAP and Planck, and is the second largest to direction 4 for WMAP and to direction 6 and 8 for Planck. While for a CVL experiment it becomes the main contributor for principal directions 5 and 6 and the second largest for direction 8.
For 4 of the 8
eigenvectors Planck obtains a better alignment of the principal
directions with the axis of the physical parameters when compared with WMAP.
This indicates that the inclusion of the reionization parameter $\tau$ already helps to break degeneracies for WMAP, when we compare the number 4 in 8 against 6 in 7 for the case without reionization.
On the other hand, only for 4 of the 8 eigenvectors CVL obtains a better alignment of the principal
directions with the axis of the physical parameters when compared with Planck, against 6 in 7 for the case without reionization.

The physical parameter $\alpha$ is the second largest contributor for principal directions 2, 5 and 7 for WMAP and for direction 4 for Planck being the main contributor for direction 3. For a CVL experiment it becomes the
 second largest for directions 1 and 7 and main contributor for direction 3 just like for Planck.

In Table \ref{fmaevTT} we display the principal directions considering Temperature only. 
Comparing Tables \ref{fmaevTT} and \ref{fmaev}, we conclude that for 4 of the 8 eigenvectors WMAP obtains a better alignment of the principal
directions with the axis of the physical parameters when polarization is included (with similar alignement for the others).
While for Planck and a CVL experiment only for 3 of the 8 eigenvectors the alignement is better (with similar alignement for the others).
When polarization is included, the largest and second largest physical parameter contributors remain the same for all but for directions 6 and 7 for WMAP, directions 2,3,6 and 7 for Planck, and directions 1,4,6 and 7 for a CVL experiment.
For Planck for the case with temperature only, the second largest contributor to direction 2 and 6, the physical parameter $\alpha$ is shifted to $\omega_{m}$  and $n_{s}$ respectively while direction 3 becomes mainly contributed by $\alpha$, when polarization is included. This indicates that when including polarization the degeneracies with $\alpha$ are indeed being broken.
The CVL case shows that the changes ocurring with inclusion of Polarization when reionization is considered is not a simple rescaling of contributions from the physical parameters to the principal directions but a rescaling by different factors for each of these physical parameters resulting in changes of the degeneracy directions. To demonstrate that this is indeed the case let us analyse both Tables in detail for the CVL case. 
For instance direction 1 remains unchanged with inclusion of polarization, while for direction 2 $\alpha$ is the third contributor by an amount similar to $\omega_{m}$ but is much reduced when polarization is included. Also this direction is better aligned with $\omega_{b}$ when temperature and polarization are considered jointly. This indicates that inclusion of polarization helped to break the degeneracy between $\omega_{b}$ and $\alpha$. Direction 3 remains aligned with $\omega_{\Lambda}$ for both cases.
Direction 5 exchanges the largest and second largest contributions from $n_{s}$ to $\tau$ when polarization is included in the analysis reducing the contribution from $\alpha$. So the inclusion of polarization helps to better define a direction of degeneracy between $\tau$ and $n_{s}$ by breaking the degeneracy with $\alpha$. The degeneracy between $Q$ and $\alpha$ is also broken by shifting the second largest contributor to direction 4 from $\alpha$ to $\omega_{m}$. 
The second largest contributor to direction 7 is shifted from $n_{s}$ to $\alpha$ when polarization is included indicating that the  degeneracy between $\mathcal R$ and $\alpha$ is now dominating over the other degeneracies with $\alpha$.


\begin{table}
\caption{\label{fmaresults}Fisher matrix analysis results for a
model with varying $\alpha$ and reionization: expected $1\sigma$
errors for the Planck satellite and for the CVL experiment (see
the text for details). The column {\it marg.} gives the error with
all other parameters being marginalized over; in the column {\it
fixed} the other parameters are held fixed at their ML value; in
the column {\it joint} all parameters are being estimated
jointly.}
\begin{ruledtabular}
\begin{tabular}{|c| c c c|c c c|}
 &  \multicolumn{6}{c}{$1\sigma$ errors (\%)} \\\hline
             & \multicolumn{3}{c}{Planck HFI} & \multicolumn{3}{c}{CVL} \\
             & marg.  & fixed & joint  & marg.  & fixed & joint           \\\hline
     & \multicolumn{6}{c}{E-Polarization Only (EE)} \\\hline
$\alpha$       &2.66       &0.06       &7.62     &0.40        &$<0.01$        &1.14 \\
$\tau$         &8.81       &2.78       &25.19    &2.26 &1.52 &6.45
\\ \hline
     & \multicolumn{6}{c}{Temperature Only (TT)} \\\hline
$\alpha$       &0.66       &0.02       &1.88     &0.41        &0.01        &1.18  \\
$\tau$         &26.93      &8.28       &77.02    &20.32 &5.89
&58.11 \\ \hline
     & \multicolumn{6}{c}{Temperature + Polarization (TT+EE)} \\\hline
$\alpha$       &0.34        &0.02      &0.97     &0.11        &$<0.01$        &0.32 \\
$\tau$         &4.48        &2.65      &12.80    &1.80       &1.48        &5.15 \\
\end{tabular}
\end{ruledtabular}
\end{table}

\begin{figure}
\includegraphics[width=3.5in]{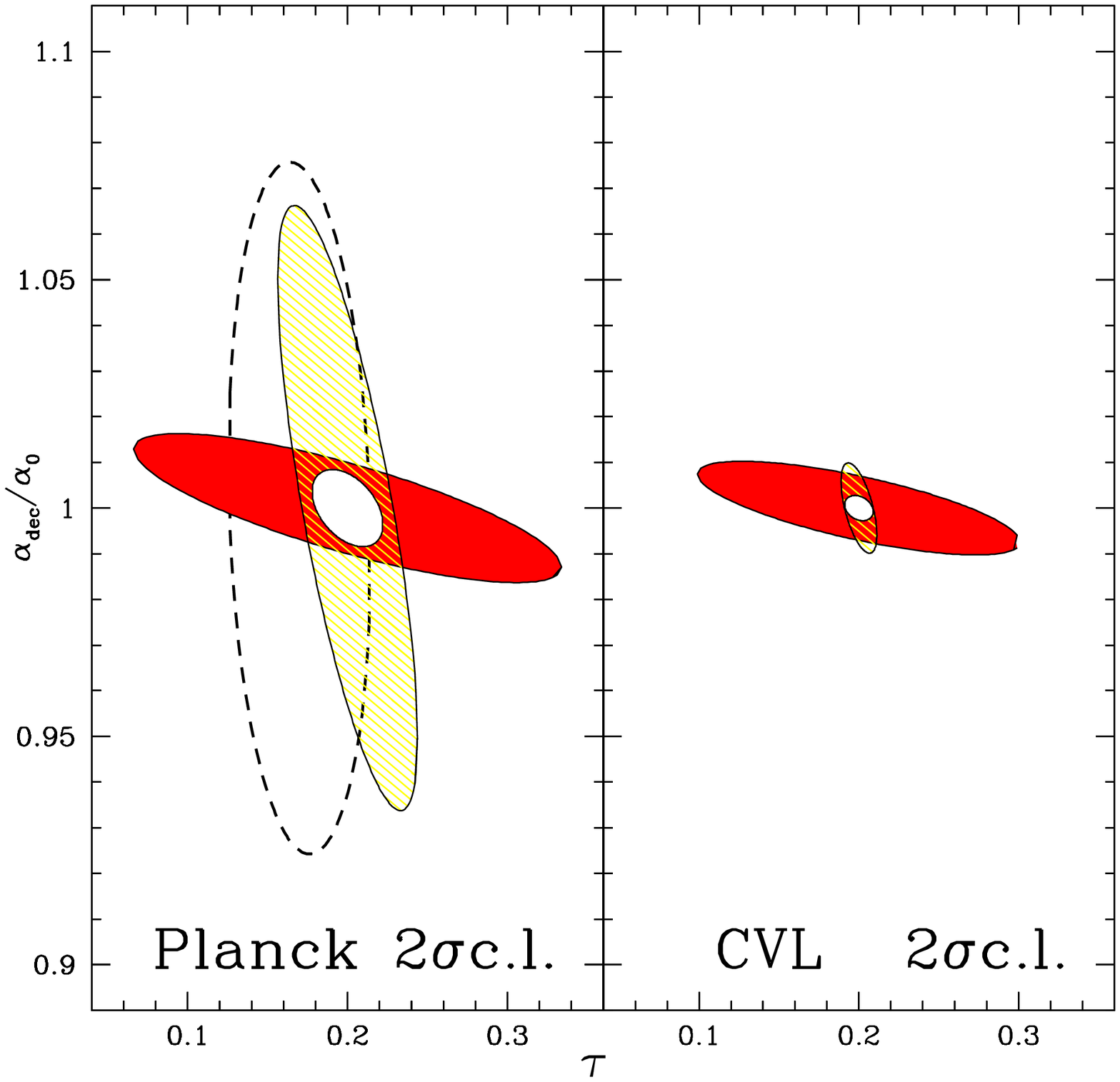}
\caption{\label{figlike} Ellipses containing $95.4\%$ ($2\sigma$)
of joint confidence in the $\alpha$ vs. $\tau$ plane (all other
parameters marginalized), for the Planck and cosmic variance
limited (CVL) experiments, using temperature alone (red),
E-polarization alone (yellow), and both jointly (white). The dashed contour represents the WMAP - 4years forecast using (TT+EE+TE) jointly. }
\end{figure}


\subsection{The $\alpha$-$\tau$ degeneracy}

Our results clearly indicate a crucial degeneracy between $\alpha$ and
$\tau$. In order to study it in more detail, we have extracted the
relevant results from Table \ref{fmaaltau} and Fig. \ref{figellipse4}
and re-displayed them in  Table \ref{fmaresults} and Fig. \ref{figlike}.
Both of these summarize the forecasts for the precision in determining both parameters with Planck and the CVL experiment. 

It is apparent from Fig. \ref{figlike} that TT and EE
suffer from degeneracies in different directions, for the
reasons explained above. Thus combining high-precision
temperature and polarization measurements one can constrain
most effectively constrain both variations of $\alpha$ and $\tau$.
Planck will be essentially cosmic variance limited for temperature but
there will still be considerable room for improvement in
polarization. This therefore argues for a
post-Planck polarization-dedicated experiment, not least because
polarization is, in itself, better at determining cosmological
parameters than temperature.

We conclude that Planck data alone will be
able to constrain variations of $\alpha$ at the epoch of
decoupling with 0.34 \% accuracy ($1\sigma$, all other parameters
marginalized), which corresponds to approximately a factor 5
improvement on the current upper bound. On the other hand, the CMB
\textit{alone} can only constrain variations of $\alpha$ up to
${\cal O}(10^{-3})$ at $z \sim 1100$. Going beyond this limit
will require additional (non-CMB) priors on some of the other
cosmological parameters. This result is to be
contrasted with the variation measured in quasar absorption
systems by Ref.\cite{Webb}, $\delta \alpha / \alpha_0 = {\cal
O}(10^{-5})$ at $z \sim 2$. Nevertheless, there are models
where deviations from the present value could be detected
using the CMB.

\section{Conclusions}

We have presented a detailed analysis of the current WMAP constraints on
the value of the fine-structure constant $\alpha$ at decoupling. We have
found that current constraints on $\alpha$, coming from WMAP alone, are as
strong as all previously existing cosmological constraints (CMB combined
with additional data, e.g. coming from type Ia supernovae or the HST Key
project) put together. On the other hand, we have also shown that the CMB
\textit{alone} can determine $\alpha$ to a maximum accuracy of $0.1\%$ -
one can only improve on this number by again combining CMB data with other
observables. Note that such combination of datasets is not without its
subtleties---see \cite{Martins} for a discussion of some specific issues
related to this case.

Hence this accuracy is well below the $10^{-5}$ detection of Webb
\textit{et al.} \cite{Webb}. However one must keep in mind that one is
dealing with much higher redshifts (about one thousand rather than a few).
Given that in the simplest, best motivated models for $\alpha$ variation,
one expects it to be a non-decreasing function of time, one finds that a
constraint of $10^{-3}$ at the epoch of decoupling can be as constraining
for these models as the Webb \textit{et al.} results. In addition, there
are also constraints on variations of $\alpha$ at the epoch of
nucleosynthesis, which are at the level of $10^{-2}$ \cite{Avelino}. The
main difference between them is that while CMB constraints are model
independent, the BBN ones are not (they rely on the assumption of the
Gasser-Leutwyler phenomenological formula for the dependence of the
neutron-proton mass difference on $\alpha$).

As discussed in the main text, we focused our analysis on model
independent constraints, and in fact explicitly avoided discussing
constraints for specific models. Nevertheless it is quite easy, given the
constraints (and forecasts) presented here, to translate them into
constraints for the specific free parameters of one's preferred model.

We have also presented a thorough analysis of future CMB constraints on
$\alpha$ and the other cosmological parameters, specifically for the WMAP
and Planck Surveyor satellites, and compared them to those for an ideal
(cosmic variance limited) experiment. Comparisons with currently published
(1 year) WMAP data indicates that our Fisher Matrix Analysis pipeline is
quantitatively robust and accurate.

By separately studying the temperature and polarization channels, we have
explicitly shown that the degeneracy directions can be quite different in
the two cases, and hence that by combining them many such degeneracies can
be broken. We have also shown that in the ideal case CMB (EE) polarization
is a much more accurate estimator of cosmological parameters than CMB
temperature.

Nevertheless, polarization measurements are much harder to do in practice.
For example, for the case of WMAP the (EE) channel will provide a quite
modest contribution for the overall parameter estimation analysis. This
situation is quite different for Planck: here the contributions of the
temperature and polarization channels are quite similar. In fact we have
also shown that Planck's temperature measurements will be almost cosmic
variance limited, while its polarization measurements will be well below
this ideal limit. (This fact was previously known, but it had never been
quantified as was done in the present paper.) Hence this, together with
the fact that polarization is intrinsically superior for the purpose of
cosmological parameter estimation, make a strong case for a post-Planck,
polarization-dedicated experiment.

Our analysis can readily be repeated for other experiments. It should be
particularly enlightening to study cases of interferometer experiments and
compare them with the WMAP and Planck satellites. On the other hand it
would also be possible to extend it to include gravity waves,
iso-curvature modes, or a dark energy component different from a
cosmological constant. However, none of these is currently required by
existing (CMB and other) data, and the latter two are in fact strongly
constrained.

To conclude, the prospects of further constraining $\alpha$ at high
redshift are definitely bright. In addition, further progress is expected
at low redshift, where at least three (to our knowledge) independent
groups are currently trying to confirm the Webb \textit{et al.}
\cite{Webb,Jenam,Murphy} claimed detection of a smaller $\alpha$. All of
these are using VLT data, while the original work \cite{Webb,Jenam,Murphy}
used Keck data. This alone will provide an important test of the
systematics of the pipeline, plus in addition the three groups are using
quite different methods. These and other completely new methods that may
be devised thus offer the real prospect of an accurate mapping of the
cosmological evolution of the fine-structure constant, $\alpha=\alpha(z)$.

Finally, a point which we have not discussed at all for reasons of space,
but which should be kept in mind in the context of forthcoming
experiments, is that any time variation of $\alpha$ will be related (in a
model-dependent way) to violations of the Einstein Equivalence principle
\cite{Will}. Thus a strong experimental and/or observational confirmation
of either of them will have revolutionary implications not just for
cosmology but for physics as a whole.


\section{acknowledgments}
We would like to thank Anthony Challinor and Anthony Lasenby 
for useful discussions.
G.R. acknowledges a Leverhulme Fellowship at the University of Cambridge, R.T. is partially supported by the Swiss 
National Science Foundation and the Schmidheiny Foundation and 
C.M. is funded by FCT (Portugal), under grant FMRH/BPD/1600/2000.
This work was done in the context of the European network CMBnet, and
was performed on COSMOS, the Origin3800 owned by the UK
Computational Cosmology Consortium, supported by Silicon Graphics/Cray
Research, HEFCE and PPARC.

\bibliography{paper}

\begin{thebibliography}{64}
\expandafter\ifx\csname natexlab\endcsname\relax\def\natexlab#1{#1}\fi
\expandafter\ifx\csname bibnamefont\endcsname\relax
  \def\bibnamefont#1{#1}\fi
\expandafter\ifx\csname bibfnamefont\endcsname\relax
  \def\bibfnamefont#1{#1}\fi
\expandafter\ifx\csname citenamefont\endcsname\relax
  \def\citenamefont#1{#1}\fi
\expandafter\ifx\csname url\endcsname\relax
  \def\url#1{\texttt{#1}}\fi
\expandafter\ifx\csname urlprefix\endcsname\relax\def\urlprefix{URL }\fi
\providecommand{\bibinfo}[2]{#2}
\providecommand{\eprint}[2][]{\url{#2}}

\bibitem[{\citenamefont{Bennett et~al.}(2003)}]{Bennett}
\bibinfo{author}{\bibfnamefont{C.~L.} \bibnamefont{Bennett}}
  \bibnamefont{et~al.} (\bibinfo{year}{2003}), \eprint{astro-ph/0302207}.

\bibitem[{\citenamefont{Hinshaw et~al.}(2003)}]{Hinshaw}
\bibinfo{author}{\bibfnamefont{G.}~\bibnamefont{Hinshaw}} \bibnamefont{et~al.}
  (\bibinfo{year}{2003}), \eprint{astro-ph/0302217}.

\bibitem[{\citenamefont{Kogut et~al.}(2003)}]{Kogut}
\bibinfo{author}{\bibfnamefont{A.}~\bibnamefont{Kogut}} \bibnamefont{et~al.}
  (\bibinfo{year}{2003}), \eprint{astro-ph/0302213}.

\bibitem[{\citenamefont{Verde et~al.}(2003)}]{Verde}
\bibinfo{author}{\bibfnamefont{L.}~\bibnamefont{Verde}} \bibnamefont{et~al.}
  (\bibinfo{year}{2003}), \eprint{astro-ph/0302218}.

\bibitem[{\citenamefont{Polchinski}(1998)}]{Polchinski}
\bibinfo{author}{\bibfnamefont{J.}~\bibnamefont{Polchinski}}
  (\bibinfo{year}{1998}), \bibinfo{note}{{ }Cambridge, U.K.: University Press}.

\bibitem[{\citenamefont{Damour}(2003{\natexlab{a}})}]{Damour1}
\bibinfo{author}{\bibfnamefont{T.}~\bibnamefont{Damour}},
  \bibinfo{journal}{Astrophys. Space Sci.} \textbf{\bibinfo{volume}{283}},
  \bibinfo{pages}{445} (\bibinfo{year}{2003}{\natexlab{a}}),
  \eprint{gr-qc/0210059}.

\bibitem[{\citenamefont{Will}(2001)}]{Will}
\bibinfo{author}{\bibfnamefont{C.~M.} \bibnamefont{Will}},
  \bibinfo{journal}{Living Rev. Rel.} \textbf{\bibinfo{volume}{4}},
  \bibinfo{pages}{4} (\bibinfo{year}{2001}), \eprint{gr-qc/0103036}.

\bibitem[{\citenamefont{Martins}(2002)}]{Essay}
\bibinfo{author}{\bibfnamefont{C.~J. A.~P.} \bibnamefont{Martins}},
  \bibinfo{journal}{Phil. Trans. Roy. Soc. Lond.}
  \textbf{\bibinfo{volume}{A360}}, \bibinfo{pages}{2681}
  (\bibinfo{year}{2002}), \eprint{astro-ph/0205504}.

\bibitem[{\citenamefont{Uzan}(2003)}]{Uzan}
\bibinfo{author}{\bibfnamefont{J.-P.} \bibnamefont{Uzan}},
  \bibinfo{journal}{Rev. Mod. Phys.} \textbf{\bibinfo{volume}{75}},
  \bibinfo{pages}{403} (\bibinfo{year}{2003}), \eprint{hep-ph/0205340}.

\bibitem[{\citenamefont{Webb et~al.}(2001)}]{Webb}
\bibinfo{author}{\bibfnamefont{J.~K.} \bibnamefont{Webb}} \bibnamefont{et~al.},
  \bibinfo{journal}{Phys. Rev. Lett.} \textbf{\bibinfo{volume}{87}},
  \bibinfo{pages}{091301} (\bibinfo{year}{2001}), \eprint{astro-ph/0012539}.

\bibitem[{\citenamefont{Webb et~al.}(2003)\citenamefont{Webb, Murphy, Flambaum,
  and Curran}}]{Jenam}
\bibinfo{author}{\bibfnamefont{J.~K.} \bibnamefont{Webb}},
  \bibinfo{author}{\bibfnamefont{M.~T.} \bibnamefont{Murphy}},
  \bibinfo{author}{\bibfnamefont{V.~V.} \bibnamefont{Flambaum}},
  \bibnamefont{and} \bibinfo{author}{\bibfnamefont{S.~J.}
  \bibnamefont{Curran}}, \bibinfo{journal}{Astrophys. J. Supp.}
  \textbf{\bibinfo{volume}{283}}, \bibinfo{pages}{565} (\bibinfo{year}{2003}),
  \eprint{astro-ph/0210531}.

\bibitem[{\citenamefont{Murphy et~al.}(2003)\citenamefont{Murphy, Webb, and
  Flambaum}}]{Murphy}
\bibinfo{author}{\bibfnamefont{M.~T.} \bibnamefont{Murphy}},
  \bibinfo{author}{\bibfnamefont{J.~K.} \bibnamefont{Webb}}, \bibnamefont{and}
  \bibinfo{author}{\bibfnamefont{V.~V.} \bibnamefont{Flambaum}}
  (\bibinfo{year}{2003}), \eprint{astro-ph/0306483}.

\bibitem[{\citenamefont{Ivanchik et~al.}(2003)\citenamefont{Ivanchik,
  Petitjean, Rodriguez, and Varshalovich}}]{Ivanchik}
\bibinfo{author}{\bibfnamefont{A.}~\bibnamefont{Ivanchik}},
  \bibinfo{author}{\bibfnamefont{P.}~\bibnamefont{Petitjean}},
  \bibinfo{author}{\bibfnamefont{E.}~\bibnamefont{Rodriguez}},
  \bibnamefont{and}
  \bibinfo{author}{\bibfnamefont{D.}~\bibnamefont{Varshalovich}},
  \bibinfo{journal}{Astrophys. Space Sci.} \textbf{\bibinfo{volume}{283}},
  \bibinfo{pages}{583} (\bibinfo{year}{2003}), \eprint{astro-ph/0210299}.

\bibitem[{\citenamefont{Damour}(2003{\natexlab{b}})}]{Damour2}
\bibinfo{author}{\bibfnamefont{T.}~\bibnamefont{Damour}}
  (\bibinfo{year}{2003}{\natexlab{b}}), \eprint{gr-qc/0306023}.

\bibitem[{\citenamefont{Fujii}(2003)}]{Fujii}
\bibinfo{author}{\bibfnamefont{Y.}~\bibnamefont{Fujii}},
  \bibinfo{journal}{Astrophys. Space Sci.} \textbf{\bibinfo{volume}{283}},
  \bibinfo{pages}{559} (\bibinfo{year}{2003}), \eprint{gr-qc/0212017}.

\bibitem[{\citenamefont{Marion et~al.}(2003)}]{Marion}
\bibinfo{author}{\bibfnamefont{H.}~\bibnamefont{Marion}} \bibnamefont{et~al.},
  \bibinfo{journal}{Phys. Rev. Lett.} \textbf{\bibinfo{volume}{90}},
  \bibinfo{pages}{150801} (\bibinfo{year}{2003}), \eprint{physics/0212112}.

\bibitem[{\citenamefont{Damour and Nordtvedt}(1993)}]{Damour}
\bibinfo{author}{\bibfnamefont{T.}~\bibnamefont{Damour}} \bibnamefont{and}
  \bibinfo{author}{\bibfnamefont{K.}~\bibnamefont{Nordtvedt}},
  \bibinfo{journal}{Phys. Rev.} \textbf{\bibinfo{volume}{D48}},
  \bibinfo{pages}{3436} (\bibinfo{year}{1993}).

\bibitem[{\citenamefont{Santiago et~al.}(1998)\citenamefont{Santiago, Kalligas,
  and Wagoner}}]{Santiago}
\bibinfo{author}{\bibfnamefont{D.~I.} \bibnamefont{Santiago}},
  \bibinfo{author}{\bibfnamefont{D.}~\bibnamefont{Kalligas}}, \bibnamefont{and}
  \bibinfo{author}{\bibfnamefont{R.~V.} \bibnamefont{Wagoner}},
  \bibinfo{journal}{Phys. Rev.} \textbf{\bibinfo{volume}{D58}},
  \bibinfo{pages}{124005} (\bibinfo{year}{1998}), \eprint{gr-qc/9805044}.

\bibitem[{\citenamefont{Barrow et~al.}(2002)\citenamefont{Barrow, Sandvik, and
  Magueijo}}]{Barrow}
\bibinfo{author}{\bibfnamefont{J.~D.} \bibnamefont{Barrow}},
  \bibinfo{author}{\bibfnamefont{H.~B.} \bibnamefont{Sandvik}},
  \bibnamefont{and} \bibinfo{author}{\bibfnamefont{J.}~\bibnamefont{Magueijo}},
  \bibinfo{journal}{Phys. Rev.} \textbf{\bibinfo{volume}{D65}},
  \bibinfo{pages}{063504} (\bibinfo{year}{2002}), \eprint{astro-ph/0109414}.

\bibitem[{\citenamefont{Sigurdson et~al.}(2003)\citenamefont{Sigurdson,
  Kurylov, and Kamionkowski}}]{Sigu}
\bibinfo{author}{\bibfnamefont{K.}~\bibnamefont{Sigurdson}},
  \bibinfo{author}{\bibfnamefont{A.}~\bibnamefont{Kurylov}}, \bibnamefont{and}
  \bibinfo{author}{\bibfnamefont{M.}~\bibnamefont{Kamionkowski}}
  (\bibinfo{year}{2003}), \eprint{astro-ph/0306372}.

\bibitem[{\citenamefont{Avelino
  et~al.}(2000{\natexlab{a}})\citenamefont{Avelino, Martins, Rocha, and
  Viana}}]{Old}
\bibinfo{author}{\bibfnamefont{P.~P.} \bibnamefont{Avelino}},
  \bibinfo{author}{\bibfnamefont{C.~J. A.~P.} \bibnamefont{Martins}},
  \bibinfo{author}{\bibfnamefont{G.}~\bibnamefont{Rocha}}, \bibnamefont{and}
  \bibinfo{author}{\bibfnamefont{P.}~\bibnamefont{Viana}},
  \bibinfo{journal}{Phys. Rev.} \textbf{\bibinfo{volume}{D62}},
  \bibinfo{pages}{123508} (\bibinfo{year}{2000}{\natexlab{a}}),
  \eprint{astro-ph/0008446}.

\bibitem[{\citenamefont{Avelino et~al.}(2001)}]{Avelino}
\bibinfo{author}{\bibfnamefont{P.~P.} \bibnamefont{Avelino}}
  \bibnamefont{et~al.}, \bibinfo{journal}{Phys. Rev.}
  \textbf{\bibinfo{volume}{D64}}, \bibinfo{pages}{103505}
  (\bibinfo{year}{2001}), \eprint{astro-ph/0102144}.

\bibitem[{\citenamefont{Martins et~al.}(2002)}]{Martins}
\bibinfo{author}{\bibfnamefont{C.~J. A.~P.} \bibnamefont{Martins}}
  \bibnamefont{et~al.}, \bibinfo{journal}{Phys. Rev.}
  \textbf{\bibinfo{volume}{D66}}, \bibinfo{pages}{023505}
  (\bibinfo{year}{2002}), \eprint{astro-ph/0203149}.

\bibitem[{\citenamefont{Martins et~al.}(2003)}]{Martinsw}
\bibinfo{author}{\bibfnamefont{C.~J. A.~P.} \bibnamefont{Martins}}
  \bibnamefont{et~al.} (\bibinfo{year}{2003}), \eprint{astro-ph/0302295}.

\bibitem[{\citenamefont{Zaldarriaga and Seljak}(1997)}]{zaldarriaga1}
\bibinfo{author}{\bibfnamefont{M.}~\bibnamefont{Zaldarriaga}} \bibnamefont{and}
  \bibinfo{author}{\bibfnamefont{U.}~\bibnamefont{Seljak}},
  \bibinfo{journal}{Phys. Rev.} \textbf{\bibinfo{volume}{D55}},
  \bibinfo{pages}{1830} (\bibinfo{year}{1997}), \eprint{astro-ph/9609170}.

\bibitem[{\citenamefont{Kosowsky}(1996)}]{kosowsky}
\bibinfo{author}{\bibfnamefont{A.}~\bibnamefont{Kosowsky}},
  \bibinfo{journal}{Ann. Phys.} \textbf{\bibinfo{volume}{246}},
  \bibinfo{pages}{49} (\bibinfo{year}{1996}), \eprint{astro-ph/9501045}.

\bibitem[{\citenamefont{Hu and White}(1997)}]{waynehu1}
\bibinfo{author}{\bibfnamefont{W.}~\bibnamefont{Hu}} \bibnamefont{and}
  \bibinfo{author}{\bibfnamefont{M.~J.} \bibnamefont{White}},
  \bibinfo{journal}{New Astron.} \textbf{\bibinfo{volume}{2}},
  \bibinfo{pages}{323} (\bibinfo{year}{1997}), \eprint{astro-ph/9706147}.

\bibitem[{\citenamefont{Hu}(2003)}]{waynehu2}
\bibinfo{author}{\bibfnamefont{W.}~\bibnamefont{Hu}}, \bibinfo{journal}{Ann.
  Phys.} \textbf{\bibinfo{volume}{303}}, \bibinfo{pages}{203}
  (\bibinfo{year}{2003}), \eprint{astro-ph/0210696}.

\bibitem[{\citenamefont{Seljak}(1997)}]{uros}
\bibinfo{author}{\bibfnamefont{U.}~\bibnamefont{Seljak}},
  \bibinfo{journal}{Astrophys. J.} \textbf{\bibinfo{volume}{482}},
  \bibinfo{pages}{6} (\bibinfo{year}{1997}), \eprint{astro-ph/9608131}.

\bibitem[{\citenamefont{Kamionkowski
  et~al.}(1997{\natexlab{a}})\citenamefont{Kamionkowski, Kosowsky, and
  Stebbins}}]{kamionkowski1}
\bibinfo{author}{\bibfnamefont{M.}~\bibnamefont{Kamionkowski}},
  \bibinfo{author}{\bibfnamefont{A.}~\bibnamefont{Kosowsky}}, \bibnamefont{and}
  \bibinfo{author}{\bibfnamefont{A.}~\bibnamefont{Stebbins}},
  \bibinfo{journal}{Phys. Rev. Lett.} \textbf{\bibinfo{volume}{78}},
  \bibinfo{pages}{2058} (\bibinfo{year}{1997}{\natexlab{a}}),
  \eprint{astro-ph/9609132}.

\bibitem[{\citenamefont{Kamionkowski
  et~al.}(1997{\natexlab{b}})\citenamefont{Kamionkowski, Kosowsky, and
  Stebbins}}]{kamionkowski2}
\bibinfo{author}{\bibfnamefont{M.}~\bibnamefont{Kamionkowski}},
  \bibinfo{author}{\bibfnamefont{A.}~\bibnamefont{Kosowsky}}, \bibnamefont{and}
  \bibinfo{author}{\bibfnamefont{A.}~\bibnamefont{Stebbins}},
  \bibinfo{journal}{Phys. Rev.} \textbf{\bibinfo{volume}{D55}},
  \bibinfo{pages}{7368} (\bibinfo{year}{1997}{\natexlab{b}}),
  \eprint{astro-ph/9611125}.

\bibitem[{\citenamefont{Challinor}(2000)}]{challinor}
\bibinfo{author}{\bibfnamefont{A.}~\bibnamefont{Challinor}},
  \bibinfo{journal}{Phys. Rev.} \textbf{\bibinfo{volume}{D62}},
  \bibinfo{pages}{043004} (\bibinfo{year}{2000}), \eprint{astro-ph/9911481}.

\bibitem[{\citenamefont{Kovac et~al.}(2002)}]{dasi}
\bibinfo{author}{\bibfnamefont{J.}~\bibnamefont{Kovac}} \bibnamefont{et~al.},
  \bibinfo{journal}{Nature} \textbf{\bibinfo{volume}{420}},
  \bibinfo{pages}{772} (\bibinfo{year}{2002}), \eprint{astro-ph/0209478}.

\bibitem[{\citenamefont{Hannestad}(1999)}]{steen}
\bibinfo{author}{\bibfnamefont{S.}~\bibnamefont{Hannestad}},
  \bibinfo{journal}{Phys. Rev.} \textbf{\bibinfo{volume}{D60}},
  \bibinfo{pages}{023515} (\bibinfo{year}{1999}), \eprint{astro-ph/9810102}.

\bibitem[{\citenamefont{Kaplinghat et~al.}(1999)\citenamefont{Kaplinghat,
  Scherrer, and Turner}}]{Kap}
\bibinfo{author}{\bibfnamefont{M.}~\bibnamefont{Kaplinghat}},
  \bibinfo{author}{\bibfnamefont{R.~J.} \bibnamefont{Scherrer}},
  \bibnamefont{and} \bibinfo{author}{\bibfnamefont{M.~S.}
  \bibnamefont{Turner}}, \bibinfo{journal}{Phys. Rev.}
  \textbf{\bibinfo{volume}{D60}}, \bibinfo{pages}{023516}
  (\bibinfo{year}{1999}), \eprint{astro-ph/9810133}.

\bibitem[{\citenamefont{Avelino
  et~al.}(2000{\natexlab{b}})\citenamefont{Avelino, Martins, and Rocha}}]{vsl}
\bibinfo{author}{\bibfnamefont{P.~P.} \bibnamefont{Avelino}},
  \bibinfo{author}{\bibfnamefont{C.~J. A.~P.} \bibnamefont{Martins}},
  \bibnamefont{and} \bibinfo{author}{\bibfnamefont{G.}~\bibnamefont{Rocha}},
  \bibinfo{journal}{Phys. Lett.} \textbf{\bibinfo{volume}{B483}},
  \bibinfo{pages}{210} (\bibinfo{year}{2000}{\natexlab{b}}),
  \eprint{astro-ph/0001292}.

\bibitem[{\citenamefont{Trotta and Hansen}(2003)}]{Trotta:2003xg}
\bibinfo{author}{\bibfnamefont{R.}~\bibnamefont{Trotta}} \bibnamefont{and}
  \bibinfo{author}{\bibfnamefont{S.~H.} \bibnamefont{Hansen}}
  (\bibinfo{year}{2003}), \eprint{astro-ph/0306588}.

\bibitem[{\citenamefont{Battye et~al.}(2001)\citenamefont{Battye, Crittenden,
  and Weller}}]{Battye}
\bibinfo{author}{\bibfnamefont{R.~A.} \bibnamefont{Battye}},
  \bibinfo{author}{\bibfnamefont{R.}~\bibnamefont{Crittenden}},
  \bibnamefont{and} \bibinfo{author}{\bibfnamefont{J.}~\bibnamefont{Weller}},
  \bibinfo{journal}{Phys. Rev.} \textbf{\bibinfo{volume}{D63}},
  \bibinfo{pages}{043505} (\bibinfo{year}{2001}), \eprint{astro-ph/0008265}.

\bibitem[{\citenamefont{Avelino and Liddle}(2003)}]{pedrolidle}
\bibinfo{author}{\bibfnamefont{P.~P.} \bibnamefont{Avelino}} \bibnamefont{and}
  \bibinfo{author}{\bibfnamefont{A.~R.} \bibnamefont{Liddle}}
  (\bibinfo{year}{2003}), \eprint{astro-ph/0305357}.

\bibitem[{\citenamefont{Bruscoli et~al.}(2002)\citenamefont{Bruscoli, Ferrara,
  and Scannapieco}}]{Bruscoli02}
\bibinfo{author}{\bibfnamefont{M.}~\bibnamefont{Bruscoli}},
  \bibinfo{author}{\bibfnamefont{A.}~\bibnamefont{Ferrara}}, \bibnamefont{and}
  \bibinfo{author}{\bibfnamefont{E.}~\bibnamefont{Scannapieco}}
  (\bibinfo{year}{2002}), \eprint{astro-ph/0201094}.

\bibitem[{\citenamefont{Hu and Holder}(2003)}]{Hu03}
\bibinfo{author}{\bibfnamefont{W.}~\bibnamefont{Hu}} \bibnamefont{and}
  \bibinfo{author}{\bibfnamefont{G.~P.} \bibnamefont{Holder}},
  \bibinfo{journal}{Phys. Rev.} \textbf{\bibinfo{volume}{D68}},
  \bibinfo{pages}{023001} (\bibinfo{year}{2003}), \eprint{astro-ph/0303400}.

\bibitem[{\citenamefont{Kaplinghat et~al.}(2003)}]{K03}
\bibinfo{author}{\bibfnamefont{M.}~\bibnamefont{Kaplinghat}}
  \bibnamefont{et~al.}, \bibinfo{journal}{Astrophys. J.}
  \textbf{\bibinfo{volume}{583}}, \bibinfo{pages}{24} (\bibinfo{year}{2003}),
  \eprint{astro-ph/0207591}.

\bibitem[{\citenamefont{Holder et~al.}(2003)\citenamefont{Holder, Haiman,
  Kaplinghat, and Knox}}]{Holder03}
\bibinfo{author}{\bibfnamefont{G.}~\bibnamefont{Holder}},
  \bibinfo{author}{\bibfnamefont{Z.}~\bibnamefont{Haiman}},
  \bibinfo{author}{\bibfnamefont{M.}~\bibnamefont{Kaplinghat}},
  \bibnamefont{and} \bibinfo{author}{\bibfnamefont{L.}~\bibnamefont{Knox}}
  (\bibinfo{year}{2003}), \eprint{astro-ph/0302404}.

\bibitem[{\citenamefont{Kolb and Turner}(1993)}]{kolbturner}
\bibinfo{author}{\bibfnamefont{E.~W.} \bibnamefont{Kolb}} \bibnamefont{and}
  \bibinfo{author}{\bibfnamefont{M.~S.} \bibnamefont{Turner}},
  \bibinfo{journal}{Addison-Wesley Publishing Company}  (\bibinfo{year}{1993}).

\bibitem[{\citenamefont{Mota and Barrow}(2003)}]{Mota}
\bibinfo{author}{\bibfnamefont{D.~F.} \bibnamefont{Mota}} \bibnamefont{and}
  \bibinfo{author}{\bibfnamefont{J.~D.} \bibnamefont{Barrow}}
  (\bibinfo{year}{2003}), \eprint{astro-ph/0306047}.

\bibitem[{\citenamefont{Spergel et~al.}(2003)}]{Spergel}
\bibinfo{author}{\bibfnamefont{D.}~\bibnamefont{Spergel}} \bibnamefont{et~al.}
  (\bibinfo{year}{2003}), \eprint{astro-ph/0302209}.

\bibitem[{\citenamefont{Melchiorri et~al.}(2003)\citenamefont{Melchiorri,
  Mersini, Odman, and Trodden}}]{mmot}
\bibinfo{author}{\bibfnamefont{A.}~\bibnamefont{Melchiorri}},
  \bibinfo{author}{\bibfnamefont{L.}~\bibnamefont{Mersini}},
  \bibinfo{author}{\bibfnamefont{C.~J.} \bibnamefont{Odman}}, \bibnamefont{and}
  \bibinfo{author}{\bibfnamefont{M.}~\bibnamefont{Trodden}},
  \bibinfo{journal}{Phys. Rev.} \textbf{\bibinfo{volume}{D68}},
  \bibinfo{pages}{043509} (\bibinfo{year}{2003}), \eprint{astro-ph/0211522}.

\bibitem[{\citenamefont{Bean et~al.}(2001)\citenamefont{Bean, Hansen, and
  Melchiorri}}]{bhm}
\bibinfo{author}{\bibfnamefont{R.}~\bibnamefont{Bean}},
  \bibinfo{author}{\bibfnamefont{S.~H.} \bibnamefont{Hansen}},
  \bibnamefont{and}
  \bibinfo{author}{\bibfnamefont{A.}~\bibnamefont{Melchiorri}},
  \bibinfo{journal}{Phys. Rev.} \textbf{\bibinfo{volume}{D64}},
  \bibinfo{pages}{103508} (\bibinfo{year}{2001}), \eprint{astro-ph/0104162}.

\bibitem[{\citenamefont{Peiris et~al.}(2003)}]{peiris}
\bibinfo{author}{\bibfnamefont{H.~V.} \bibnamefont{Peiris}}
  \bibnamefont{et~al.} (\bibinfo{year}{2003}), \eprint{astro-ph/0302225}.

\bibitem[{\citenamefont{Kinney et~al.}(2003)\citenamefont{Kinney, Kolb,
  Melchiorri, and Riotto}}]{kkmr}
\bibinfo{author}{\bibfnamefont{W.~H.} \bibnamefont{Kinney}},
  \bibinfo{author}{\bibfnamefont{E.~W.} \bibnamefont{Kolb}},
  \bibinfo{author}{\bibfnamefont{A.}~\bibnamefont{Melchiorri}},
  \bibnamefont{and} \bibinfo{author}{\bibfnamefont{A.}~\bibnamefont{Riotto}}
  (\bibinfo{year}{2003}), \eprint{hep-ph/0305130}.

\bibitem[{\citenamefont{Bean et~al.}(2003)\citenamefont{Bean, Melchiorri, and
  Silk}}]{bms}
\bibinfo{author}{\bibfnamefont{R.}~\bibnamefont{Bean}},
  \bibinfo{author}{\bibfnamefont{A.}~\bibnamefont{Melchiorri}},
  \bibnamefont{and} \bibinfo{author}{\bibfnamefont{J.}~\bibnamefont{Silk}}
  (\bibinfo{year}{2003}), \eprint{astro-ph/0306357}.

\bibitem[{\citenamefont{Fisher}(1935)}]{fisher}
\bibinfo{author}{\bibfnamefont{R.}~\bibnamefont{Fisher}}, \bibinfo{journal}{J.
  Roy. Stat. Soc.} \textbf{\bibinfo{volume}{98}}, \bibinfo{pages}{39}
  (\bibinfo{year}{1935}).

\bibitem[{\citenamefont{Tegmark et~al.}(1997)\citenamefont{Tegmark, Taylor, and
  Heavens}}]{tegmark}
\bibinfo{author}{\bibfnamefont{M.}~\bibnamefont{Tegmark}},
  \bibinfo{author}{\bibfnamefont{A.}~\bibnamefont{Taylor}}, \bibnamefont{and}
  \bibinfo{author}{\bibfnamefont{A.}~\bibnamefont{Heavens}},
  \bibinfo{journal}{Astrophys. J.} \textbf{\bibinfo{volume}{480}},
  \bibinfo{pages}{22T} (\bibinfo{year}{1997}), \eprint{astro-ph/9603021}.

\bibitem[{\citenamefont{Jungman
  et~al.}(1996{\natexlab{a}})\citenamefont{Jungman, Kamionkowski, Kosowsky, and
  Spergel}}]{jungman1}
\bibinfo{author}{\bibfnamefont{G.}~\bibnamefont{Jungman}},
  \bibinfo{author}{\bibfnamefont{M.}~\bibnamefont{Kamionkowski}},
  \bibinfo{author}{\bibfnamefont{A.}~\bibnamefont{Kosowsky}}, \bibnamefont{and}
  \bibinfo{author}{\bibfnamefont{D.~N.} \bibnamefont{Spergel}},
  \bibinfo{journal}{Phys. Rev. Lett.} \textbf{\bibinfo{volume}{76}},
  \bibinfo{pages}{1007} (\bibinfo{year}{1996}{\natexlab{a}}),
  \eprint{astro-ph/9507080}.

\bibitem[{\citenamefont{Jungman
  et~al.}(1996{\natexlab{b}})\citenamefont{Jungman, Kamionkowski, Kosowsky, and
  Spergel}}]{jungman2}
\bibinfo{author}{\bibfnamefont{G.}~\bibnamefont{Jungman}},
  \bibinfo{author}{\bibfnamefont{M.}~\bibnamefont{Kamionkowski}},
  \bibinfo{author}{\bibfnamefont{A.}~\bibnamefont{Kosowsky}}, \bibnamefont{and}
  \bibinfo{author}{\bibfnamefont{D.~N.} \bibnamefont{Spergel}},
  \bibinfo{journal}{Phys. Rev.} \textbf{\bibinfo{volume}{D54}},
  \bibinfo{pages}{1332} (\bibinfo{year}{1996}{\natexlab{b}}),
  \eprint{astro-ph/9512139}.

\bibitem[{\citenamefont{Knox}(1995)}]{knox}
\bibinfo{author}{\bibfnamefont{L.}~\bibnamefont{Knox}}, \bibinfo{journal}{Phys.
  Rev.} \textbf{\bibinfo{volume}{D52}}, \bibinfo{pages}{4307}
  (\bibinfo{year}{1995}), \eprint{astro-ph/9504054}.

\bibitem[{\citenamefont{Zaldarriaga et~al.}(1997)\citenamefont{Zaldarriaga,
  Spergel, and Seljak}}]{zaldarriaga2}
\bibinfo{author}{\bibfnamefont{M.}~\bibnamefont{Zaldarriaga}},
  \bibinfo{author}{\bibfnamefont{D.~N.} \bibnamefont{Spergel}},
  \bibnamefont{and} \bibinfo{author}{\bibfnamefont{U.}~\bibnamefont{Seljak}},
  \bibinfo{journal}{Astrophys. J.} \textbf{\bibinfo{volume}{488}},
  \bibinfo{pages}{1} (\bibinfo{year}{1997}), \eprint{astro-ph/9702157}.

\bibitem[{\citenamefont{Bond et~al.}(1997)\citenamefont{Bond, Efstathiou, and
  Tegmark}}]{bond}
\bibinfo{author}{\bibfnamefont{J.~R.} \bibnamefont{Bond}},
  \bibinfo{author}{\bibfnamefont{G.}~\bibnamefont{Efstathiou}},
  \bibnamefont{and} \bibinfo{author}{\bibfnamefont{M.}~\bibnamefont{Tegmark}},
  \bibinfo{journal}{Mon. Not. Roy. Astron. Soc.}
  \textbf{\bibinfo{volume}{291}}, \bibinfo{pages}{L33} (\bibinfo{year}{1997}),
  \eprint{astro-ph/9702100}.

\bibitem[{\citenamefont{Efstathiou and Bond}(1999)}]{efstathiou1}
\bibinfo{author}{\bibfnamefont{G.}~\bibnamefont{Efstathiou}} \bibnamefont{and}
  \bibinfo{author}{\bibfnamefont{J.~R.} \bibnamefont{Bond}},
  \bibinfo{journal}{MNRAS} \textbf{\bibinfo{volume}{304}}, \bibinfo{pages}{75}
  (\bibinfo{year}{1999}), \eprint{astro-ph/9807103}.

\bibitem[{\citenamefont{Efstathiou}(2002)}]{efstathiou2}
\bibinfo{author}{\bibfnamefont{G.}~\bibnamefont{Efstathiou}},
  \bibinfo{journal}{MNRAS} \textbf{\bibinfo{volume}{332}}, \bibinfo{pages}{193}
  (\bibinfo{year}{2002}), \eprint{astro-ph/0109151}.

\bibitem[{\citenamefont{Melchiorri and Griffiths}(2001)}]{melch:01}
\bibinfo{author}{\bibfnamefont{A.}~\bibnamefont{Melchiorri}} \bibnamefont{and}
  \bibinfo{author}{\bibfnamefont{L.~M.} \bibnamefont{Griffiths}},
  \bibinfo{journal}{New Astron. Rev.} \textbf{\bibinfo{volume}{45}},
  \bibinfo{pages}{321} (\bibinfo{year}{2001}), \eprint{astro-ph/0011147}.

\bibitem[{\citenamefont{Bowen et~al.}(2002)\citenamefont{Bowen, Hansen,
  Melchiorri, Silk, and Trotta}}]{Bowen:01}
\bibinfo{author}{\bibfnamefont{R.}~\bibnamefont{Bowen}},
  \bibinfo{author}{\bibfnamefont{S.~H.} \bibnamefont{Hansen}},
  \bibinfo{author}{\bibfnamefont{A.}~\bibnamefont{Melchiorri}},
  \bibinfo{author}{\bibfnamefont{J.}~\bibnamefont{Silk}}, \bibnamefont{and}
  \bibinfo{author}{\bibfnamefont{R.}~\bibnamefont{Trotta}},
  \bibinfo{journal}{Mon. Not. Roy. Astron. Soc.}
  \textbf{\bibinfo{volume}{334}}, \bibinfo{pages}{760} (\bibinfo{year}{2002}),
  \eprint{astro-ph/0110636}.

\bibitem[{\citenamefont{Hu and Sugiyama}(1995)}]{HuandSugiyama}
\bibinfo{author}{\bibfnamefont{W.}~\bibnamefont{Hu}} \bibnamefont{and}
  \bibinfo{author}{\bibfnamefont{N.}~\bibnamefont{Sugiyama}},
  \bibinfo{journal}{Phys. Rev.} \textbf{\bibinfo{volume}{D51}},
  \bibinfo{pages}{2599} (\bibinfo{year}{1995}), \eprint{astro-ph/9411008}.

\bibitem[{\citenamefont{Press et~al.}(1992)}]{Numerical:92}
\bibinfo{author}{\bibfnamefont{W.~H.} \bibnamefont{Press}} \bibnamefont{et~al.}
  (\bibinfo{year}{1992}).

\end{thebibliography}

\end{document}